\documentclass[]{aiaa-tc}

\usepackage{amsmath}
\usepackage{amssymb}
\usepackage{bm}
\usepackage{mathrsfs}
\usepackage{verbatim}
\usepackage{subfigure}
\usepackage{placeins}
\usepackage{overpic}
\usepackage{multicol}

\usepackage[pdftex]{hyperref}
\hypersetup{
pdfauthor   = {},
pdftitle    = {},
pdfsubject  = {},
pdfkeywords = {},
pdfcreator  = {PDFLaTeX with hyperref package},
pdfproducer = {pdflatex},
colorlinks  = true,
linkcolor   = blue,
anchorcolor = red,
citecolor   = blue,
filecolor   = red,
urlcolor    = red
}

\title{Influence of Different Subgrid Scale Models in LES of Supersonic Jet Flows}

\author{
Carlos Junqueira-Junior\thanks{Postdoctoral Research Fellow, 
Aerodynamics Division, Departamento de Ci\^{e}ncia e Tecnologia 
Aeroespacial, DCTA/IAE/ALA;
E-mail: junior.hmg@gmail.com.}\\
{\normalsize\itshape Instituto de Aeron\'{a}utica e Espa\c{c}o, 12228-904
S\~{a}o Jos\'{e} dos Campos, SP, Brazil}\\
\and
Sami Yamouni\thanks{Postdoctoral Reasearch Fellow,
Graduate Program on Computer Sciences and Electrical Engineering,
Departamento de Ci\^{e}ncia e Tecnologia Aeroespacial, DCTA/ITA;
E-mail: sami.yamouni@gmail.com.}\\
{\normalsize\itshape Instituto Tecnol\'{o}gico de Aeron\'{a}utica, 12228-900
S\~{a}o Jos\'{e} dos Campos, SP, Brazil}\\
\and
Jo\~{a}o Luiz F. Azevedo
\thanks{Senior Research Engineer, Aerodynamics Division,
Departamento de Ci\^{e}ncia e Tecnologia Aeroespacial, DCTA/IAE/ALA;
E-mail: joaoluiz.azevedo@gmail.com. AIAA Fellow.}\\
{\normalsize\itshape Instituto de Aeron\'{a}utica e Espa\c{c}o, 12228-904
S\~{a}o Jos\'{e} dos Campos, SP, Brazil}
\and
William R. Wolf
\thanks{Assistant Professor, Faculty of Mechanical Engineering;
E-mail: wolf@fem.unicamp.br. AIAA Member}\\
{\normalsize\itshape Universidade Estadual de Campinas, 13083-970
Campinas, SP, Brazil}
}

 \AIAApapernumber{2015-NUMBER}
 \AIAAconference{22nd AIAA Computational Fluid Dynamic Conference, June, 2015, Dallas, Texas}
 \AIAAcopyright{\AIAAcopyrightD{2015}}

\begin{document}

\maketitle


\begin{abstract}

\section*{Abstract}

Current design constraints have encouraged the studies of 
aeroacoustics fields around compressible jet flows. The 
present work addresses the numerical study of subgrid scale modeling 
for unsteady turbulent jet flows as a preliminary step for future aeroacoustic analyses 
of main engine rocket plumes. 
An in-house large eddy simulation (LES) tool is developed in order to 
reproduce high fidelity results of compressible jet flows. In the present study, 
perfectly expanded jets are considered because the authors want to emphasize the 
effects of the jet mixing phenomena. 
The large eddy simulation formulation is written using the finite 
difference approach, with an explicit time integration and using a 
second order spatial discretization. The energy equation is carefully 
discretized in order to model the energy equation of the filtered Navier-Stokes 
formulation. The classical Smagorinsky model, the dynamic Smagorinsky 
model and the Vreman models are the chosen subgrid scale closures 
for the present work. Numerical simulations of perfectly expanded 
jets are performed and compared with the literature in order to 
validate and compare the performance of each subgrid closure in the 
solver. 

\end{abstract}




\section{Introduction}

One of the main design issues related to launch vehicles 
lies on noise emission originated from the complex 
interaction between the high-temperature/high-velocity 
exhaustion gases and the atmospheric air. These emissions 
yield very high noise levels, which must be minimized due
to several design constraints. For instance, the 
resulting pressure fluctuations can damage the solid 
structure of different parts of the launcher by 
vibrational acoustic stress. Therefore, it is a design
constraint to consider the loads resulting from acoustic 
sources in the structural dimensioning of large launch 
vehicles during the take off and also during the transonic 
flight. Moreover, one cannot neglect the energy dissipation 
effect caused by the acoustic waves generated even if the 
vehicles is far from the ground. Theoretically, all 
chemical energy should be converted into kinectic
energy. However, in reallity, the noise generation 
consumes part of the chemical energy.

The acoustic design constraints have encouraged the 
studies of aeroacoustic fields around compressible jet 
flows. Instituto de Aeronautica e Espa\c{c}o (IAE) 
in Brazil is interested in this flow configuration for 
rocket design applications. Unsteady property fields of 
the flow are necessary for the aerocoustic studies. 
Therefore, the present work addresses the numerical study 
of unsteady turbulent compressible jet flows for such 
aeroacoustic applications. More precisely, on the effects 
of subgrid scale modeling using second order centered schemes
for compressible LES. An in-house computational tool is developed 
regarding the study of unsteady turbulent compressible flow. 
JAZzY is a novel large eddy simulation tool which is developed 
in order to reproduce high fidelity results of compressible 
jet flows which are used for aeroacoustic studies using the 
Ffowcs Williams and Hawkings approach \cite{Wolf2012}.

The LES formulation is written using the finite difference 
approach. Inviscid numerical fluxes are calculated using a 
second order accurate centered scheme with the explicit 
addition of artificial dissipation. A five steps second order 
accurate Runge-Kutta is the chosen time marching method. A 
formulation based on the System I set of equations 
\cite{Vreman1995} is used here in order to model the filtered 
terms of the energy equation. The classical Smagorinsky model 
\cite{Smagorinsky63,Lilly65,Lilly67}, the dynamic Smagorinsky 
model \cite{Germano91,moin91} and the Vreman model 
\cite{vreman2004} are the subgrid scale (SGS) turbulence 
closures used in the present work. Numerical simulation of 
perfectly expanded jets are performed and compared with 
numerical \cite{Mendez10} and experimental 
\cite{bridges2008turbulence} data.


  \section{Large Eddy Simulation Filtering}

The large eddy simulation is based on the principle 
of scale separation, which is addressed as a 
filtering procedure in a mathematical formalism. A 
modified version of the the System I filtering 
approach \cite{Vreman1995} is used in present work
which is given by
\begin{equation}
\begin{array}{c}
\displaystyle \frac{\partial \overline{\rho} }{\partial t} + \frac{\partial}{\partial x_{j}} 
\left( \overline{\rho} \widetilde{  u_{j} } \right) = 0 \, \mbox{,}\\
\displaystyle \frac{\partial}{\partial t} \left( \overline{ \rho } \widetilde{ u_{i} } \right) 
+ \frac{\partial}{\partial x_{j}} 
\left( \overline{ \rho } \widetilde{ u_{i} } \widetilde{ u_{j} } \right)
+ \frac{\partial \overline{p}}{\partial x_{i}} 
- \frac{\partial {\tau}_{ij}}{\partial x_{j}}  
+ \frac{1}{3} \frac{\partial}{\partial x_{j}}\left( {\delta}_{ij} \sigma_{ii}\right)
= 0 \, \mbox{,} \\ 
\displaystyle \frac{\partial \overline{e}}{\partial t} 
+ \frac{\partial}{\partial x_{j}} 
\left[ \left( \overline{e} + \overline{p} \right)\widetilde{u_{j}} \right]
- \frac{\partial}{\partial x_{j}}\left({\tau}_{ij} \widetilde{u_{i}} \right)
+ \frac{1}{3} \frac{\partial}{\partial x_{j}}
\left[ \left( \delta_{ij}{\sigma}_{ii} \right) \widetilde{u_{i}} \right]
+ \frac{\partial {q}_{j}}{\partial x_{j}} = 0 \, \mbox{,}
\end{array}
\label{eq:modified_system_I}
\end{equation}
in which $t$ and $x_{i}$ are independent variables 
representing time and spatial coordinates of a 
Cartesian coordinate system $\textbf{x}$, respectively. 
The components of the velocity vector $\textbf{u}$ are 
written as $u_{i}$, and $i=1,2,3$. Density, pressure and 
total energy per mass unit are denoted by $\rho$, $p$ and 
$e$, respectively. The $\left( \overline{\cdot} \right)$ and
$\left( \tilde{\cdot} \right)$ operators are used in order 
to represent filtered and Favre averaged properties, 
respectively. The System I formulation neglectes the double 
correlation term and the total energy per mass unit is written 
as 
\begin{equation}
	\overline{e} = \frac{\overline{p}}{\gamma - 1} 
	+ \frac{1}{2} \rho \widetilde{u}_{i} \widetilde{u}_{i} \, \mbox{.} 
\end{equation}
The heat flux, $q_{j}$, is given by
\begin{equation}
	{q}_{j} = \left(\kappa+{\kappa}_{sgs}\right) 
	\frac{\partial \widetilde{T}}{\partial x_{j}} 
	\, \mbox{.}
	\label{eq:q_mod}
\end{equation}
where $T$ is the static temperature and $\kappa$ is the thermal 
conductivity, which can by expressed by
\begin{equation}
\kappa = \frac{\mu C_{p}}{Pr} \, \mbox{,}
\end{equation}
The thermal conductivity is a function of the specific heat at 
constant pressure, $Cp$, of the Prandtl number, $Pr$, which is 
equal to $0.72$ for air, and of the dynamic viscosity, $\mu$.
The SGS thermal conductivity, $\kappa_{sgs}$, is written as
\begin{equation}
	\kappa_{sgs} = \frac{\mu_{sgs} C_{p}}{ {Pr}_{sgs} } 
	\, \mbox{,}
	\label{eq:kappa_sgs}
\end{equation}
where ${Pr}_{sgs}$ is the SGS Prandtl number, which is
equal to $0.9$ for static SGS models and $\mu_{sgs}$
is the eddy viscosity which is calculated by the SGS
closure. The dynamic viscosity, $\mu$ can be calculated 
using the Sutherland Law,
\begin{eqnarray}
	\mu \left( \widetilde{T} \right) = \mu_{\infty} 
	\left( \frac{\widetilde{T}}{\widetilde{T}_{\infty}}
	\right)^{\frac{3}{2}} 
	\frac{\widetilde{T}_{0}+S_{1}}{\widetilde{T}+S_{1}} &
\mbox{with} \: S_{1} = 110.4K \, \mbox{.}
\label{eq:sutherland}
\end{eqnarray}
Density, static pressure and static temperature are correlated 
by the equation of state given by
\begin{equation}
	\overline{p} = \rho R \widetilde{T} \, \mbox{,}
\end{equation}
where $R$ is the gas constant, written as
\begin{equation}
R = C_{p} - C_{v} \, \mbox{,}
\end{equation}
and $C_{v}$ is the specif heat at constant volume.
The shear-stress tensor, $\tau_{ij}$, is written 
according to the Stokes hypothesis and includes
the eddy viscosity, $\mu_{sgs}$,
\begin{equation}
	{\tau}_{ij} = 2 \left(\mu+{\mu}_{sgs}\right) 
	\left( \tilde{S}_{ij} - \frac{1}{3} \delta_{ij} \tilde{S}_{kk} \right) \,
	\label{eq:tau_mod}
\end{equation}
in which $\tilde{S}_{ij}$, components of rate-of-strain tensor, are 
given by
\begin{equation}
	\tilde{S}_{ij} = \frac{1}{2} 
	\left( \frac{\partial \tilde{u}_{i}}{\partial x_{j}} 
	+ \frac{\partial \tilde{u}_{j}}{\partial x_{i}} 
\right) \, \mbox{.}
\end{equation}
The SGS stress tensor components are written using the eddy 
viscosity \cite{Sagaut05},
\begin{equation}
    \sigma_{ij} = - 2 \mu_{sgs} \left( \tilde{S}_{ij} 
	- \frac{1}{3} \tilde{S}_{kk} \right)
    + \frac{1}{3} \delta_{ij} \sigma_{kk}
    \, \mbox{.}
    \label{eq:sgs_visc}
\end{equation}
The eddy viscosity, $\mu_{sgs}$, and the components of the 
isotropic part of the SGS stress tensor, $\sigma_{kk}$, are
modeled by the SGS closure.

\section{Subgrid Scale Modeling}

The present section toward the description of the turbulence modeling and the 
theoretical formulation of subgrid scales closures inclued in the present work. 
The closures models presented here are founded on the homogeneous turbulence theory, 
which is usually developed in the spectral space as an atempt to quantify the 
interaction between the different scales of turbulence.

\subsection{Smagorinky Model}

The Smagorinsky model \cite{Smagorinsky63} is one of the simplest algebric models 
for the deviatoric part of the SGS tensor used in large-eddy simulations. The 
isotropic part of the SGS tensor is neglected for Smagorinsky model in the
current work. This SGS closure is a classical model based the large scales 
properties and is written as
\begin{equation}
\mu_{sgs} = \left( \rho C_{s} \Delta \right)^{2} | \widetilde{S} | \, \mbox{,}
\end{equation}
where
\begin{equation}
| \tilde{S} | = \left( 2 \tilde{S}_{ij} \tilde{S}_{ij} \right)^{\frac{1}{2}} \, \mbox{,}
\end{equation}
$\Delta$ is the filter size and $C_{s}$ is the Smagorinsky constant. 
Several attempts can be found in the literature regarding the 
evaluation of the Smagorinsky constant. The value of this constant 
is adjusted to improve the results of different flow 
configurations. In pratical terms, the Smagorinsky subgrid 
model has a flow dependency of the constant which takes value 
ranging from 0.1 to 0.2 depending on the flow. The suggestion 
of Lilly \cite{Lilly67}, $C_{s}=0.148$, is used in the current 
work.


This model is generally over-dissipative in regions of large mean strain. 
This is particulary true in the trasitional region between laminar and 
turbulent flows. Moreover, the limiting behavior near the wall is not
correct, and the model predictions correlate poorly with the exact subgrid 
scale tensor \cite{Garnier09}. However, it is a very simple model and, 
with the use of damping function and good calibration, can be successfully 
applied on large-eddy simulations.

\subsection{Vreman Model}

Vreman \cite{vreman2004} proposed a turbulence model that can correctly predict
inhomogeneous turbulent flows. For such flows, the eddy viscosity should become 
small in laminar and transitional regions. This requirement is unfortunately not 
satisfied by existing simple eddy-viscosity closures such as the classic 
Smagorinsky model \cite{Smagorinsky63,Lilly65,Deardorff70}. The Vreman SGS model
is very simple and is given by
\begin{equation}
	\mu_{sgs} = \rho \, \bm{c} \, 
	\sqrt{\frac{B_{\beta}}{\alpha_{ij} \alpha_{ij}}} 
	\,\mbox{,}
\end{equation}
with
\begin{equation}
	\alpha_{ij} = \frac{\partial \tilde{u}_{j}}{\partial x_{i}} 
	\, \mbox{,}
\end{equation}
\begin{equation}
	\beta_{ij} = \Delta^{2}_{m}\alpha_{mi}\alpha_{mj}
\end{equation}
and 
\begin{equation}
	B_{\beta} = \beta_{11}\beta_{22} - \beta_{12}^{2} 
	          + \beta_{11}\beta_{33} - \beta_{13}^{2}
			  + \beta_{22}\beta_{33} - \beta_{23}^{2}
			  \, \mbox{.}
\end{equation}
The constant $\bm{c}$ is related to the Smagorinsky constant, $C_{s}$, 
and it is given by
\begin{equation}
	\bm{c} = 2.5 \, C_{s}^{2} 
	\, \mbox{,}
\end{equation}
and $\Delta_{m}$ is the filter width in each direction. In the present work,
the isotropic part of the SGS tensor is neglected for the Vreman model.
The $\alpha$ symbol represents the matrix of first oder derivatives of the 
filtered components of velocity, $\tilde{u}_{i}$. The SGS eddy-viscosity is 
defined as zero when $\alpha_{ij}\alpha_{ij}$ equals zero. Vreman
\cite{vreman2004} affirms that the tensor $\beta$ is proportional to the 
gradient model \cite{Leonard74,Clark79} in its general anisotropic form
\cite{vreman1996}.

The Vreman model can be classified as very simple model because it is expresed 
in first-order derivatives and it dos not involves explicit filtering, averaging, 
clipping procedures and is rotationally invariant for isotropic filter widths. The 
model is originally created for incompressible flows and it has presented good 
results for two incompressible flows configurations: the transitional and turbulent 
mixing layer at high Reynolds number and the turbulent channel flow \cite{vreman1996}. 
In both cases, the Vreman model is found to be more accurate than the classical 
Smagorinsky model and as good as the dynamic Smagorinsky model.

\subsection{Dynamic Smagorinsky Model}

Germano {\it et al.} \cite{germano90} developed a dynamic SGS model in order to 
overcome the issues of the classical Smagorinsky closure. The model uses the strain 
rate fields at two different scales and thus extracts spectral information in 
the large-scale field to extrapolate the small stresses \cite{moin91}. 
The coefficients of the model are computed instantaneously in the dynamic model. 
They are function of the positioning in space and time rather than being specified 
a priori. Moin {\it et al.} \cite{moin91} extended the work of Germano for compressible 
flows. The dynamic Smagorinsky model for compressible flow configurations is detailed in 
the present section.

The Dynamic model introduces the test filter, $\widehat{\left( \cdot \right)}$, which
has a larger filter width, $\widehat{\Delta}$, than the one of the resolved grid filter, 
$\overline{\left( \cdot \right)}$. The use of test filters generates a second field 
with larger scales than the resolved field. The Yoshizawa model \cite{Yoshizawa86} is 
used for the isotropic portion of the SGS tensor and it is written as
\begin{equation}
	\sigma_{ll} = 2 C_{I} \overline{\rho} {\Delta}^{2}|\tilde{S}|^{2}
	\, \mbox{,}
	\label{eq:yoshizawa}
\end{equation}
where $C_{I}$ is defined by
\begin{equation}
	C_{I} = \frac{\biggl\langle \widehat{\overline{\rho} \tilde{u}_{l} \tilde{u}_{l}} -
	        \left( \widehat{\overline{\rho}\tilde{u}_{l}}
	               \widehat{\overline{\rho}\tilde{u}_{l}}/
				   \widehat{\overline{\rho}} \right)\biggr\rangle }
		   {\biggl\langle 2 \widehat{\Delta}^{2} \widehat{\overline{\rho}}
		   |\widehat{\overline{S}}|^{2} - 
	        2 {\Delta}^{2} \widehat{ \overline{\rho}
		   |\overline{S}|^{2}}\biggr\rangle}
		   \, \mbox{.}
		   \label{eq:av_ci}
\end{equation}
A volume averaging, here indicated by $\langle \,\, \rangle$, is suggest by 
Moin {\it et al} \cite{moin91} and by Garnier {\it et al} in order to avoid 
numerical issues. The eddy viscosity, $\mu_{sgs}$, is calculated using the 
same approach used by static Smagorinsky model,
\begin{equation}
\mu_{sgs} = \left( \rho C_{ds} \Delta \right)^{2} | \tilde{S} | \, \mbox{,}
\end{equation}
where
\begin{equation}
| \tilde{S} | = \left( 2 \tilde{S}_{ij} \tilde{S}_{ij} \right)^{\frac{1}{2}} 
\, \mbox{,}
\end{equation}
and $C_{ds}$ is the dynamic constant of the model, which is given by
\begin{equation}
	C_{ds} = \frac{\biggl\langle
	        \left[ \widehat{\overline{\rho} \tilde{u}_{i} \tilde{u}_{j}} -
	        \left( \widehat{\overline{\rho}\tilde{u}_{i}}
	        \widehat{\overline{\rho}\tilde{u}_{j}}/
			\widehat{\overline{\rho}} \right) \right]\tilde{S}_{ij} - 
            \frac{1}{3}\tilde{S}_{mm}
	        \left(\mathscr{T}_{ll} - \widehat{\sigma}_{ll}\right)
			\biggr\rangle}{\biggl\langle
			2 {\Delta}^{2}\left[
			\widehat{\overline{\rho}|\tilde{S}|\tilde{S}_{ij}}\tilde{S}_{ij}
			- \frac{1}{3}\left(\overline{\rho}|\tilde{S}|\tilde{S}_{mm}\right)^{\widehat{ }}
			\tilde{S}_{ll}\right] -
            2 \widehat{\Delta}^{2}\left(
			\widehat{\overline{\rho}}|\widehat{\tilde{S}}|
			\widehat{\tilde{S}}_{ij}\tilde{S}_{ij} -
            \frac{1}{3}\widehat{\overline{\rho}}|\widehat{\tilde{S}}|
            \widehat{\tilde{S}}_{mm} \tilde{S}_{ll}\right)
			\biggr\rangle}
	\, \mbox{.}
    \label{eq:av_cd}
\end{equation}
The SGS Prandtl number is computed using the dynamic constant, $C_{ds}$, and 
written as
\begin{equation}
	{Pr}_{sgs} = C_{ds} \frac{\biggl\langle \Delta^{2} 
	\biggl(\overline{\rho}|\tilde{S}|\frac{\partial \overset{\sim}{T}}{\partial x_{j}}
	\biggr)^{\widehat{ }} \,
	\frac{\partial\overset{\sim}{T}}{\partial x_{j}} -
	\widehat{\Delta}^{2}\widehat{\overline{\rho}}|\widehat{\tilde{S}}|
	\frac{\partial \overset{\sim}{T}}{\partial x_{j}}
	\frac{\partial \overset{\sim}{T}}{\partial x_{j}} \biggr\rangle}
	{\biggl\langle\left[ \widehat{\overline{\rho} \tilde{u}_{j} \overset{\sim}{T}} - 
	\left( \widehat{\overline{\rho} \tilde{u_{j}}} 
	\widehat{\overline{\rho} \overset{\sim}{T}} \right)/
		   \widehat{\overline{\rho}}\right] 
		   \frac{\partial \overset{\sim}{T}}{\partial x_{j}}\biggr\rangle}
	\, \mbox{.}
	\label{eq:av_Pr_sgs}
\end{equation}
%












  \section{Transformation of Coordinates}

The formulation is written in the a general curvilinear coordinate 
system in order to facilitate the implementation and add more 
generality for the CFD tool. Hence, the filtered Navier-Stokes 
equations can be written in strong conservation form for a 
3-D general curvilinear coordinate system as
\begin{equation}
	\frac{\partial \hat{Q}}{\partial t} 
	+ \frac{\partial }{\partial \xi}\left(\hat{\mathbf{E}_{e}}-\hat{\mathbf{E}_{v}}\right) 
	+ \frac{\partial}{\partial \eta}\left(\hat{\mathbf{F}_{e}}-\hat{\mathbf{F}_{v}}\right)
	+ \frac{\partial}{\partial \zeta}\left(\hat{\mathbf{G}_{e}}-\hat{\mathbf{G}_{v}}\right) 
	= 0 \, \mbox{.}
	\label{eq:vec-LES}
\end{equation}
In the present work, the chosen general coordinate transformation is given by
\begin{eqnarray}
	\xi & = & \xi \left(x,y,z,t \right)  \, \mbox{,} \nonumber\\
	\eta & = & \eta \left(x,y,z,t \right)  \, \mbox{,} \\
	\zeta & = & \zeta \left(x,y,z,t \right)\ \, \mbox{.} \nonumber
\end{eqnarray}
In the jet flow configuration, $\xi$ is the axial jet flow direction, $\eta$ is 
the radial direction and $\zeta$ is the azimuthal direction. The vector of
conserved properties is written as
\begin{equation}
	\hat{Q} = J^{-1} \left[ \overline{\rho} \quad \overline{\rho}\tilde{u} \quad 
	\overline{\rho}\tilde{v} \quad \overline{\rho}\tilde{w} \quad \overline{e} \right]^{T} 
	\quad \mbox{,}
	\label{eq:hat_Q_vec}
\end{equation}
where the Jacobian of the transformation, $J$, is given by
\begin{equation}
	J = \left( x_{\xi} y_{\eta} z_{\zeta} + x_{\eta}y_{\zeta}z_{\xi} +
	           x_{\zeta} y_{\xi} z_{\eta} - x_{\xi}y_{\zeta}z_{\eta} -
			   x_{\eta} y_{\xi} z_{\zeta} - x_{\zeta}y_{\eta}z_{\xi} 
	    \right)^{-1} \, \mbox{,}
\end{equation}
and
\begin{eqnarray}
	\displaystyle x_{\xi}   = \frac{\partial x}{\partial \xi}  \, \mbox{,} & 
	\displaystyle x_{\eta}  = \frac{\partial x}{\partial \eta} \, \mbox{,} & 
	\displaystyle x_{\zeta} = \frac{\partial x}{\partial \zeta}\, \mbox{,} \nonumber \\
	\displaystyle y_{\xi}   = \frac{\partial y}{\partial \xi}  \, \mbox{,} & 
	\displaystyle y_{\eta}  = \frac{\partial y}{\partial \eta} \, \mbox{,} & 
	\displaystyle y_{\zeta} = \frac{\partial y}{\partial \zeta}\, \mbox{,} \\
	\displaystyle z_{\xi}   = \frac{\partial z}{\partial \xi}  \, \mbox{,} & 
	\displaystyle z_{\eta}  = \frac{\partial z}{\partial \eta} \, \mbox{,} & 
	\displaystyle z_{\zeta} = \frac{\partial z}{\partial \zeta}\, \mbox{.} \nonumber
\end{eqnarray}

The inviscid flux vectors, $\hat{\mathbf{E}}_{e}$, $\hat{\mathbf{F}}_{e}$ and 
$\hat{\mathbf{G}}_{e}$, are given by
{\small
\begin{eqnarray}
	\hat{\mathbf{E}}_{e} = J^{-1} \left\{\begin{array}{c}
		\overline{\rho} U \\
		\overline{\rho}\tilde{u} U + \overline{p} \xi_{x} \\
		\overline{\rho}\tilde{v} U + \overline{p} \xi_{y} \\
		\overline{\rho}\tilde{w} U + \overline{p} \xi_{z} \\
		\left( \overline{e} + \overline{p} \right) U - \overline{p} \xi_{t}
\end{array}\right\} \, \mbox{,} &
%
	\hat{\mathbf{F}}_{e} = J^{-1} \left\{\begin{array}{c}
		\overline{\rho} V \\
		\overline{\rho}\tilde{u} V + \overline{p} \eta_{x} \\
		\overline{\rho}\tilde{v} V + \overline{p} \eta_{y} \\
		\overline{\rho}\tilde{w} V + \overline{p} \eta_{z} \\
		\left( \overline{e} + \overline{p} \right) V - \overline{p} \eta_{t}
\end{array}\right\} \, \mbox{,} &
%
	\hat{\mathbf{G}}_{e} = J^{-1} \left\{\begin{array}{c}
		\overline{\rho} W \\
		\overline{\rho}\tilde{u} W + \overline{p} \zeta_{x} \\
		\overline{\rho}\tilde{v} W + \overline{p} \zeta_{y} \\
		\overline{\rho}\tilde{w} W + \overline{p} \zeta_{z} \\
		\left( \overline{e} + \overline{p} \right) W - \overline{p} \zeta_{t}
	\end{array}\right\} \, \mbox{.}
	\label{eq:hat-flux-G}
\end{eqnarray}
}
The contravariant velocity components, $U$, $V$ and $W$, are calculated as
%
%
\begin{eqnarray}
  U = \xi_{x}\overline{u} + \xi_{y}\overline{v} + \xi_{z}\overline{w} 
  \, \mbox{,} \nonumber \\
  V = \eta_{x}\overline{u} + \eta_{y}\overline{v} + \eta_{z}\overline{w} 
  \, \mbox{,} \\
  W = \zeta_{x}\overline{u} + \zeta_{y}\overline{v} + \zeta_{z}\overline{w} 
  \, \mbox{.} \nonumber
  \label{eq:vel_contrv}
\end{eqnarray}
The metric terms are given by
%
%
\begin{eqnarray}
	\xi_{x} = J \left( y_{\eta}z_{\zeta} - y_{\zeta}z_{\eta} \right) \, \mbox{,} & 
	\xi_{y} = J \left( z_{\eta}x_{\zeta} - z_{\zeta}x_{\eta} \right) \, \mbox{,} & 
	\xi_{z} = J \left( x_{\eta}y_{\zeta} - x_{\zeta}y_{\eta} \right) \, \mbox{,} \nonumber \\
	\eta_{x} = J \left( y_{\eta}z_{\xi} - y_{\xi}z_{\eta} \right) \, \mbox{,} & 
	\eta_{y} = J \left( z_{\eta}x_{\xi} - z_{\xi}x_{\eta} \right) \, \mbox{,} & 
	\eta_{z} = J \left( x_{\eta}y_{\xi} - x_{\xi}y_{\eta} \right) \, \mbox{,} \\
	\zeta_{x} = J \left( y_{\xi}z_{\eta} - y_{\eta}z_{\xi} \right) \, \mbox{,} & 
	\zeta_{y} = J \left( z_{\xi}x_{\eta} - z_{\eta}x_{\xi} \right) \, \mbox{,} & 
	\zeta_{z} = J \left( x_{\xi}y_{\eta} - x_{\eta}y_{\xi} \right) \, \mbox{.} \nonumber \\
\end{eqnarray}

The viscous flux vectors, $\hat{\mathbf{E}}_{v}$, $\hat{\mathbf{F}}_{v}$ and 
$\hat{\mathbf{G}}_{v}$, are written as
\begin{equation}
	\hat{\mathbf{E}}_{v} = J^{-1} \left\{\begin{array}{c}
		0 \\
		\xi_{x}{\tau}_{xx} +  \xi_{y}{\tau}_{xy} + \xi_{z}{\tau}_{xz} \\
		\xi_{x}{\tau}_{xy} +  \xi_{y}{\tau}_{yy} + \xi_{z}{\tau}_{yz} \\
		\xi_{x}{\tau}_{xz} +  \xi_{y}{\tau}_{yz} + \xi_{z}{\tau}_{zz} \\
		\xi_{x}{\beta}_{x} +  \xi_{y}{\beta}_{y} + \xi_{z}{\beta}_{z} 
	\end{array}\right\} \, \mbox{,}
	\label{eq:hat-flux-Ev}
\end{equation}
\begin{equation}
	\hat{\mathbf{F}}_{v} = J^{-1} \left\{\begin{array}{c}
		0 \\
		\eta_{x}{\tau}_{xx} +  \eta_{y}{\tau}_{xy} + \eta_{z}{\tau}_{xz} \\
		\eta_{x}{\tau}_{xy} +  \eta_{y}{\tau}_{yy} + \eta_{z}{\tau}_{yz} \\
		\eta_{x}{\tau}_{xz} +  \eta_{y}{\tau}_{yz} + \eta_{z}{\tau}_{zz} \\
		\eta_{x}{\beta}_{x} +  \eta_{y}{\beta}_{y} + \eta_{z}{\beta}_{z} 
	\end{array}\right\} \, \mbox{,}
	\label{eq:hat-flux-Fv}
\end{equation}
\begin{equation}
	\hat{\mathbf{G}}_{v} = J^{-1} \left\{\begin{array}{c}
		0 \\
		\zeta_{x}{\tau}_{xx} +  \zeta_{y}{\tau}_{xy} + \zeta_{z}{\tau}_{xz} \\
		\zeta_{x}{\tau}_{xy} +  \zeta_{y}{\tau}_{yy} + \zeta_{z}{\tau}_{yz} \\
		\zeta_{x}{\tau}_{xz} +  \zeta_{y}{\tau}_{yz} + \zeta_{z}{\tau}_{zz} \\
		\zeta_{x}{\beta}_{x} +  \zeta_{y}{\beta}_{y} + \zeta_{z}{\beta}_{z} 
	\end{array}\right\} \, \mbox{,}
	\label{eq:hat-flux-Gv}
\end{equation}
where $\beta_{x}$, $\beta_{y}$ and $\beta_{z}$ are defined as
\begin{eqnarray}
	\beta_{x} = {\tau}_{xx}\tilde{u} + {\tau}_{xy}\tilde{v} +
	{\tau}_{xz}\tilde{w} - \overline{q}_{x} \, \mbox{,} \nonumber \\
	\beta_{y} = {\tau}_{xy}\tilde{u} + {\tau}_{yy}\tilde{v} +
	{\tau}_{yz}\tilde{w} - \overline{q}_{y} \, \mbox{,} \\
	\beta_{z} = {\tau}_{xz}\tilde{u} + {\tau}_{yz}\tilde{v} +
	{\tau}_{zz}\tilde{w} - \overline{q}_{z} \mbox{.} \nonumber
\end{eqnarray}
%

\section{Dimensionless Formulation}

A convenient nondimensionalization is necessary in to order to achieve a consistent 
implementation of the governing equations of motion. Dimensionless formulation 
yields to a more general numerical tool. There is no need to change the formulation 
for each configuration intended to be simulated. Moreover, dimensionless formulation 
scales all the necessary properties to the same order of magnitude which is a 
computational advantage \cite{BIGA02}. Dimensionless variables are presented in the 
present section in order perform the nondimensionalization of Eq.\ 
\eqref{eq:vec-LES}

The dimensionless time, $\underline{t}$, is written as function of the 
speed of sound of the jet at the inlet, $a_{j}$, and of a reference lenght, $l$,
\begin{equation}
	\underline{t} = t \frac{a_{j}}{l} \, \mbox{.}
	\label{eq:non-dim-time}
\end{equation}
%
%
The dimensionless velocity components are obtained using the speed of sound of the 
jet at the inlet,
\begin{equation}
	\underline{\textbf{u}} = \frac{\textbf{u}}{a_{j}} \, \mbox{.}
	\label{eq:non-dim-vel}
\end{equation}
Dimensionless pressure and energy are calculated using density and speed of the sound
of the jet at the inlet as
\begin{equation}
	\underline{p} = \frac{p}{\rho_{j}a_{j}^{2}} \, \mbox{,}
	\label{eq:non-dim-press}
\end{equation}
\begin{equation}
	\underline{E} = \frac{E}{\rho_{j}a_{j}^{2}} \, \mbox{.}
	\label{eq:non-dim-energy}
\end{equation}
Dimensionless density, $\underline{\rho}$, temperature, $\underline{T}$ and 
viscosity, $\underline{\mu}$, are calculated using freestream properties
\begin{equation}
	\underline{\rho} = \frac{\rho}{\rho_{j}} \, \mbox{.}
	\label{eq:non-dim-rho}
\end{equation}

One can use the dimensionless properties described above in order to write the 
dimensionless form of the RANS equations as
\begin{equation}
	\frac{\partial \underline{Q}}{\partial t} + 
	\frac{\partial \underline{\mathbf{E}}_{e}}{\partial \xi} +
	\frac{\partial \underline{\mathbf{F}}_{e}}{\partial \eta} + 
	\frac{\partial \underline{\mathbf{G}}_{e}}{\partial \zeta} =
	\frac{M_{j}}{Re} \left( \frac{\partial \underline{\mathbf{E}}_{v}}{\partial \xi} 
	+ \frac{\partial \underline{\mathbf{F}}_{v}}{\partial \eta} 
	+ \frac{\partial \underline{\mathbf{G}}_{v}}{\partial \zeta} \right)	\, \mbox{,}
	\label{eq:vec-underline-split-RANS}
\end{equation}
where the underlined terms are calculated using dimensionless properties.
The Mach number of the jet, $M_{j}$, and the Reynolds number are based on 
the mean inlet velocity of the jet, $U_{j}$, diamenter of the inlet, $D$,
and freestream properties such as speed of sound, $a_{\infty}$, density, 
$\rho_{\infty}$ and viscosity, $\mu_{\infty}$,
\begin{eqnarray}
	M_{j}=\frac{U_j}{a_{\infty}} & \mbox{and} & 
	Re = \frac{\rho_{j}U_{j}D}{\mu_{j}} \, \mbox{.}
\end{eqnarray}

\section{Numerical Formulation}

The governing equations previously described are discretized in a 
structured finite difference context for general curvilinear 
coordinate system \cite{BIGA02}. The numerical flux is calculated 
through a central difference scheme with the explicit addition 
of the anisotropic scalar artificial dissipation of Turkel and Vatsa
\cite{Turkel_Vatsa_1994}. The time integration is performed by an 
explicit, 2nd-order, 5-stage Runge-Kutta scheme 
\cite{jameson_mavriplis_86, Jameson81}.  Conserved properties
and artificial dissipation terms are properly treated near boundaries in order
to assure the physical correctness of the numerical formulation. 
 
\subsection{Spatial Discretization}

For the sake of simplicity the formulation discussed in the present section
is no longer written using bars. However, the reader should notice that the 
equations are dimensionless and filtered. The Navier-Stokes equations, 
presented in Eq.\ \eqref{eq:vec-underline-split-RANS}, are discretized in 
space in a finite difference fashion and, then, rewritten as
\begin{equation}
	\left(\frac{\partial Q}{\partial t}\right)_{i,j,k} \  
	= \  - RHS_{i,j,k} \, \mbox{,}	
	\label{eq:spatial_discret}
\end{equation}
where $RHS$ is the right hand side of the equation and it is written as function of 
the numerical flux vectors at the interfaces between grid points,
\begin{eqnarray}
	{RHS}_{i,j,k} & = & 
	\frac{1}{\Delta \xi} \left( 
	{\mathbf{E}_{e}}_{(i+\frac{1}{2},j,k)} - {\mathbf{E}_{e}}_{(i-\frac{1}{2},j,k)} - 
	{\mathbf{E}_{v}}_{(i+\frac{1}{2},j,k)} + {\mathbf{E}_{v}}_{(i-\frac{1}{2},j,k)} 
	\right) \nonumber \\
	& & \frac{1}{\Delta \eta} \left( 
	{\mathbf{F}_{e}}_{(i,j+\frac{1}{2},k)} - {\mathbf{F}_{e}}_{(i,j-\frac{1}{2},k)} - 
	{\mathbf{F}_{v}}_{(i,j+\frac{1}{2},k)} + {\mathbf{F}_{v}}_{(i,j-\frac{1}{2},k)} 
	\right) \\
	& & \frac{1}{\Delta \zeta} \left( 
	{\mathbf{G}_{e}}_{(i,j,k+\frac{1}{2})} - {\mathbf{G}_{e}}_{(i,j,k-\frac{1}{2})} - 
	{\mathbf{G}_{v}}_{(i,j,k+\frac{1}{2})} + {\mathbf{G}_{v}}_{(i,j,k-\frac{1}{2})} 
	\right) \, \mbox{.} \nonumber
\end{eqnarray}
For the general curvilinear coordinate case 
$\Delta \xi = \Delta \eta = \Delta \zeta = 1$. The anisotropic scalar 
artificial dissipation method of Turkel and Vatsa \cite{Turkel_Vatsa_1994}
is implemented through the modification of the inviscid flux vectors, 
$\mathbf{E}_{e}$, $\mathbf{F}_{e}$ and $\mathbf{G}_{e}$. The numerical scheme 
is nonlinear and allows the selection between artificial dissipation terms of 
second and fourth differences, which is very important for capturing discontinuities 
in the flow. The numerical fluxes are calculated at interfaces in order to reduce 
the size of the calculation cell and, therefore, facilitate the implementation 
of second derivatives since the the concept of numerical fluxes vectors is 
used for flux differencing. Only internal interfaces receive the corresponding 
artificial dissipation terms, and differences of the viscous flux vectors 
use two neighboring points of the interface. 

The inviscid flux vectors, with the addition of the artificial dissipation
contribution, can be written as
\begin{eqnarray}
	{\mathbf{E}_{e}}_{(i \pm \frac{1}{2},j,k)} 
	= \frac{1}{2} \left( {\mathbf{E}_{e}}_{(i,j,k)} + {\mathbf{E}_{e}}_{(i \pm 1,j,k)} \right)
	- J^{-1} \mathbf{d}_{(i \pm \frac{1}{2},j,k)} \, \mbox{,} \nonumber \\
	{\mathbf{F}_{e}}_{(i,j\pm \frac{1}{2},k)} 
	= \frac{1}{2} \left( {\mathbf{F}_{e}}_{(i,j,k)} + {\mathbf{F}_{e}}_{(i,j \pm 1,k)} \right)
	- J^{-1} \mathbf{d}_{(i,j \pm \frac{1}{2},k)} \, \mbox{,} \label{eq:inv_flux_vec}\\
	{\mathbf{G}_{e}}_{(i,j,k\pm \frac{1}{2})} 
	= \frac{1}{2} \left( {\mathbf{G}_{e}}_{(i,j,k)} + {\mathbf{G}_{e}}_{(i,j,k \pm 1)} \right)
	- J^{-1} \mathbf{d}_{(i,j,k \pm \frac{1}{2})} \, \mbox{,} \nonumber
\end{eqnarray}
in which the $\mathbf{d}_{(i\pm 1,j,k)}$,$\mathbf{d}_{(i,j\pm 1,k)}$ and $\mathbf{d}_{(i,j,k\pm 1)}$ terms
are the Turkel and Vatsa \cite{Turkel_Vatsa_1994} artificial dissipation terms
in the $i$, $j$, and $k$ directions respectively. The scaling of the artificial
dissipation operator in each coordinate direction is weighted by its own spectral 
radius of the corresponding flux Jacobian matrix, which gives the non-isotropic 
characteristics of the method \cite{BIGA02}. The artificial dissipation contribution
in the $\xi$ direction is given by
\begin{eqnarray}
	\mathbf{d}_{(i + \frac{1}{2},j,k)} & = & 
	\lambda_{(i + \frac{1}{2},j,k)} \left[ \epsilon_{(i + \frac{1}{2},j,k)}^{(2)}
	\left( \mathcal{W}_{(i+1,j,k)} - \mathcal{W}_{(i,j,k)} \right) \right. \label{eq:dissip_term}\\
	& & \epsilon_{(i + \frac{1}{2},j,k)}^{(4)} \left( \mathcal{W}_{(i+2,j,k)} 
	- 3 \mathcal{W}_{(i+1,j,k)} + 3 \mathcal{W}_{(i,j,k)} 
	- \mathcal{W}_{(i-1,j,k)} \right) \left. \right] \, \mbox{,} \nonumber
\end{eqnarray}
in which
\begin{eqnarray}
	\epsilon_{(i + \frac{1}{2},j,k)}^{(2)} & = &
	k^{(2)} \mbox{max} \left( \nu_{(i+1,j,k)}^{d}, 
	\nu_{(i,j,k)}^{d} \right) \, \mbox{,} \label{eq:eps_2_dissip} \\
	\epsilon_{(i + \frac{1}{2},j,k)}^{(4)} & = &
	\mbox{max} \left[ 0, k^{(4)} - \epsilon_{(i + \frac{1}{2},j,k)}^{(2)} \right] 
	\, \mbox{.} \label{eq:eps_4_dissip}
\end{eqnarray}
The original article \cite{Turkel_Vatsa_1994} recomends using $k^{(2)}=0.25$ and 
$k^{(4)}=0.016$ for the dissipation artificial constants. The pressure 
gradient sensor, $\nu_{(i,j,k)}^{d}$, for the $\xi$ direction is written as
\begin{equation}
	\nu_{(i,j,k)}^{d} = \frac{|p_{(i+1,j,k)} - 2 p_{(i,j,k)} + p_{(i-1,j,k)}|}
	                          {p_{(i+1,j,k)} - 2 p_{(i,j,k)} + p_{(i-1,j,k)}} 
	\, \mbox{.}
\label{eq:p_grad_sensor}
\end{equation}
The $\mathcal{W}$ vector from Eq.\ \eqref{eq:dissip_term} is calculated as a function of the
conserved variable vector, $\hat{Q}$, written in Eq.\ \eqref{eq:hat_Q_vec}.
The formulation intends to keep the total enthalpy constant in the final converged 
solution, which is the correct result for the Navier-Stokes equations with 
$Re \rightarrow \infty$. This approach is also valid for the viscous formulation 
because the dissipation terms are added to the inviscid flux terms, in which they 
are really necessary to avoid nonlinear instabilities of the numerical formulation. 
The $\mathcal{W}$ vector is given by
\begin{equation}
	\mathcal{W} = \hat{Q} + \left[0 \,\, 0 \,\, 0 \,\, 0 \,\, p \right]^{T} \, \mbox{.}
	\label{eq:W_dissip}
\end{equation}
The spectral radius-based scaling factor, $\lambda$, for the $i-\mbox{th}$ 
direction is written
\begin{equation}
	\lambda_{(i+\frac{1}{2},j,k)} = \frac{1}{2} \left[ 
	\left( \overline{\lambda_{\xi}}\right)_{(i,j,k)} + 
	\left( \overline{\lambda_{\xi}}\right)_{(i+1,j,k)}
	\right] \, \mbox{,} 
\end{equation}
where
\begin{equation}
    \overline{\lambda_{\xi}}_{(i,j,k)} = \lambda_{\xi} \left[ 1 + 
	\left(\frac{\lambda_{\eta}}{\lambda_{\xi}} \right)^{0.5} + 
	\left(\frac{\lambda_{\zeta}}{\lambda_{\xi}} \right)^{0.5} \right] 
	\, \mbox{.}
\end{equation}
The spectral radii, $\lambda_{\xi}$, $\lambda_{\eta}$ and $\lambda_{\zeta}$ are given
by
\begin{eqnarray}
	\lambda_{\xi} & = & 
	|U| + a \sqrt{\xi_{x}^{2} + \eta_{y}^{2} + \zeta_{z}^{2}} 
	\, \mbox{,} \nonumber \\
	\lambda_{\xi} & = & 
	|V| + a \sqrt{\xi_{x}^{2} + \eta_{y}^{2} + \zeta_{z}^{2}} 
	\, \mbox{,} \\
	\lambda_{\xi} & = & 
	|W| + a \sqrt{\xi_{x}^{2} + \eta_{y}^{2} + \zeta_{z}^{2}} 
	\, \mbox{,} \nonumber
\end{eqnarray}
in which, $U$, $V$ and $W$ are the contravariants velocities in the $\xi$, $\eta$
and $\zeta$, previously written in Eq.\ \eqref{eq:vel_contrv}, and $a$ is the local 
speed of sound, which can be written as
\begin{equation}
	a = \sqrt{\frac{\gamma p}{\rho}} \, \mbox{.}
\end{equation}
The calculation of artificial dissipation terms for the other coordinate directions
are completely similar and, therefore, they are not discussed in the present work.

\subsection{Time Marching Method}

The time marching method used in the present work is a 2nd-order, 5-step Runge-Kutta
scheme based on the work of Jameson \cite{Jameson81, jameson_mavriplis_86}. 
The time integration can be written as
\begin{equation}
	\begin{array}{ccccc}
	Q_{(i,jk,)}^{(0)} & = & Q_{(i,jk,)}^{(n)} \, \mbox{,} & & \\
	Q_{(i,jk,)}^{(l)} & = & Q_{(i,jk,)}^{(0)} -  
	& \alpha_{l} {\Delta t}_{(i,j,k)} {RHS}_{(i,j,k)}^{(l-1)} \, & 
	\,\,\,\, l = 1,2 \cdots 5, \\
	Q_{(i,jk,)}^{(n+1)} & = & Q_{(i,jk,)}^{(5)} \, \mbox{,} & &
	\end{array}
	\label{eq:localdt}
\end{equation}
in which $\Delta t$ is the time step and $n$ and $n+1$ indicate the property
values at the current and at the next time step, respectively. The literature
\cite{Jameson81, jameson_mavriplis_86} recommends 
\begin{equation}
	\begin{array}{ccccc}
		\alpha_{1} = \frac{1}{4} \,\mbox{,} & \alpha_{2} = \frac{1}{6} \,\mbox{,} &
		\alpha_{3} = \frac{3}{8} \,\mbox{,} & \alpha_{4} = \frac{1}{2} \,\mbox{,} & 
		\alpha_{5} = 1 \,\mbox{,} 
	\end{array}
\end{equation}
in order to improve the numerical stability of the time integration. The present
scheme is theoretically stable for $CFL \leq 2\sqrt{2}$, under a linear analysis
\cite{BIGA02}.

\section{Boundary Conditions} \label{sec:BC}

\begin{figure}[ht]
       \begin{center}
		   \subfigure[Lateral view of boundary conditions.]{
           \includegraphics[width=0.475\textwidth]
		   {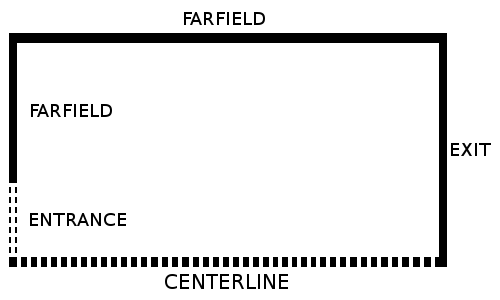} 
		   \label{fig:bc-1}
		   }
		   \subfigure[Frontal view of boundary conditions.]{
           \includegraphics[width=0.475\textwidth]
		   {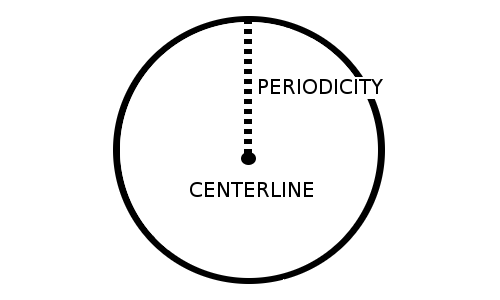} 
		   \label{fig:bc-2}
		   }
		   \caption{Lateral and frontal views of the computational domain 
		   indicating boundary conditions.}
		   \label{fig:bc}
	   \end{center}
\end{figure}

The geometry used in the present work presents a
cylindrical shape which is gererated by the rotation of 
a 2-D plan around a centerline. Figure \ref{fig:bc} 
presents a lateral view and a frontal view of the 
computational domain used in the present work and 
the positioning of the entrance, exit, centerline, 
far field and periodic boundary conditions. A discussion
on all boundary conditions is performed in the following 
subsections.

\subsection{Far Field Boundary}

Riemann invariants \cite{Long91} are used to implement far field boundary conditions.
They are derived from the characteristic relations for the Euler equations.
At the interface of the outer boundary, the following expressions apply
\begin{eqnarray}
	\mathbf{R}^{-} = {\mathbf{R}}_{\infty}^{-} & = & q_{n_\infty}-\frac{2}{\gamma-1}a_\infty\, \mbox{,}  \\
	\mathbf{R}^{+} = {\mathbf{R}}_{e}^{+} & = & q_{n_e}-\frac{2}{\gamma-1}a_e \, \mbox{,}
	\label{eq:R-farfield}
\end{eqnarray}
where $\infty$ and $e$ indexes stand for the property in the freestream and in the 
internal region, respectively. $q_n$ is the velocity component normal to the outer surface,
defined as
\begin{equation}
	q_n={\bf u} \cdot \vec{n} \, \mbox{,}
	\label{eq:qn-farfield}
\end{equation}
and $\vec{n}$ is the unit outward normal vector 
\begin{equation}
	\vec{n}=\frac{1}{\sqrt{\eta_{x}^2+\eta_{y}^2+\eta_{z}^2}}
	[\eta_x \ \eta_y \ \eta_z ]^T \, \mbox{.}
	\label{eq:norm-vec}
\end{equation}
Equation \eqref{eq:qn-farfield} assumes that the $\eta$ direction is pointing from the jet to the 
external boundary. Solving for $q_n$ and $a$, one can obtain
\begin{eqnarray}
	q_{n f} = \frac{\mathbf{R}^+ + \mathbf{R}^-}{2} \, \mbox{,} & \ & 
	a_f = \frac{\gamma-1}{4}(\mathbf{R}^+ - \mathbf{R}^-) \, \mbox{.}
	\label{eq: qn2-farfield}
\end{eqnarray}
The index $f$ is linked to the property at the boundary surface and will be used to update 
the solution at this boundary. For a subsonic exit boundary, $0<q_{n_e}/a_e<1$, the 
velocity components are derived from internal properties as
 \begin{eqnarray}
	 u_f&=&u_e+(q_{n f}-q_{n_e})\eta_x \, \mbox{,} \nonumber \\ 
	 v_f&=&v_e+(q_{n f}-q_{n_e})\eta_y \, \mbox{,} \\ 
	 w_f&=&w_e+(q_{n f}-q_{n_e})\eta_z \, \mbox{.} \nonumber
	 \label{eq:vel-farfield}
 \end{eqnarray}
Density and pressure properties are obtained by extrapolating the entropy from 
the adjacent grid node,
\begin{eqnarray}
	\rho_f = 
	\left(\frac{\rho_{e}^{\gamma}a_{f}^2}{\gamma p_e} \right)^{\frac{1}{\gamma-1}}
	\, \mbox{,} & \ &
	p_{f} = \frac{\rho_{f} a_{f}^2}{\gamma} \, \mbox{.} \nonumber
	 \label{eq:rhop-farfield}
\end{eqnarray}
For a subsonic entrance, $-1<q_{n_e}/a_e<0$, properties are obtained similarly 
from the freestream variables as
\begin{eqnarray}
	u_f&=&u_\infty+(q_{n f}-q_{n_\infty})\eta_x \, \mbox{,} \nonumber \\
	v_f&=&v_\infty+(q_{n f}-q_{n_\infty})\eta_y \, \mbox{,} \\
	w_f&=&w_\infty+(q_{n f}-q_{n_\infty})\eta_z \, \mbox{,} \nonumber
	\label{eq:vel2-farfield}
\end{eqnarray}
\begin{equation}
	\rho_f = 
	\left(\frac{\rho_{\infty}^{\gamma}a_{f}^2}{\gamma p_\infty} \right)^{\frac{1}{\gamma-1}}
	\, \mbox{.}
	\label{eq:rhop2-farfield}
\end{equation}
For a supersonic exit boundary, $q_{n_e}/a_e>1$, the properties are extrapolated 
from the interior of the domain as
\begin{eqnarray}
	\rho_f&=&\rho_e \, \mbox{,} \nonumber\\
	u_f&=&u_e \, \mbox{,} \nonumber\\
	v_f&=&v_e \, \mbox{,} \\
	w_f&=&w_e \, \mbox{,} \nonumber\\
	e_f&=&e_e \, \mbox{,} \nonumber   
	\label{eq:supso-farfield}
\end{eqnarray}
and for a supersonic entrance, $q_{n_e}/a_e<-1$, the properties are extrapolated 
from the freestream variables as
\begin{eqnarray}
	\rho_f&=&\rho_\infty \, \mbox{,}  \nonumber\\
	u_f&=&u_\infty \, \mbox{,}  \nonumber\\
	v_f&=&v_\infty \, \mbox{,} \\
	w_f&=&w_\infty \, \mbox{,} \nonumber\\
	e_f&=&e_\infty \, \mbox{.} \nonumber
	\label{eq:supso2-farfield}
\end{eqnarray}

\subsection{Entrance Boundary}

For a jet-like configuration, the entrance boundary is divided in two areas: the
jet and the area above it. The jet entrance boundary condition is implemented through 
the use of the 1-D characteristic relations for the 3-D Euler equations for a flat
velocity profile. The set of properties then determined is computed from within and 
from outside the computational domain. For the subsonic entrance, the $v$ and $w$ components
of the velocity are extrapolated by a zero-order extrapolation from inside the 
computational domain and the angle of flow entrance is assumed fixed. The rest of the properties 
are obtained as a function of the jet Mach number, which is a known variable. 
\begin{eqnarray}
	\left( u \right)_{1,j,k} & = & u_{j} \, \mbox{,} \nonumber \\
	\left( v \right)_{1,j,k} & = & \left( v \right)_{2,j,k} \,\mbox{,} \\
	\left( w \right)_{1,j,k} & = & \left( w \right)_{2,j,k} \, \mbox{.} \nonumber
	\label{eq:vel-entry}
\end{eqnarray}
The dimensionless total temperature and total pressure are defined with the isentropic relations:
\begin{eqnarray}
	T_t = 1+\frac{1}{2}(\gamma-1)M_{\infty}^{2} \, & \mbox{and} & 
	P_t = \frac{1}{\gamma}(T_t)^{\frac{\gamma}{\gamma-1}} \, \mbox{.}
	\label{eq:Tot-entry}
\end{eqnarray}
The dimensionless static temperature and pressure are deduced from Eq.\ \eqref{eq:Tot-entry},
resulting in
\begin{eqnarray}
	\left( T \right)_{1,j,k}=\frac{T_t}{1+\frac{1}{2}(\gamma-1)(u^2+v^2+w^2)_{1,j,k}} \, 
	& \mbox{and} & 
	\left( p \right)_{1,j,k}=\frac{1}{\gamma}(T)_{1,j,k}^{\frac{\gamma}{\gamma-1}} \, \mbox{.}
	\label{eq:Stat-entry}
\end{eqnarray}
For the supersonic case, all conserved variables receive jet property values.

The far field boundary conditions are implemented outside of the jet area in order to correctly
propagate information comming from the inner domain of the flow to the outter region of 
the simulation. However, in the present case, $\xi$, instead of $\eta$, as presented in 
the previous subsection, is the normal direction used to define the Riemann invariants.

\subsection{Exit Boundary Condition}

At the exit plane, the same reasoning of the jet entrance boundary is applied. This time, 
for a subsonic exit, the pressure is obtained from the outside and all other variables are 
extrapolated from the interior of the computational domain by a zero-order extrapolation. The 
conserved variables are obtained as
\begin{eqnarray}
	(\rho)_{I_{MAX},j,k} &=& \frac{(p)_{I_{MAX},j,k}}{(\gamma-1)(e)_{I_{MAX}-1,j,k}} \mbox{,} \\
	(\vec{u})_{I_{MAX},j,k} &=& (\vec{u})_{I_{MAX}-1,j,k}\mbox{,} \\
	(e_i)_{I_{MAX},j,k} &=& 
	(\rho)_{I_{MAX},j,k}\left[ (e)_{I_{MAX}-1,j,k}+
	\frac{1}{2}(\vec{u})_{I_{MAX},j,k}\cdot(\vec{u})_{I_{MAX},j,k} \right] \, \mbox{,}
	\label{eq:exit}
\end{eqnarray}
in which $I_{MAX}$ stands for the last point of the mesh in the axial direction. For 
the supersonic exit, all properties are extrapolated from the interior domain.

\subsection{Centerline Boundary Condition}

The centerline boundary is a singularity of the coordinate transformation, and, hence, 
an adequate treatment of this boundary must be provided. The conserved properties 
are extrapolated from the ajacent longitudinal plane and are averaged in the azimuthal 
direction in order to define the updated properties at the centerline of the jet.

\begin{figure}[ht]
       \begin{center}
       {\includegraphics[width=0.5\textwidth]{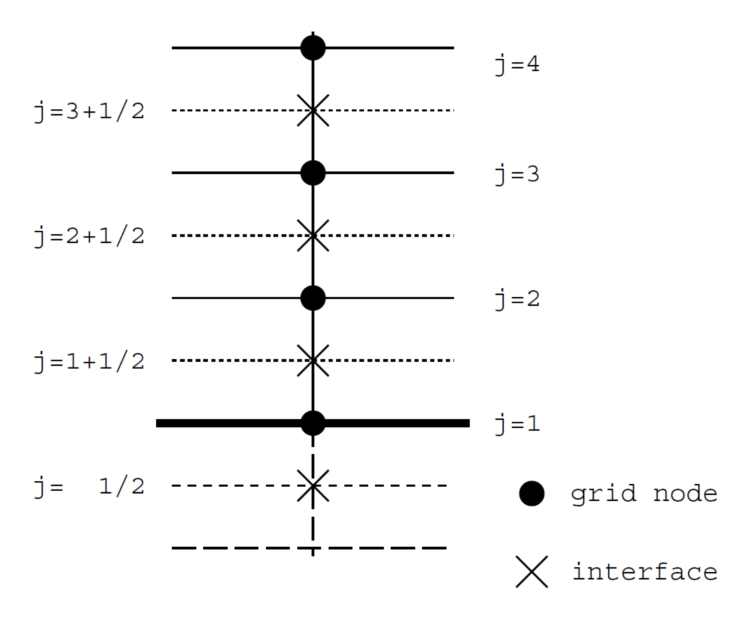}}\\
	   \caption{Boundary points dissipation \cite{BIGA02}.}\label{fig:centerline}
       \end{center}
\end{figure}

The fourth-difference terms of the artificial dissipation scheme, used in the present 
work, are carefully treated in order to avoid the five-point difference stencils at 
the centerline singularity. 
If one considers the flux balance at one grid point near the centerline boundary in 
a certain coordinate direction, let $w_{j}$ denote a component of the $\mathcal{W}$ 
vector from Eq.\ \eqref{eq:W_dissip} and $\mathbf{d}_{j}$ denote the corresponding artificial
dissipation term at the mesh point $j$. In the present example, 
$\left(\Delta w\right)_{j+\frac{1}{2}}$ stands for the difference between the solution
at the interface for the points $j+1$ and $j$. The fouth-difference of the dissipative
fluxes from Eq.\ \eqref{eq:dissip_term} can be written as
\begin{equation}
	\mathbf{d}_{j+\frac{1}{2}} = \left( \Delta w \right)_{j+\frac{3}{2}} 
	- 2 \left( \Delta w \right)_{j+\frac{1}{2}}
	+ \left( \Delta w \right)_{j-\frac{1}{2}} \, \mbox{.}
\end{equation}
Considering the centerline and the point $j=1$, as presented in Fig.\ 
\ref{fig:centerline}, the calculation of $\mathbf{d}_{1+\frac{1}{2}}$ demands the 
$\left( \Delta w \right)_{\frac{1}{2}}$ term, which is unknown since it is outside the
computation domain. In the present work a extrapolation is performed and given by
\begin{equation}
	\left( \Delta w \right)_{\frac{1}{2}} =
	- \left( \Delta w \right)_{1+\frac{1}{2}} \, \mbox{.}
\end{equation}
This extrapolation modifies the calculation of $\mathbf{d}_{1+\frac{1}{2}}$ that can be written as
\begin{equation}
	\mathbf{d}_{j+\frac{1}{2}} = \left( \Delta w \right)_{j+\frac{3}{2}} 
	- 3 \left( \Delta w \right)_{j+\frac{1}{2}} \, \mbox{.}
\end{equation}
The approach is plausible since the centerline region is smooth and does not have high
gradient of properties.

\subsection{Periodic Boundary Condition}

A periodic condition is implemented between the first ($K=1$) and the last point in the 
azimutal direction ($K=K_{MAX}$) in order to close the 3-D computational domain. There 
are no boundaries in this direction, since all the points are inside the domain. The first
and the last points, in the azimuthal direction, are superposed in order to facilitate
the boundary condition implementation which is given by
\begin{eqnarray}
	(\rho)_{i,j,K_{MAX}} &=& (\rho)_{i,j,1} \, \mbox{,} \nonumber\\
	(u)_{i,j,K_{MAX}} &=& (u)_{i,j,1} \, \mbox{,} \nonumber\\
	(v)_{i,j,K_{MAX}} &=& (v)_{i,j,1} \, \mbox{,} \\
	(w)_{i,j,K_{MAX}} &=& (w)_{i,j,1} \, \mbox{,} \nonumber\\
	(e)_{i,j,K_{MAX}} &=& (e)_{i,j,1} \, \mbox{.} \nonumber
	\label{eq:periodicity}
\end{eqnarray}

  \section{Study of Supersonic Jet Flow}

Four numerical studies are performed in the present research in 
order to study the use of 2nd-order spatial discretization on large 
eddy simulations of a perfectly expanded jet flow configuration.
The effects of mesh refinement and SGS models are compared in the 
present work. Two different meshes are created for the refinement 
study. The three SGS models implemented in the code, classic 
Smagorinsky, dynamic Smagorinsky and Vreman, are compared in the 
current section. Results are compared with analytical, numerical 
and experimental data from the literature \cite{Mendez10,Mendez12,
bridges2008turbulence}. 

\subsection{Geometry Characteristics}

Two different geometries are created for the simulations discussed
in the current work. One geometry presents a cylindrical shape 
and the other one presents a divergent conical shape.
For the sake of simplicity, the round geometry is named geometry 
A and the other one is named geometry B in present text.
The computational domains are created in two steps. First, a 2-D 
region is generated. In the sequence, this region is rotated in 
order to generate a fully 3-D geometry. An in-house code is used 
for the generation of the 2-D domain of geometry A. The commercial 
mesh generator ANSYS\textsuperscript{\textregistered} ICEM CFD 
\cite{ICEM} is used for the 2-D domain of geometry B. 

\begin{figure}[htb!]
       \begin{center}
		   \subfigure[3-D view of two XZ plans of geometry A.]{
           \includegraphics[width=0.475\textwidth]
		   {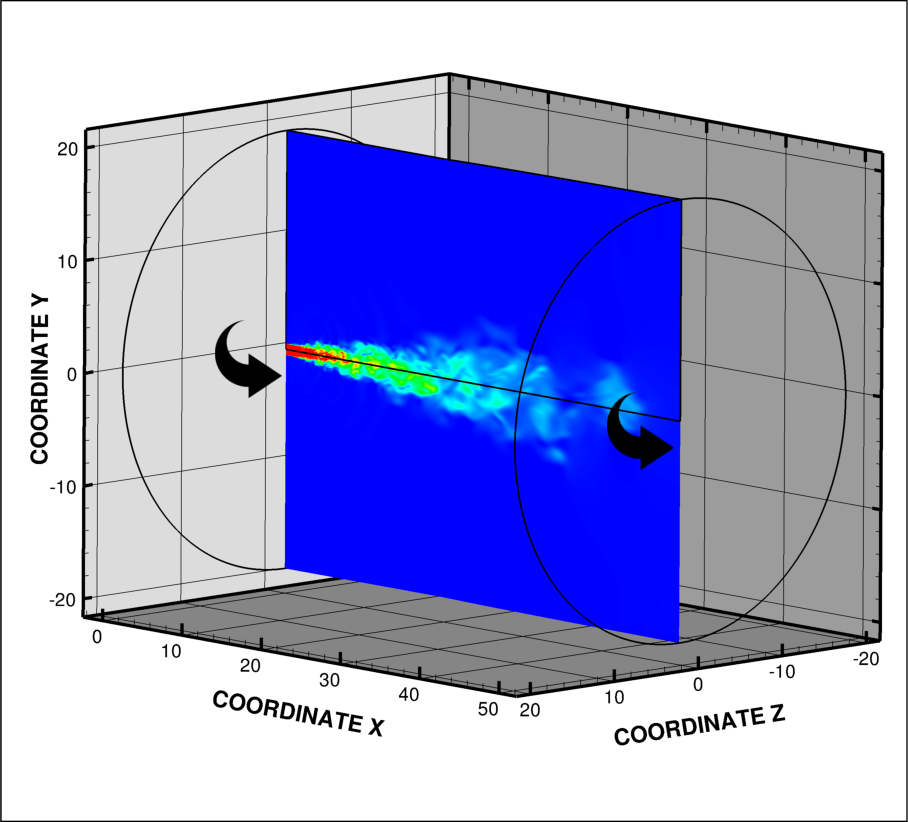} 
		   \label{fig:3d-geom-14m}
		   }
		   \subfigure[3-D view of two XZ plans of geometry B.]{
           \includegraphics[width=0.475\textwidth]
		   {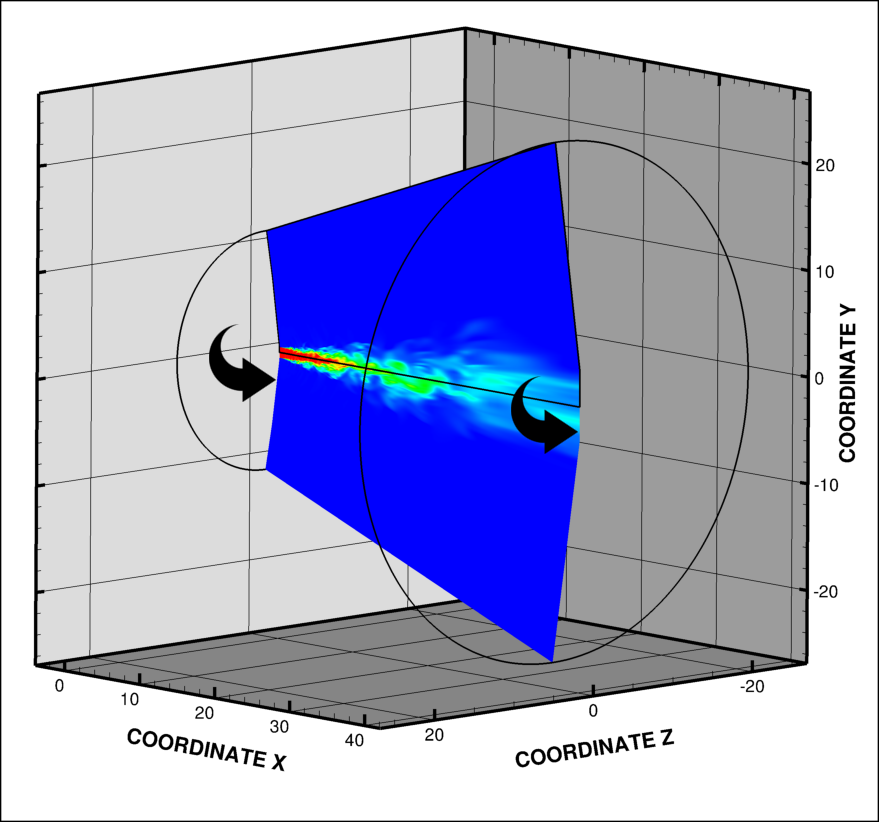} 
		   \label{fig:3d-geom-50m}
		   }
		   \caption{3-D view of geometries used for the LES.}
		   \label{fig:geom}
	   \end{center}
\end{figure}

The geometry A is a cylindrical domain with radius of $20D$ and and length 
of $50D$. 
%
Geometry B presents a divergent form whose axis length is $40D$\@. The minimum 
and maximum heights of geometry B are $\approx 16D$ and $25D$, respectively.
The zones of this geometry are created based on results from simulations using
geometry A in order to refine the mesh in the shear layer region of the 
flow. 
%
Geometry A and geometry B are illustrated in Fig.\ \ref{fig:geom} which 
presents a 3-D view of the two computational domains used in the 
current work. The geometries are colored by a time solution of the axial 
component of velocity of the flow. 
%
%

\subsection{Mesh Configurations}

\begin{figure}[htb!]
       \begin{center}
		   \subfigure[2-D view of mesh A in the XZ plan.]{
           \includegraphics[width=0.475\textwidth]
		   {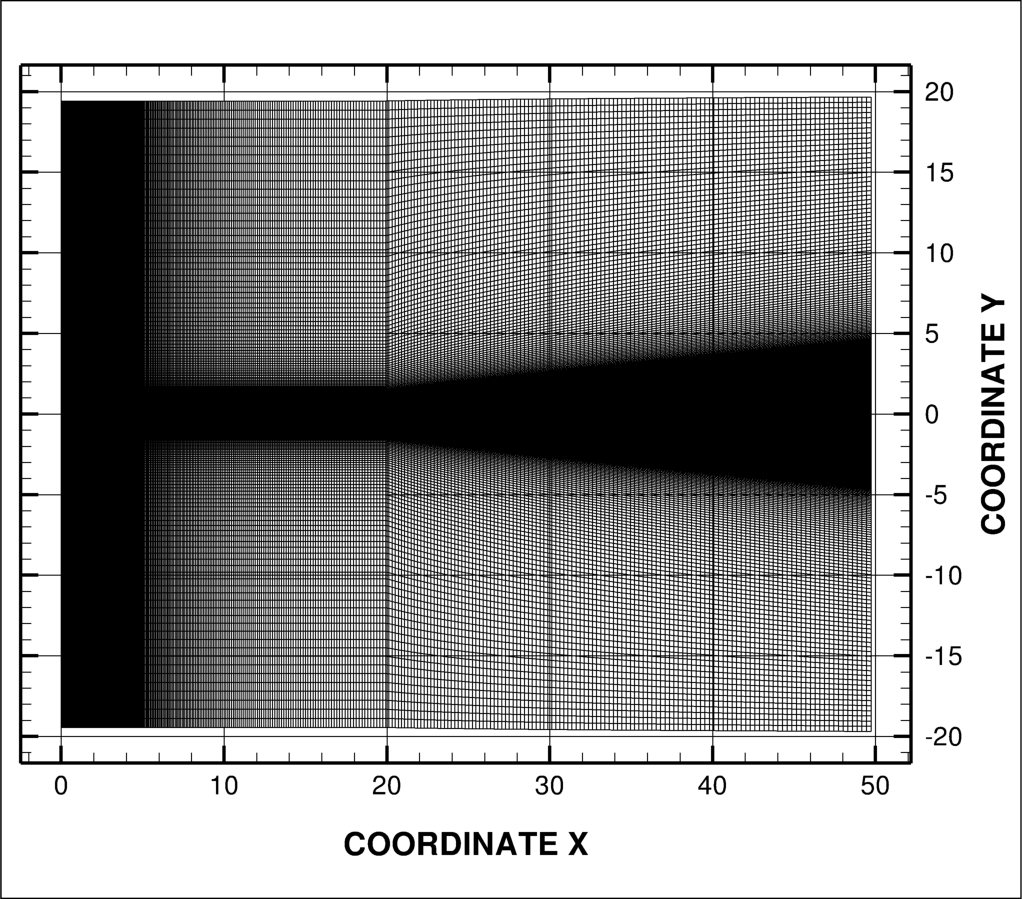} 
		   \label{fig:2d-mesh-14m}
		   }
		   \subfigure[2-D view of mesh B in the XZ plan.]{
           \includegraphics[width=0.475\textwidth]
		   {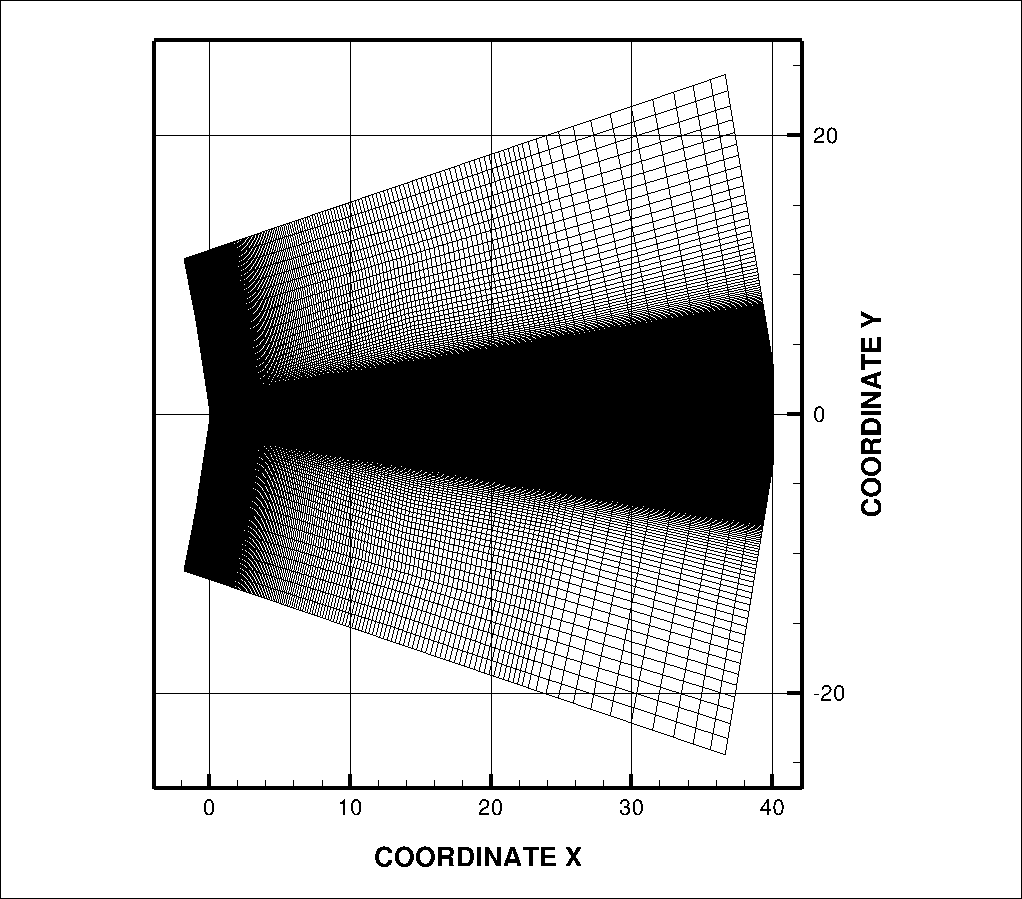} 
		   \label{fig:2d-mesh-50m}
		   }
		   \caption{2-D view of the computational meshes 
		   used in the current work.}
		   \label{fig:mesh-ab}
	   \end{center}
\end{figure}

One grid is generated for each geometry used in the present study. These
computational grids are named mesh A and mesh B. The second mesh is created 
based on results using mesh A. One illustration of the computational 
grids is presented in Fig.\ \ref{fig:mesh-ab}. 
Mesh A is created using a mesh 
generator developed by the research group for the cylindrical shape 
configuration. This computational mesh is composed by 400 points in the axial 
direction, 200 points in the radial direction and 180 points in the azimuthal 
direction, which originates 14.4 million grid points. Hyperbolic tangent 
functions are used for the points distribution in the radial and axial 
directions. Grid points are clustered near the shear layer of the jet. The 
mesh is coarsened towards the outer regions of the domain in order to dissipate 
properties of the flow far from the jet. Such mesh refinement approach can avoid 
reflection of information into the domain. The radial and longitudinal dimensions 
of the smallest distance between mesh points of the computational grid are 
given by $(\Delta \underline{r})_{min}=0.002D$ and $(\Delta 
\underline{x})_{min}=0.0126D$, respectively. This minimal spacing occurs at the 
shear layer of the jet and at the entrance of the computational domain. Mesh A 
is created based on a reference grid of Mendez {\it et al.} \cite{Mendez10,
Mendez12}.

The refined computational grid is composed by 343 points in the axial 
direction, 398 points in the radial direction and 360 points in the 
azimuthal direction, which yields approximately 50 million grid 
points. The 2-D mesh is generated with ANSYS\textsuperscript{\textregistered} 
ICEM CFD \cite{ICEM}. The points are allocated using different 
distributions in eight edges of the 2-D domain. The same coarsening 
approach used for mesh A is also applied for mesh B. The distance 
between mesh points increase towards the outer region of the domain. 
This procedure force the dissipation of properties far from the jet 
in order to avoid reflection of data into the domain. The reader can 
find more details about the mesh generation on the work of Junqueira-Junior
\cite{Junior16}.



\subsection{Flow Configuration and Boundary Conditions}

An unheated perfectly expanded jet flow is chosen to validate the
LES tool. The flow is characterized by an unheated perfectly 
expanded inlet jet with a Mach number of $1.4$ at the domain 
entrance. Therefore, the pressure ratio, $PR=P_{j}/P_\infty$, 
and the temperature ratio, $TR=T_{j}/T_\infty$, between the jet 
exit and the ambient freestream conditions, are equal to one, {\em i.e.}, $PR = 1$ and $TR=1$. 
The Reynolds number of the jet is $Re=1.57 \times 10^{6}$, based on the jet exit diameter. This 
flow configuration is chosen due to the absence of strong shocks 
waves. Strong discontinuities must be carefully treated using 
numerical approaches which are not yet implemented into the 
solver. Moreover, numerical and experimental data of this flow
configuration are available in the literature such as the work 
of Mendez {\it et al.} \cite{Mendez10, Mendez12} and the work of 
Bridges and Wernet \cite{bridges2008turbulence}.

The boundary conditions discussed in section \ref{sec:BC} are used 
in the simulations performed in the current thesis. Figure 
\ref{fig:bc} presented a lateral view and a frontal view of the 
computational domain used by the simulation in where the positioning 
of each boundary condition is indicated. A flat-hat velocity profile, 
with $M=1.4$, is used at the entrance boundary. Riemann invariants are 
used at the farfield regions. A special singularity treatment is 
performed at the centerline. A periodicity is imposed in the azimuthal 
direction in order to create a transparency for the flow. 
%

Properties of flow at the inlet and at the farfield regions have to be 
provided to the code in order to impose the boundary conditions. Density, 
$\rho$, temperature, $T$, velocity, $U$, Reynolds number $Re$, and specific 
heat at constant volume, $C_{v}$, are provided in the dimensionless form 
to the simulation. These properties are given by
\begin{eqnarray}
	\rho_{j} = 1.00 \, \mbox{,} & 
	\rho_{\infty} = 1.00 \, \mbox{,} \nonumber \\
	T_{j} = 1.00 \, \mbox{,} & 
	T_{\infty} = 1.00 \, \mbox{,} \\
	U_{j} = 1.4 \, \mbox{,} & 
	U_{\infty} = 0.00 \nonumber \, \mbox{,}\\
	Re_{j}=1.57 \times 10^{6} \, \mbox{,}& 
	C_{v} = 1.786 \, \mbox{,} \nonumber
\end{eqnarray}
where the $j$ subscript stands for property at the jet entrance and the 
$\infty$ subscript stands for property at the farfield region. 

\subsection{Large Eddy Simulations}

Four simulations are performed in the present. The objective is to 
study the effects of mesh refinement and to evaluate the three 
different SGS models included into the code. The calculations are 
performed in two steps. First a preliminary simulation is performed 
in order to achieve a statistically steady state condition. In the 
sequence, the simulations are run for another period in order to 
collect enough data for the calculation of time averaged properties 
of the flow and its fluctuations. 


The configurations of all simulations are discussed in the current section,
towards the description of the preliminary calculations which are performed
in order to drive the flow to a statistically steady flow condition. Table 
\ref{tab:simu} presents the operating conditions of all four numerical 
studies performed in the current research. Mesh A is only used on S1. 
The other calculations are performed using the refined grid, Mesh B. The 
stagnated flow condition is used as initial condition for all simulations 
but S3, which uses the solution of S2 after 10.15 flow through times (FTT). 
One flow through time is the necessary time for a particle to cross all
the domain considering the inlet velocity of the jet. The dimensionless 
time increment used for all configurations is the biggest one which the 
solver can handle without diverging the solution. The static Smagorinsky
model \cite{Smagorinsky63,Lilly65,Lilly67} is used on S1 and S2. The
dynamic Smagorinsky model \cite{Germano91,moin91} and the Vreman model 
\cite{vreman2004} are used on S3 and S4, respectively.

The last column of Tab.\ \ref{tab:simu} represents the period simulated 
by all numerical studies in order to achieve the statistically steady state 
flow condition. The choice of this period is related to the computational 
cost of each study. S1 is the least expensive test case studied. It uses a 
14 million point mesh while the other simulations use the 50 million point grid. 
Therefore, S1 has been run for a longer period in order to achieve the statistically 
steady state condition. On the other hand, S3 is the most expensive numerical test 
case. The dynamic Smagorinsky SGS model, which is used by S3, needs more time per iteration 
when compared with the other SGS models implemented in the code. Hence, S3 has only been run 
for 5.86 FTT for this preliminary simulation. 
\begin{table}[htbp]
\begin{center}
	\caption{Configuration of large eddy simulations performed 
	in the present work}
\label{tab:simu}
\begin{tabular}{|c|c|c|c|c|c|}
\hline
Simulation & Mesh & SGS & $\Delta t$ & Initial condition & FTT \\
\hline
S1 & A & Static Smagorinsky  & $2.5 \times 10^{-5}$ & \small Stagnated flow & 37.8 \\
S2 & B & Static Smagorinsky  & $1 \times 10^{-4}$ & \small Stagnated flow  & 10.15 \\
S3 & B & Dynamic Smagorinsky & $5 \times 10^{-5}$ & \small Stagnated flow  & 5.86  \\
S4 & B & Vreman & $1 \times 10^{-4}$ & \small S2 -- 10.15 FTT & 13.65\\
\hline
\end{tabular}
\end{center}
\end{table}


The simulations are restarted and run for another period in which 
data of the flow are extracted and recorded in a fixed frequency
after the preliminary study. The collected data are time averaged 
in order to calculate mean properties of the flow and compare with 
the results of the numerical and experimental references.

In the present work, time averaged properties are notated as 
$\langle \cdot \rangle$. Table \ref{tab:mean-simu} presents the 
configuration of simulations performed in order to calculate mean 
flow properties. The second column presents the number of extractions 
performed during the simulations. Data are extracted each 0.02 
dimensionless time in the present work which is equivalent to a 
dimensionless frequency of 50\@. The choice of this frequency is based 
on the numerical work reported in Refs.\ \citen{Mendez10} and \citen{Mendez12}. The last two 
columns of Tab.\ \ref{tab:mean-simu} present the total dimensionless 
time simulated to calculate the mean properties.
\begin{table}[htbp]
\begin{center}
	\caption{Time average configuration}
\label{tab:mean-simu}
\begin{tabular}{|c|c|c|c|}
\hline
Simulation & Nb. Extractions & Frequency & Total time \\
\hline
S1 & 2048 & 50 & 40.96 (1.14 FTT)\\
S2 & 3365 & 50 & 67.3  (2.36 FTT)\\
S3 & 2841 & 50 & 56.6  (1.98 FTT)\\
S4 & 1543 & 50 & 30.86 (1.08 FTT)\\
\hline
\end{tabular}
\end{center}
\end{table}



A power spectral density (PSD) of the time fluctuation of the 
axial component of velocity, $u^{*}$, is calculated in order to 
study the transient part of the flow. The PSD computation is 
performed using the following methodology: first, sensors are 
included at three different positions along the lipline of the 
jet ($r/D=0.5$). For each position along the lipline, 120 sensors 
are allocated in the azimuthal direction. Information at this 
direction are averaged in order eliminate azimuthal dependence. 
Table \ref{tab:sensor} presents the positioning of the sensors 
in the axial and radial directions. The choice of the positioning 
is based on the numerical reference of Mendez {\it et. al.} 
\cite{Mendez10,Mendez12}. In the sequence, data are extracted 
from the sensors in order to generate a time-dependent signal.
The signal is partitioned into three equal parts and $u^{*}$ is 
calculated for all three partitioned signals. In the next step
of the methodology, the time fluctuation signals are multiplied 
by a window function in order to create periodic distribution. 
The Hamming window function \cite{harris1978} is used in the 
present work and it is written as
\begin{equation}
	w(n) = \alpha - \beta cos\left(\frac{2 \pi n}{N-1}\right)
	\, \mbox{,}
	\label{eq:window}
\end{equation}
where $\alpha=0.54$, $\beta=0.46$, $n$ stands for the time index 
and $N$ stands for the size of the sample. After applying the fast 
Fourier transformation (FFT) on the signals one can calculate the 
PSD of $u^{*}$. In the end, a simple average is applied on the three 
signals in order to have a final PSD of $u^{*}$ distribution.
\begin{table}[htbp]
\begin{center}
\caption{Positionig of the sensors used to collect fluctuation data}
\label{tab:sensor}
\begin{tabular}{|c|c|}
\hline
Signal & Positioning \\
\hline
(a) & $(X/D=0.10,r/D=0.5)$\\
(b) & $(X/D=0.25,r/D=0.5)$\\
(c) & $(X/D=1.25,r/D=0.5)$\\
\hline
\end{tabular}
\end{center}
\end{table}

In the present work the transient part is studied by the PSD of 
$u^{*}$ distribution as function of the number of Strouhal which 
is given by
\begin{equation}
	St(t) = \frac{\mathit{f}(t) D}{U_{j}} \, \mbox{,}
	\label{eq:strouhal}
\end{equation}
where $\mathit{f}$ stands for the frequency as a function of the time,
$D$ stands for the inlet diameter and $U_{j}$ stands for the velocity of
the jet at the entrance of the domain. Data are collected from the 
sensors using the same informations provided by Tab.\ \ref{tab:mean-simu}.
The minimum and maximum values of the Strouhal number for all simulations
are presented in Tab.\ \ref{tab:psd}.
\begin{table}[htbp]
\begin{center}
	\caption{Strouhal limits for all simulations}
\label{tab:psd}
\begin{tabular}{|c|c|c|}
\hline
Simulation & ${St}_{min}$ & ${St}_{max}$ \\
\hline
S1 & $1.74$ & 17.86\\
S2 & $1.06\times 10^{-2}$ & 17.86\\
S3 & $1.26\times 10^{-2}$ & 17.86\\
S4 & $2.31\times 10^{-2}$ & 17.86\\
\hline
\end{tabular}
\end{center}
\end{table}

\subsection{Study of Mesh Refinement Effects} \label{sec:mesh-study}

\begin{figure}[htb!]
  \centering
    {\includegraphics[width=0.7\textwidth]
	{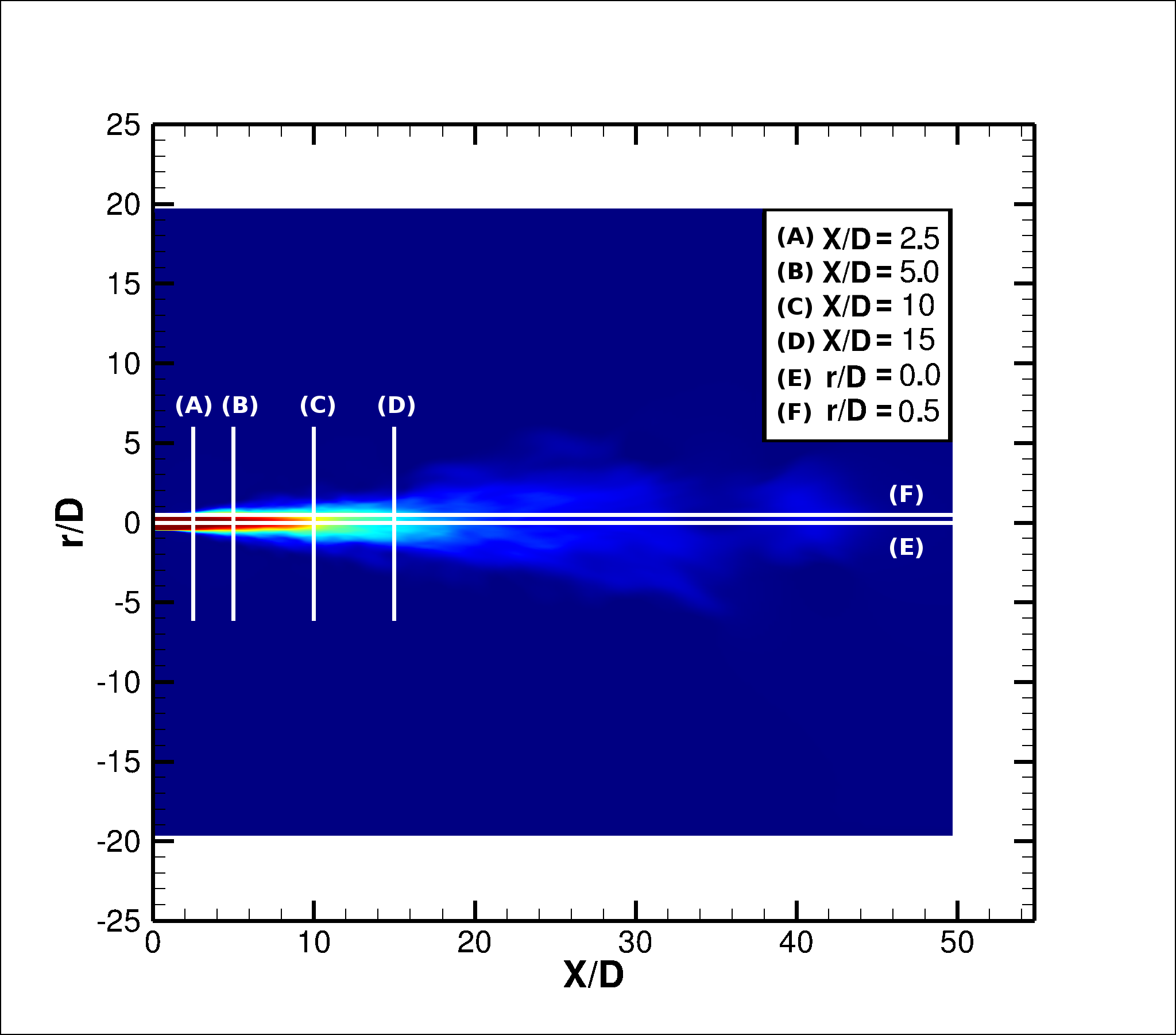}}
	\caption{Positioning in the computational domain of surfaces 
	studied in the present work}
	\label{fig:extract}
\end{figure}

Effects of mesh refinement on compressible LES using the JAZzY 
solver are discussed in the present section. 2-D distribution
of properties and profiles of S1 and S2 are collected and 
compared with numerical and experimental results from the 
literature \cite{Mendez10,Mendez12,bridges2008turbulence}. 
Both simulations use the same SGS model, the static Smagorinsky
model \cite{Smagorinsky63,Lilly65,Lilly67}. Mesh A is used on S1 
and Mesh B is used on S2. Time averaged distributions of the axial 
component of velocity, density and eddy viscosity are presented in 
the subsection along with the RMS distribution of all three 
components of velocity, distributions of the $\langle u^{*}v^{*}
\rangle$ component of the Reynolds stress tensor and distributions
of the turbulent kinetic energy, $k$. Figure \ref{fig:extract} 
illustrates the positioning of surfaces and profiles extracted 
for all simulations performed in the current work.

\subsubsection*{Time Averaged Axial Component of Velocity}

One important characteristic of a round jet flow configurations is 
the potential core length, $\delta_{j}^{95\%}$. The potential 
core, $U_{j}^{95\%}$, is defined as $95\%$ of the velocity of 
the jet at the inlet,
\begin{equation}
	U_{j}^{95\%} = 0.95 \cdot U_{j}
	\, \mbox{.}
	\label{eq:pot-core}
\end{equation}
Therefore, the potential core length can be defined as the 
positioning in the centerline where $U_{j}^{95\%}$ is located.

\begin{figure}[htb!]
  \centering
  \subfigure[Lateral view of $\langle U \rangle$ for S1.]
    {\includegraphics[width=0.495\textwidth]
	{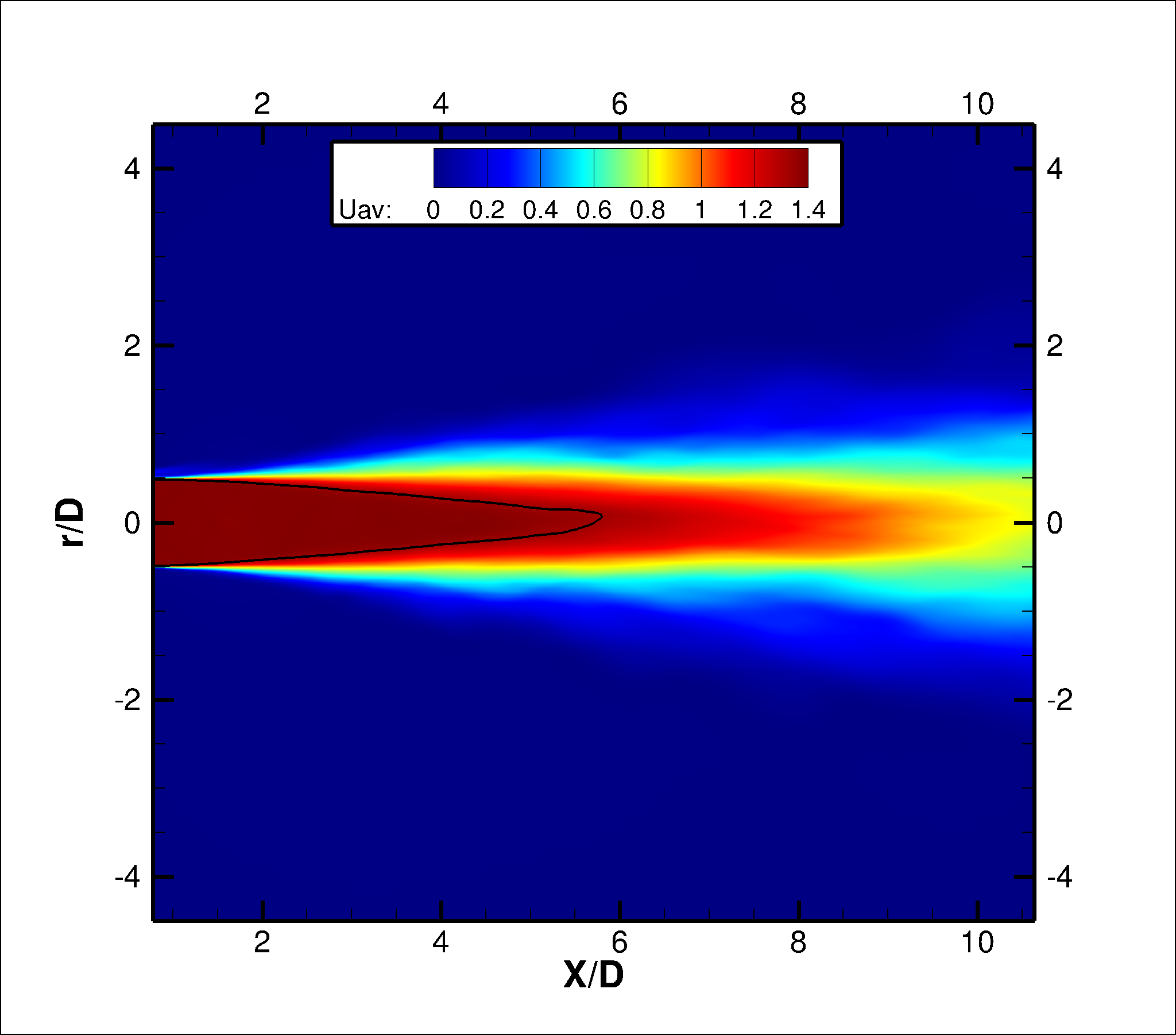}}
  \subfigure[Lateral view of $\langle U \rangle$ for S2.]
    {\includegraphics[width=0.495\textwidth]
	{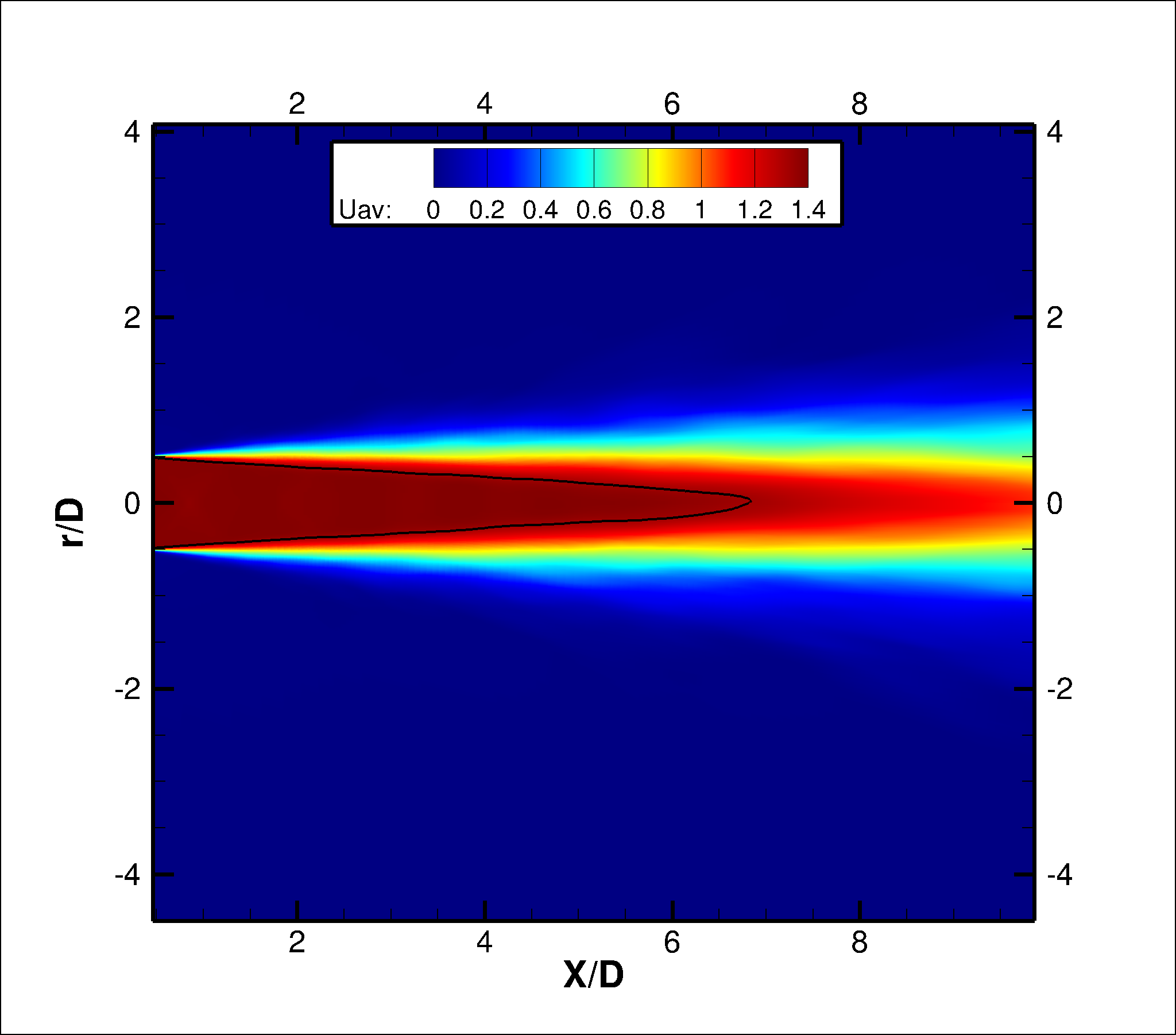}}
	\caption{Lateral view of the averaged axial component of 
	velocity, $\langle U \rangle$, for S1 and S2. 
	(\LARGE \textbf{--}\scriptsize) indicates the 
	potential core of the jet, $U_{j}^{95\%}$.} 
	\label{fig:lat-u-av-mesh}
\end{figure}

Time averaged results of the axial component of velocity are 
presented in the subsection. A lateral view of $\langle U \rangle$ 
for S1 and S2, side by side, are presented in Fig.\ \ref{fig:lat-u-av-mesh}, 
where $U_{j}^{95\%}$ is indicated by the solid line. The positioning of surfaces is 
indicated in Fig.\ \ref{fig:extract}. Table \ref{tab:core-mesh} presents the size of 
the potential core of S1, S2 and the numerical results from Refs.\ \citen{Mendez10} 
and \citen{Mendez12}, along with the relative error compared with the experimental
data \cite{bridges2008turbulence}.
\begin{table}[htbp]
\begin{center}
  \caption{Potential core length and relative error of S1 and S2.}
  \label{tab:core-mesh}
  \begin{tabular}{|c|c|c|}
  \hline
  Simulation & $\delta_{j}^{95\%}$ & Relative error\\
  \hline
  S1 & 5.57 & 40\%\\
  S2 & 6.84 & 26\%\\
  Mendez {\it et al.} & 8.35 & 8\%\\
  \hline
  \end{tabular}
\end{center}
\end{table}
%
%

Comparing the results, one can observe the difference in the 
potential core length between S1 and S2. The results of the 
first case present a smaller $\delta_{j}^{95\%}$ when compared to 
results of S2, {\em i.e.}, 5,57 and 6.84, respectively. One can say that 
the S1 solution is over dissipative when compared to the S2 
results. The jet vanishes earlier in S1\@. The mesh which is 
used in the S1 test case is very coarse when compared with the grid used 
for S2\@. This lack of resolution can generate very 
dissipative solutions which yield the under prediction of 
the potential core length. The mesh refinement reduced in 
14\% the relative error of S2 when compared to the 
experimental data. 

\begin{figure}[htb!]
  \centering
  \subfigure[$\langle U \rangle$ - X=2.5D ; $-1.5D\leq Y\leq 1.5D$]
    {\includegraphics[width=0.45\textwidth]
	{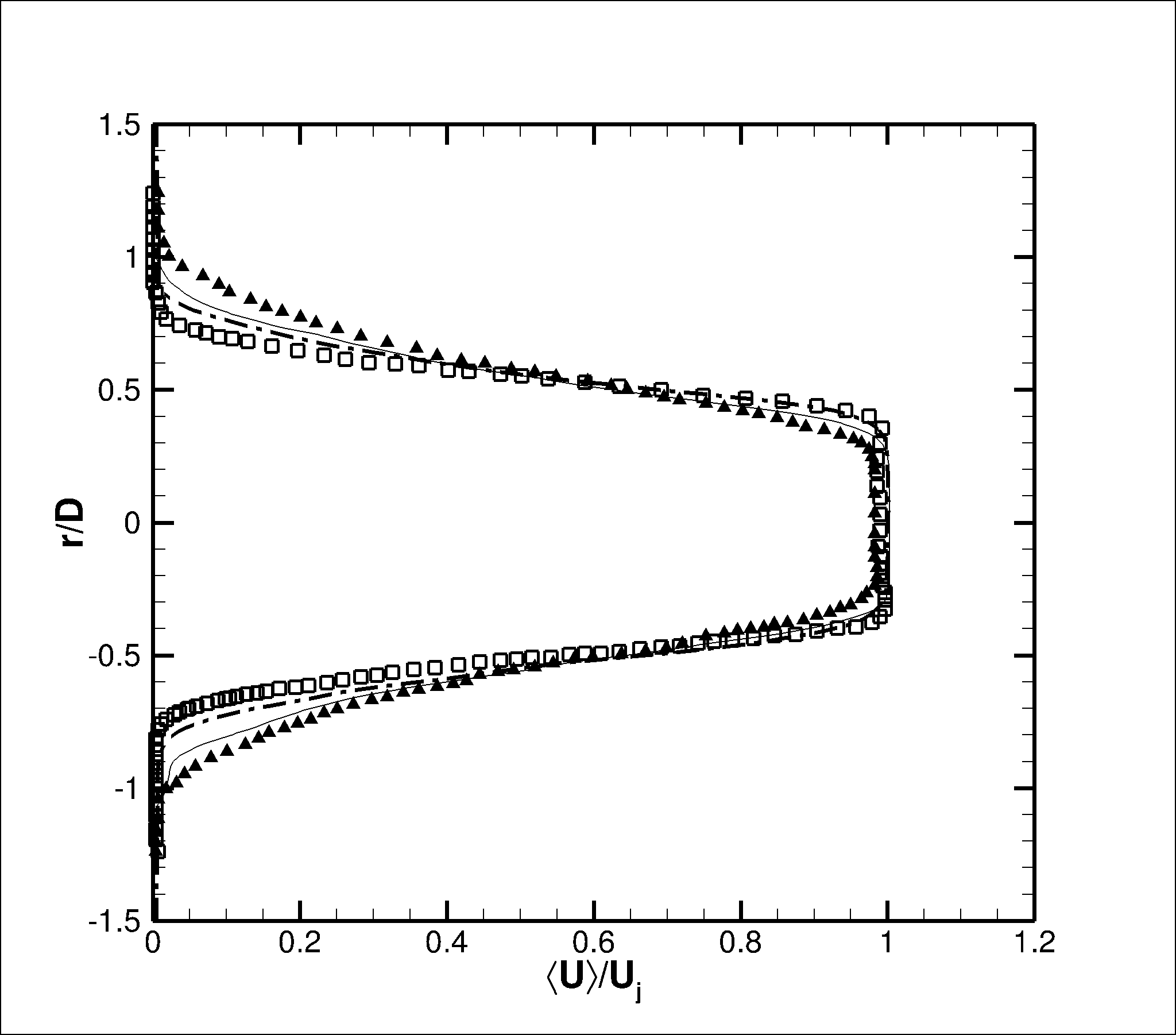}}
  \subfigure[$\langle U \rangle$ - X=5.0D ; $-1.5D\leq Y\leq 1.5D$]
    {\includegraphics[width=0.45\textwidth]
	{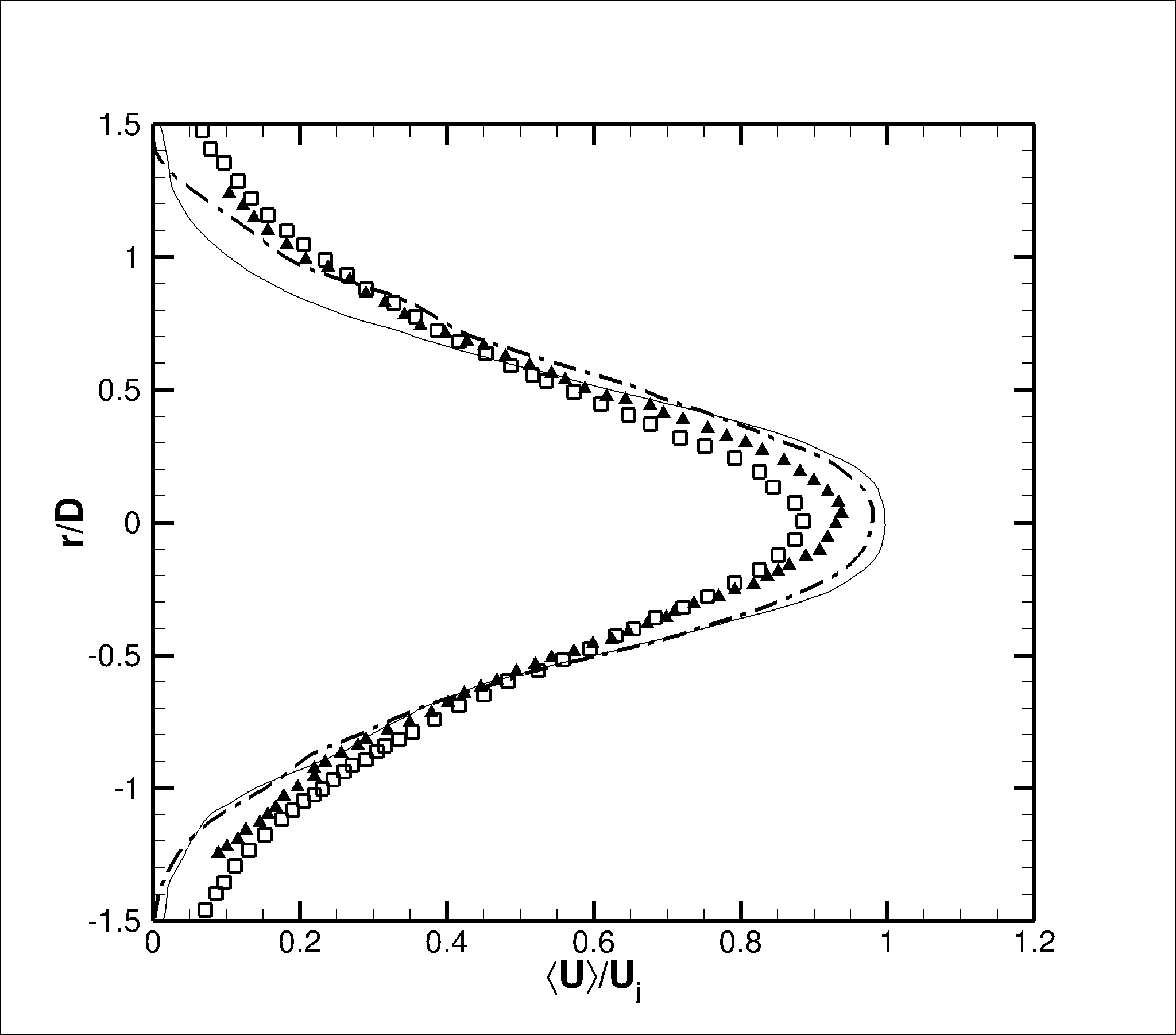}}
  \subfigure[$\langle U \rangle$ - X=10D ; $-1.5D\leq Y\leq 1.5D$]
    {\includegraphics[width=0.45\textwidth]
	{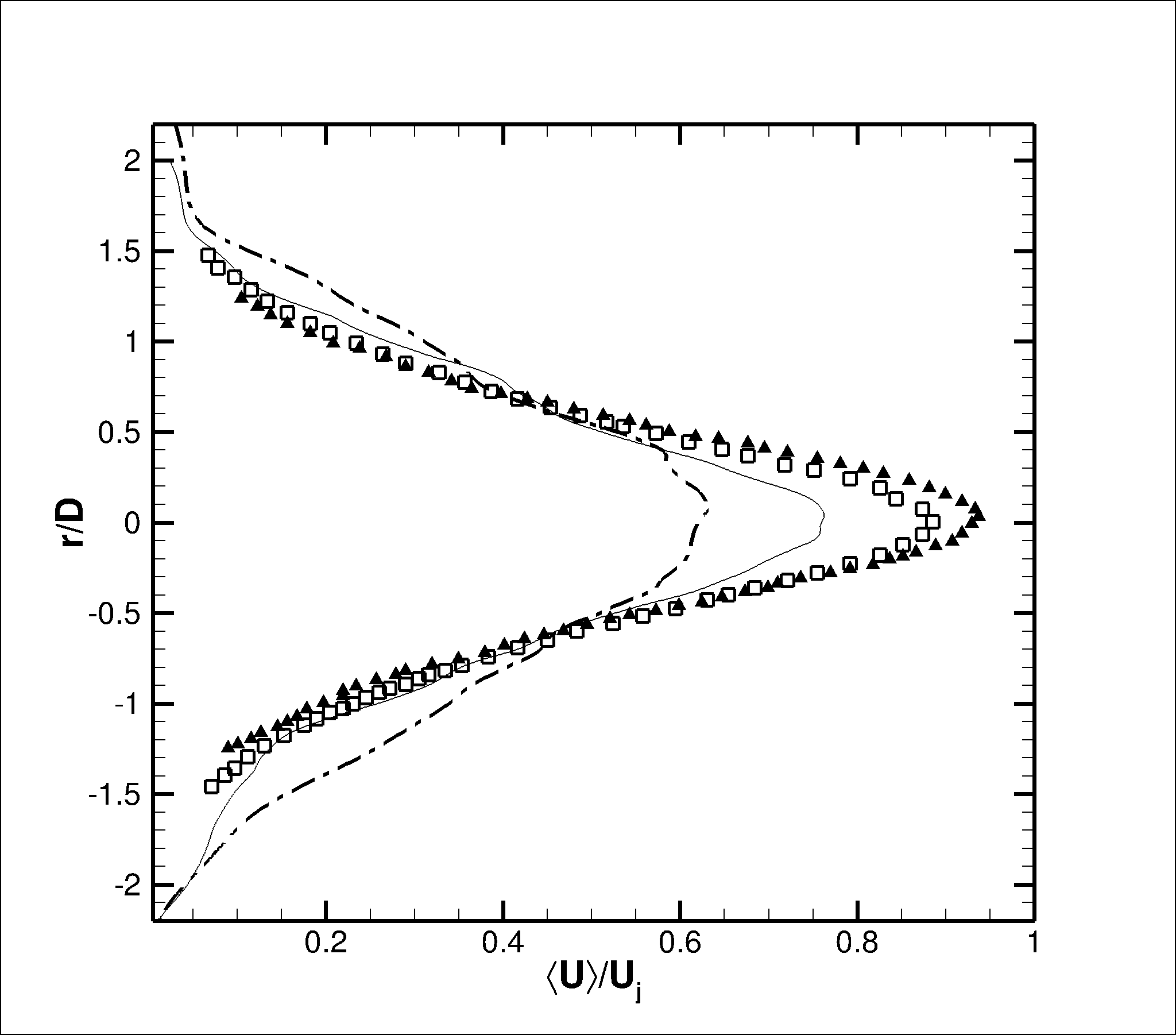}}
  \subfigure[$\langle U \rangle$ - X=15D ; $-1.5D\leq Y\leq 1.5D$]
    {\includegraphics[width=0.45\textwidth]
	{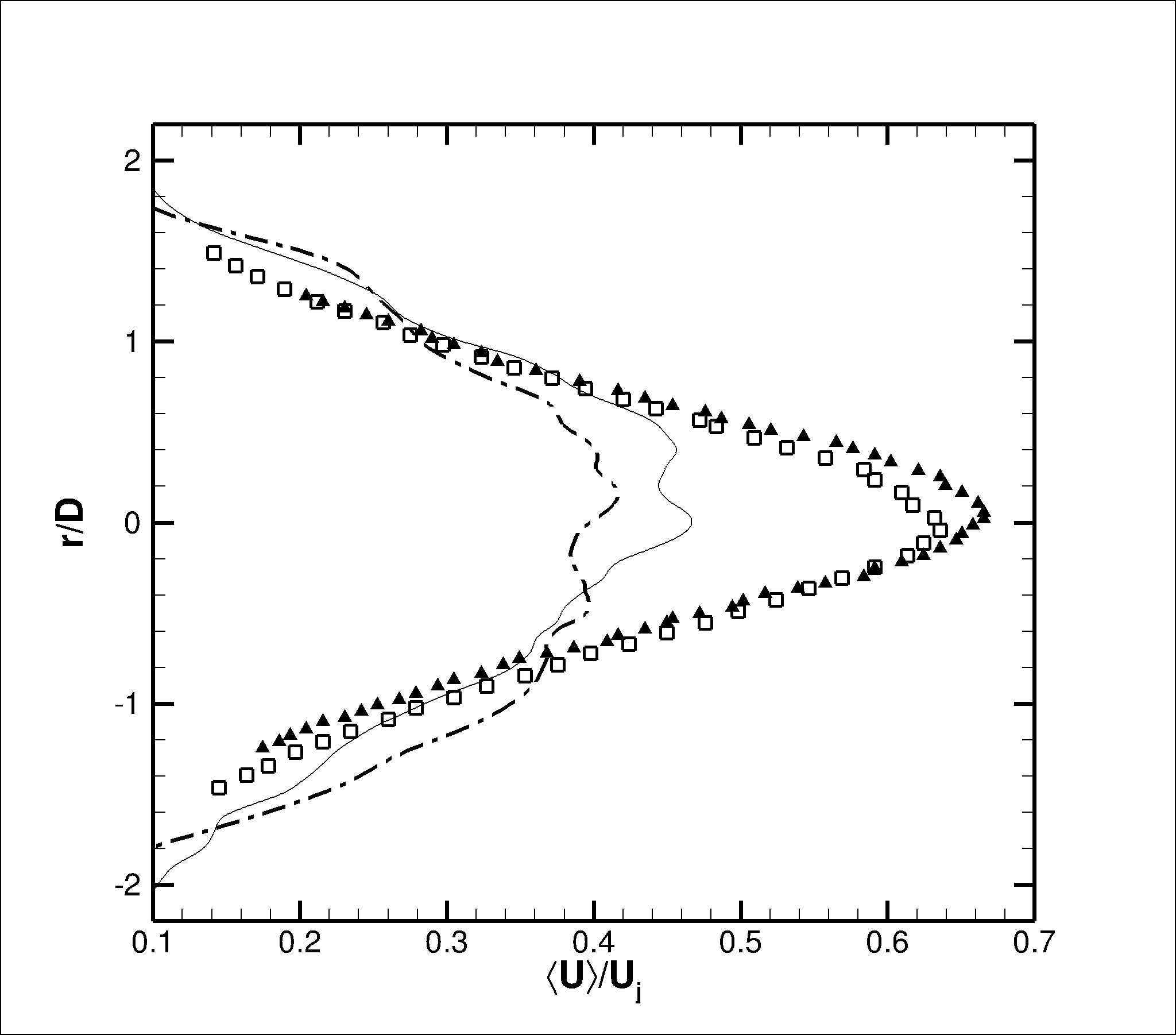}}
  \subfigure[$\langle U \rangle$ - Centerline - Y=0 ; $0\leq X \leq 20D$ ]
    {\includegraphics[width=0.45\textwidth]
	{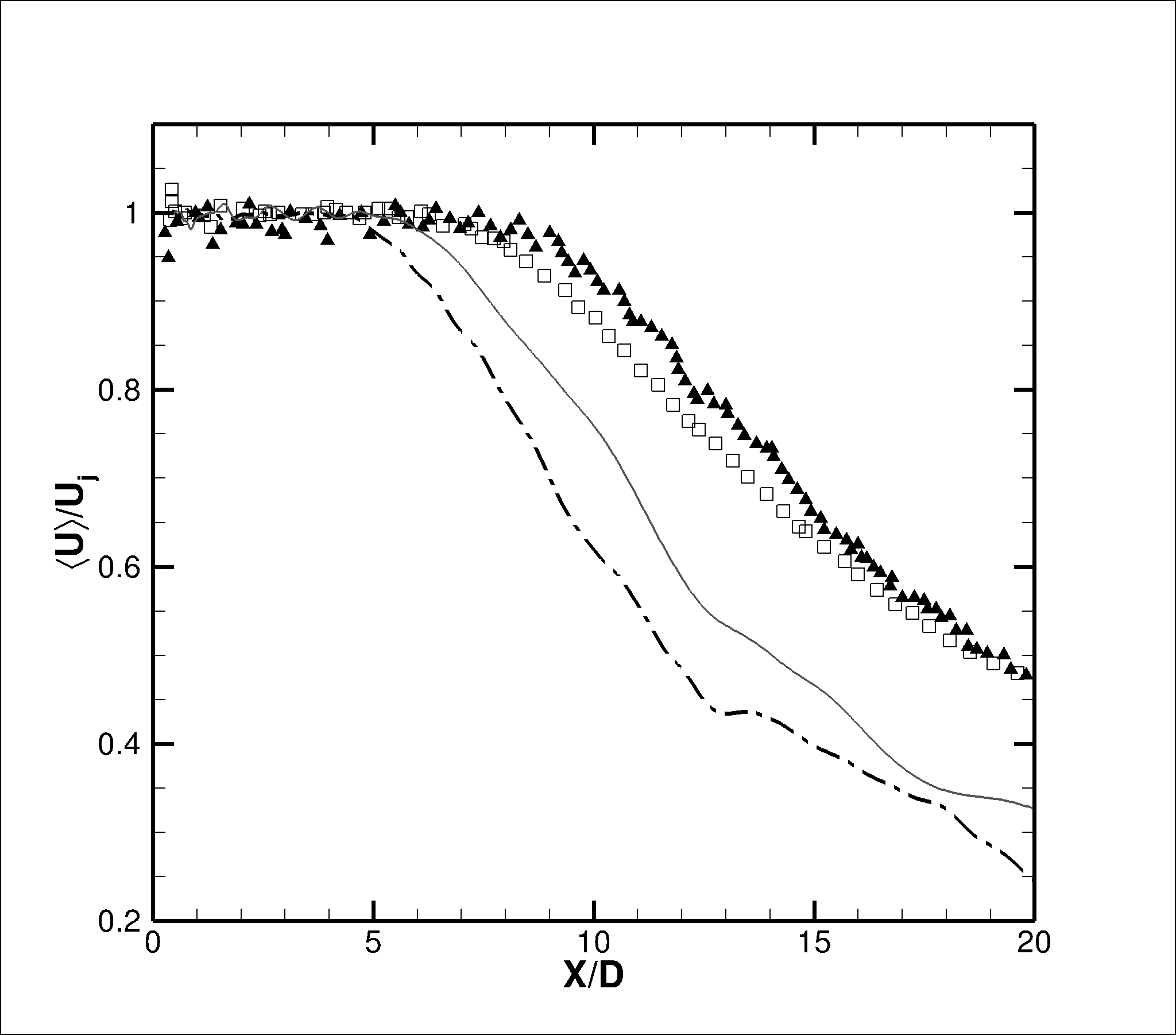}}
  \subfigure[$\langle U \rangle$ - Lipline - Y=0.5D ; $0\leq X \leq 20D$ ]
    {\includegraphics[width=0.45\textwidth]
	{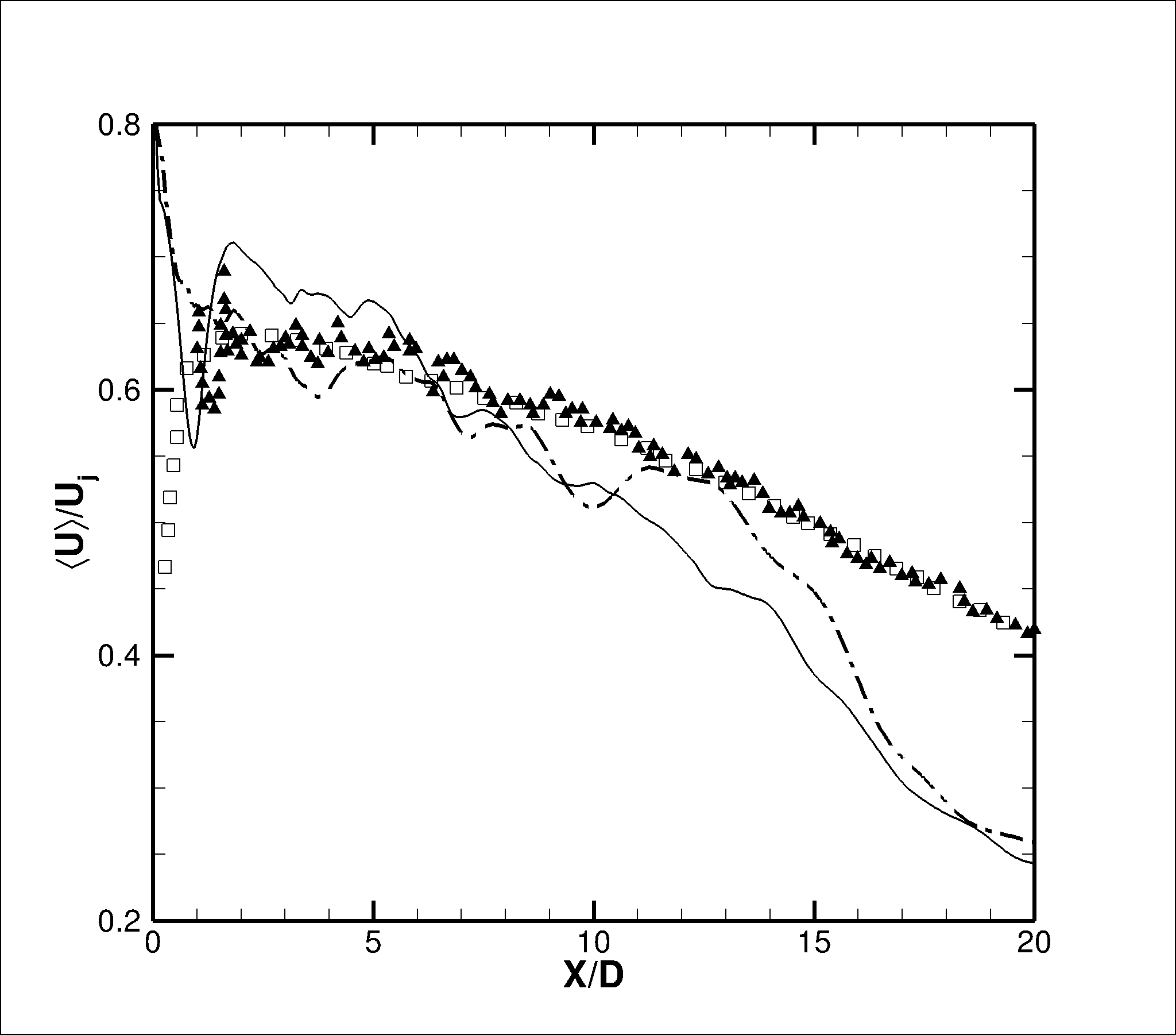}}
	\caption{Profiles of averaged axial component of velocity at 
	different positions within the computational domain.
	(\LARGE \textbf{-}$\cdot$\textbf{-}\scriptsize), S1;
	(\LARGE \textbf{--}\scriptsize), S2; 
	($\square$), numerical data; ($\blacktriangle$), experimental data.}
	\label{fig:prof-u-av-mesh}
\end{figure}

Profiles of $\langle U \rangle$ from S1 and S2, along the
mainstream direction, and the evolution of $\langle U \rangle$
along the centerline and along the lipline are compared 
with numerical and experimental results in Fig.\ \ref{fig:prof-u-av-mesh}. 
The centerline and the lipline are 
indicated as (E) and (F) in Fig.\ \ref{fig:extract}. The dash-point 
line and the solid line stand for the results of the S1 and S2 test cases,
respectively, in Fig.\ \ref{fig:prof-u-av-mesh}. The square symbols 
stand for the LES results of Mendez {\it et al.} \cite{Mendez10,
Mendez12}, while the triangular symbols stand for the experimental 
data of Bridges and Wernet \cite{bridges2008turbulence}.

The comparison of profiles indicates that distributions of 
$\langle U \rangle$ calculated on S1 and S2 correlates well 
with the references until $X=5.0D$. The $\langle U \rangle$ 
profile calculated with S2 at $X=10.0D$ is under predicted 
when compared with the reference profiles. However, it is 
closer to the reference when compared with the S1 results. 
One can notice that S1 and S2 $\langle U \rangle$ distributions 
along the centerline correlates with the references in the 
regions which the grid presents a good resolution. When the 
mesh spacing increases, due to the mesh coarsening in the 
streamwise direction, the time average axial component of 
velocity start to correlate poorly with the reference. The 
time averaged axial component of velocity calculated by S1, 
along the lipline, correlates better with the reference 
than the same property calculated on S2. The second case
overestimates the magnitude of $\langle U \rangle$ until 
$X\approx 6.0D$.


\subsubsection*{Root Mean Square Distribution of Time Fluctuations 
of Axial Velocity Component}

\begin{figure}[htb!]
  \centering
  \subfigure[Lateral view of $u^{*}_{RMS}$ for S1.]
    {\includegraphics[width=0.45\textwidth]
	{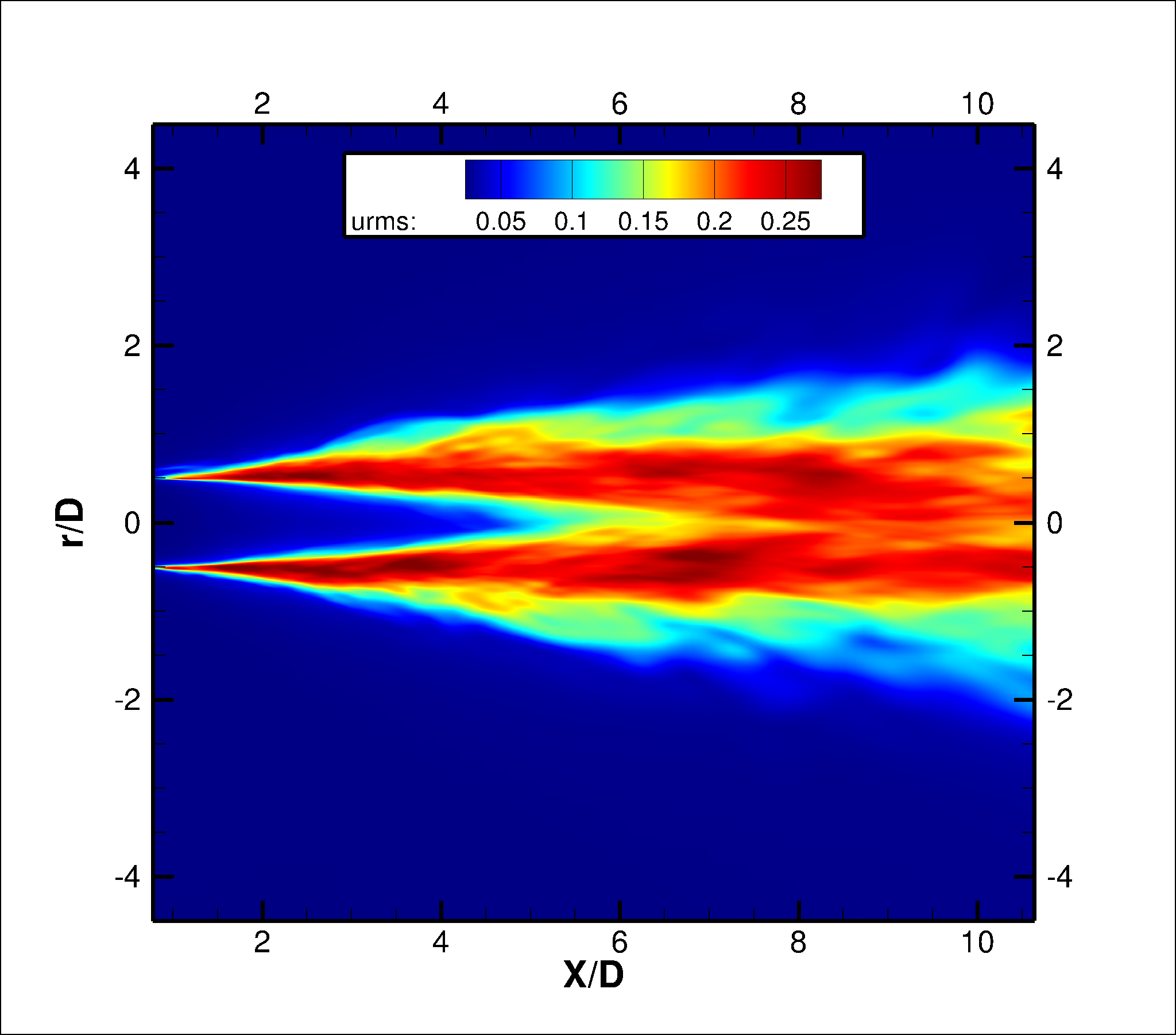}\label{subfig:urms-det-s1}}
  \subfigure[Lateral view of $u^{*}_{RMS}$ for S2.]
    {\includegraphics[width=0.45\textwidth]
	{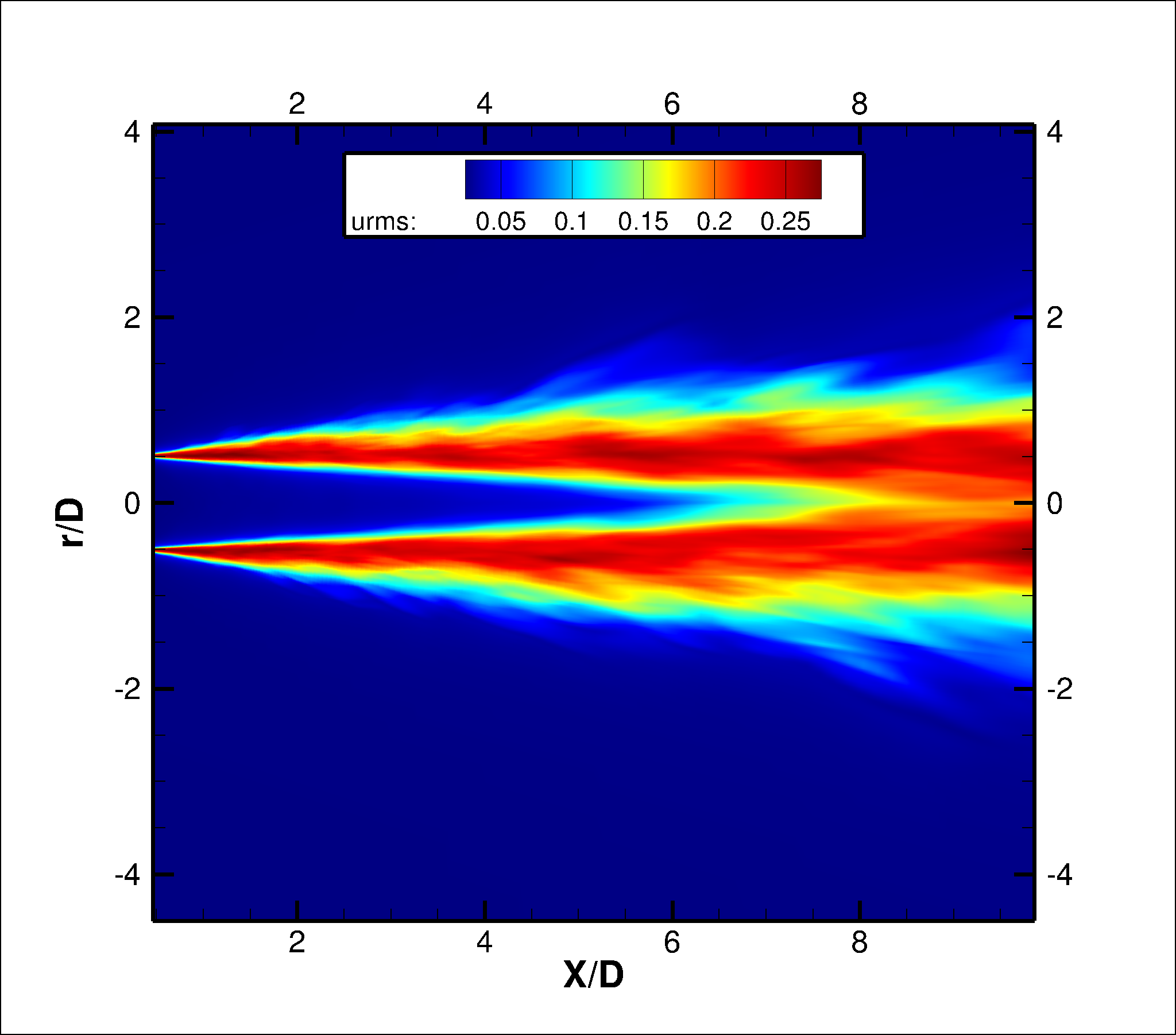}\label{subfig:urms-det-s2}}
  \subfigure[Lateral view of $v^{*}_{RMS}$ for S1.]
    {\includegraphics[width=0.45\textwidth]
	{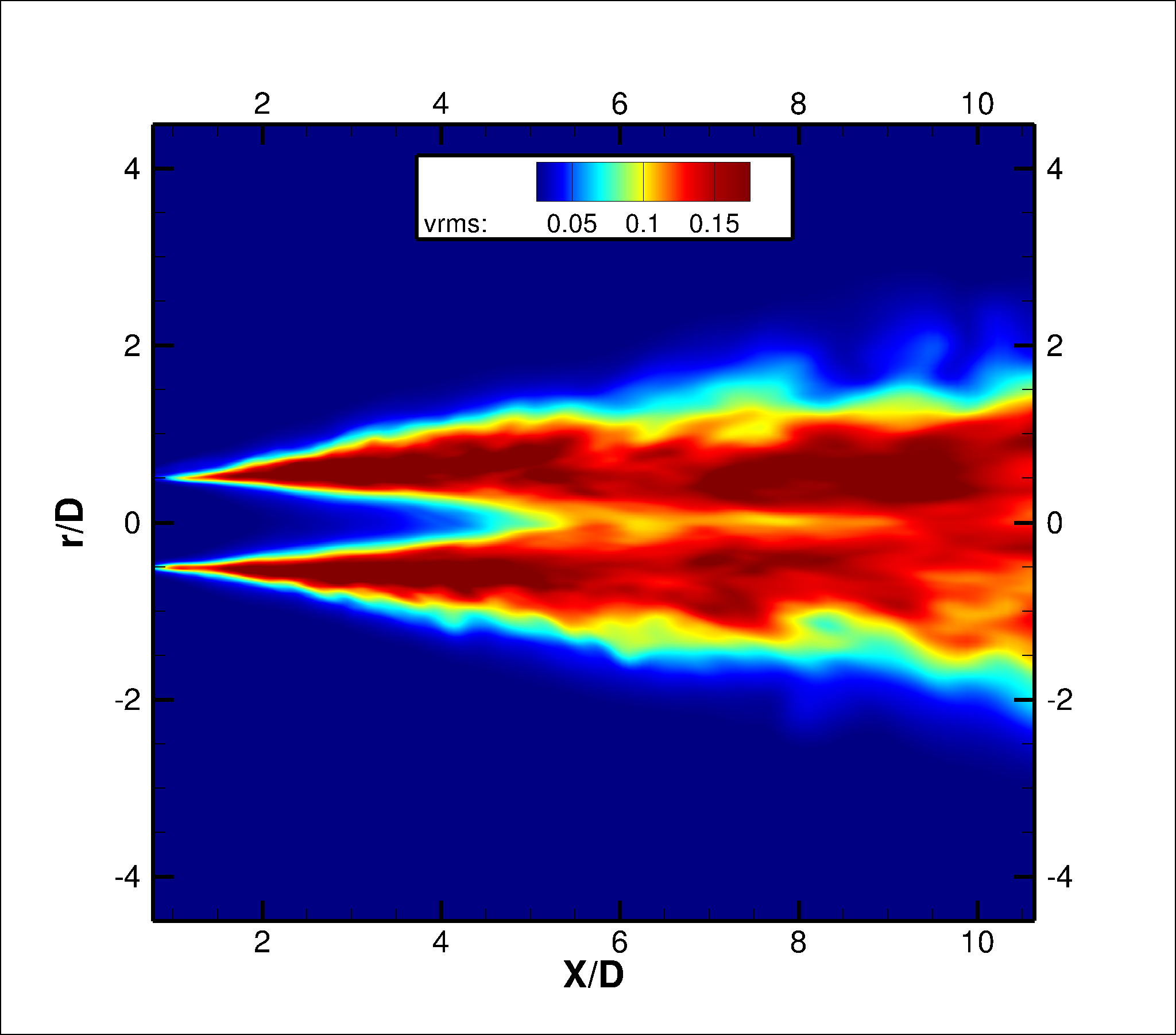}\label{subfig:vrms-det-s1}}
  \subfigure[Lateral view of $v^{*}_{RMS}$ for S2.]
    {\includegraphics[width=0.45\textwidth]
	{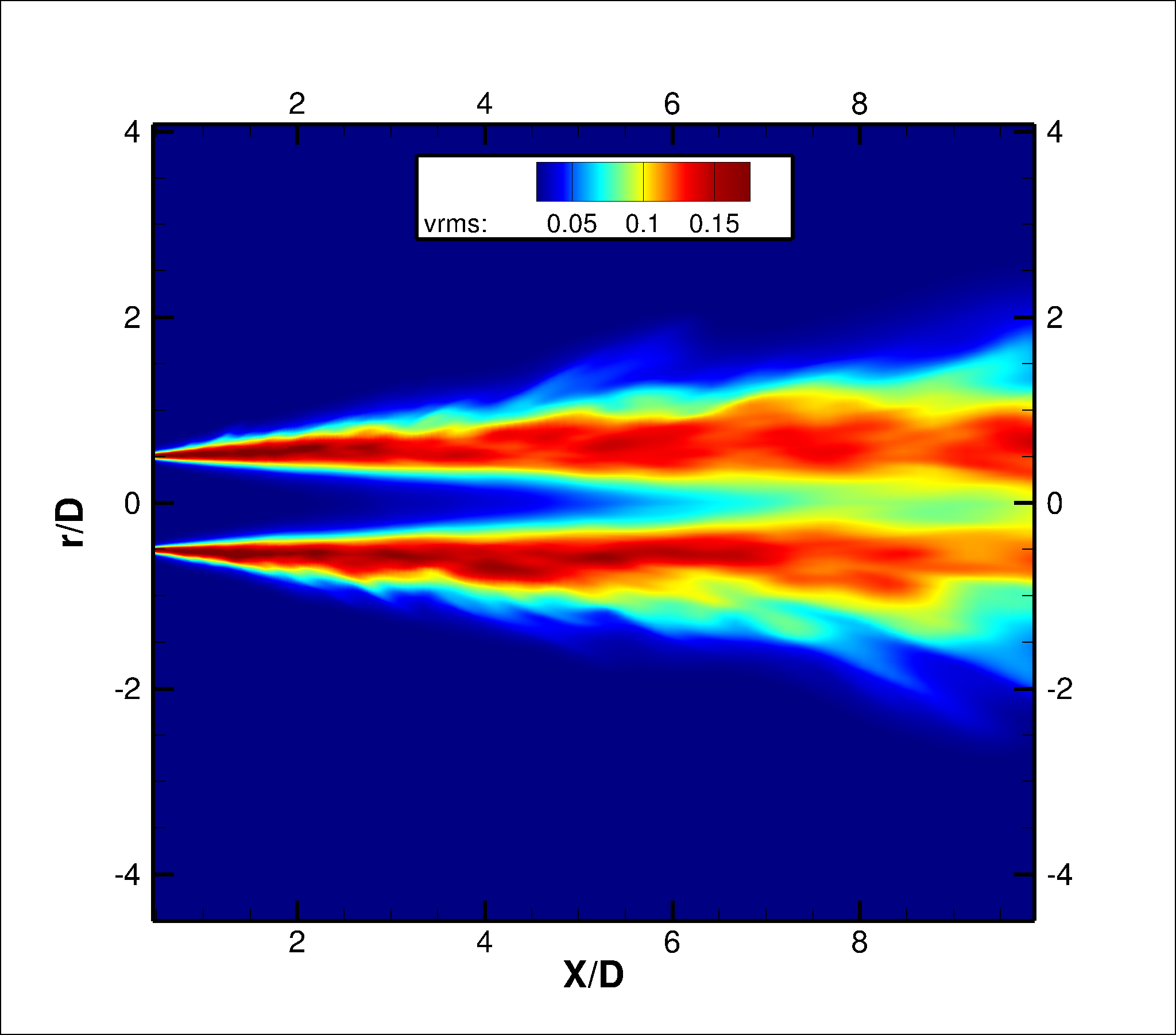}\label{subfig:vrms-det-s2}}
  \subfigure[Lateral view of $\langle u^{*}v^{*}\rangle$ for S1.]
    {\includegraphics[width=0.45\textwidth]
	{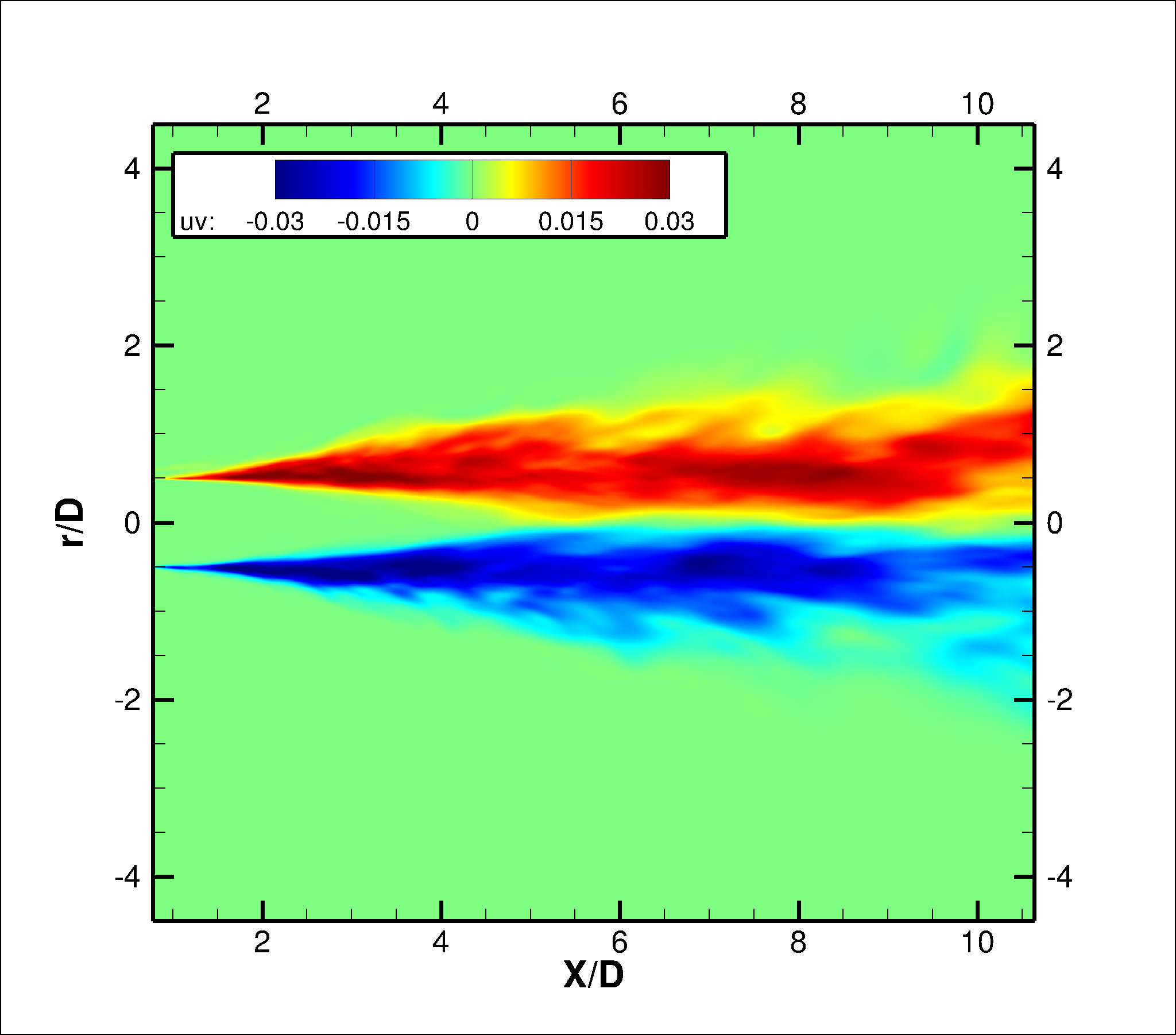}\label{subfig:uv-det-s1}}
  \subfigure[Lateral view of $\langle u^{*}v^{*}\rangle$ for S2.]
    {\includegraphics[width=0.45\textwidth]
	{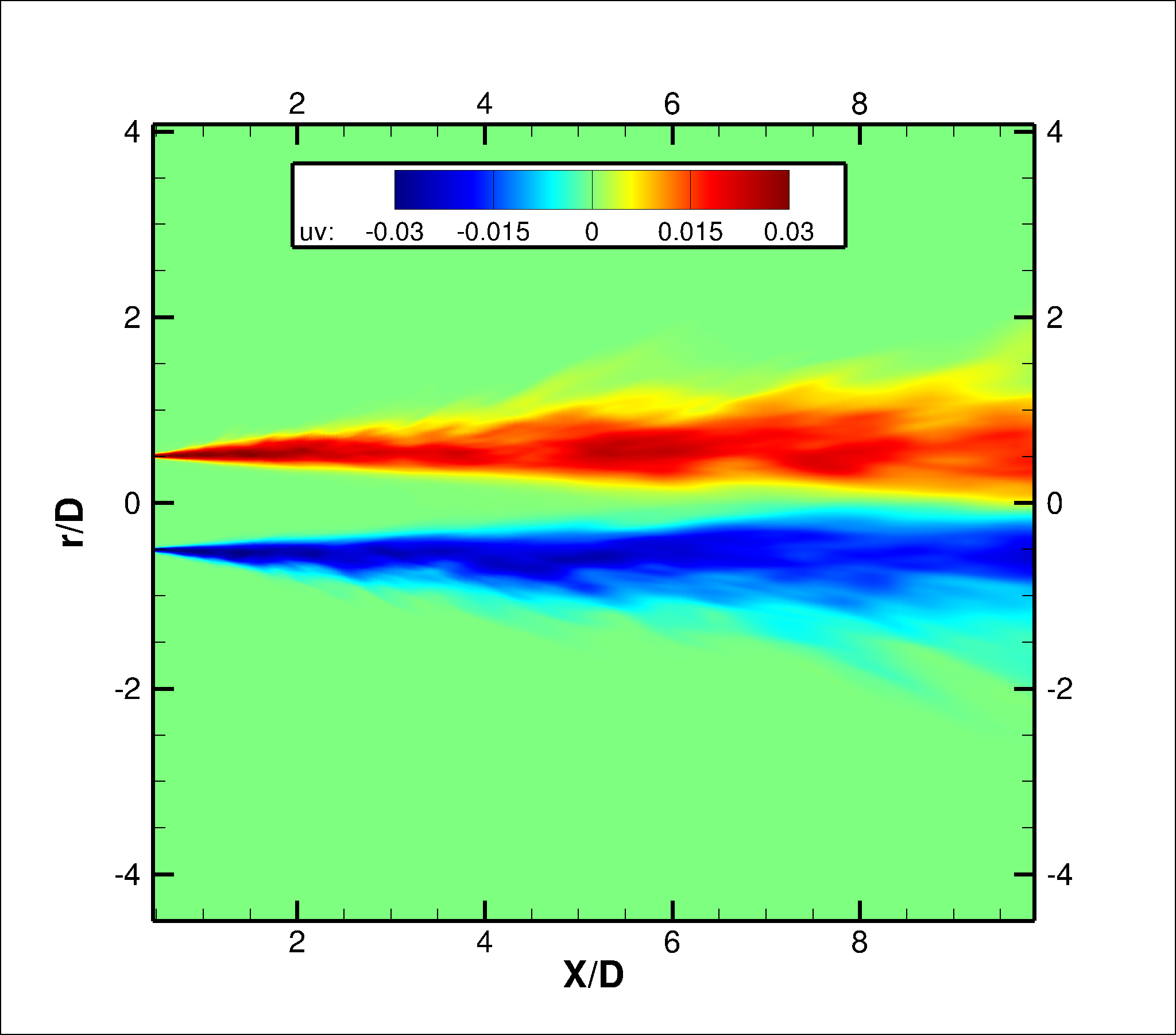}\label{subfig:uv-det-s2}}
	\caption{Lateral view of RMS of time fluctuation
	of axial component of velocity, $u_{RMS}^{*}$, RMS 
	of time fluctuation of radial component of velocity, 
	$v_{RMS}^{*}$ and $\langle u^{*}v^{*}\rangle$ Reynolds 
	shear stress tensor component, for S1 and S2.} 
	\label{fig:lat-u-v-uv-mesh}
\end{figure}

The time fluctuation part of the flow is also important to be studied.
The present work evaluates the axial and radial velocity components 
using the root mean square. A lateral view of $u^{*}_{RMS}$ computed 
by S1 and S2 simulations are presented in Figs.\ \ref{subfig:urms-det-s1}
and \ref{subfig:urms-det-s2}, respectively. The figures indicate that 
the property calculated by S1 is more spread when compared with the 
same property computed by S2. The mesh A refinement along with the 
spatial discretization can generate a more dissipative solution which 
creates the spread effect of $u^{*}_{RMS}$ calculated by S1 when compared
to the same property calculated by S2.
%
%

\begin{figure}[htb!]
  \centering
  \subfigure[$u^{*}_{RMS}$ - X=2.5D ; $-1.5D\leq Y\leq 1.5D$]
    {\includegraphics[width=0.45\textwidth]
	{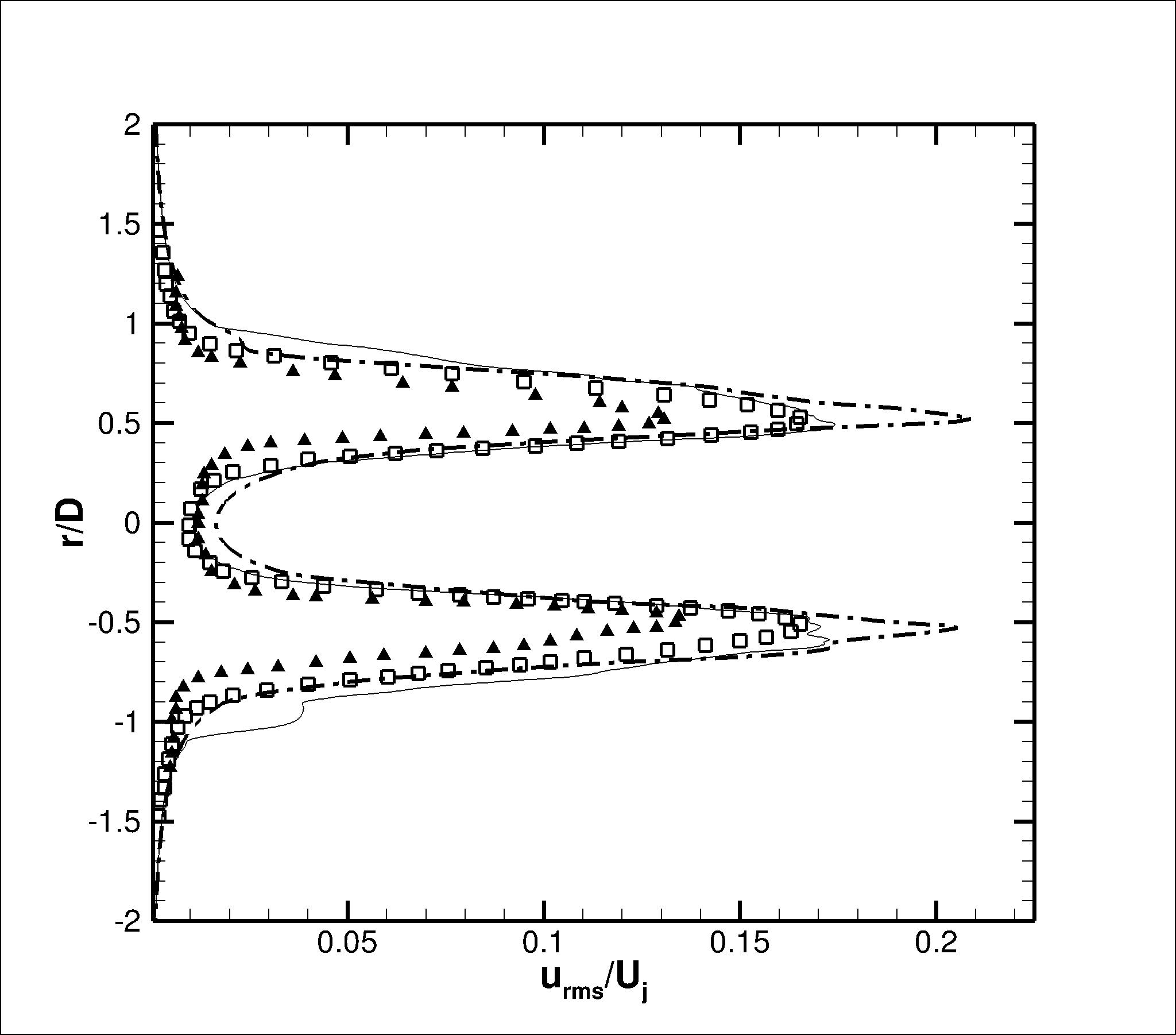}}
  \subfigure[$u^{*}_{RMS}$ - X=5.0D ; $-1.5D\leq Y\leq 1.5D$]
    {\includegraphics[width=0.45\textwidth]
	{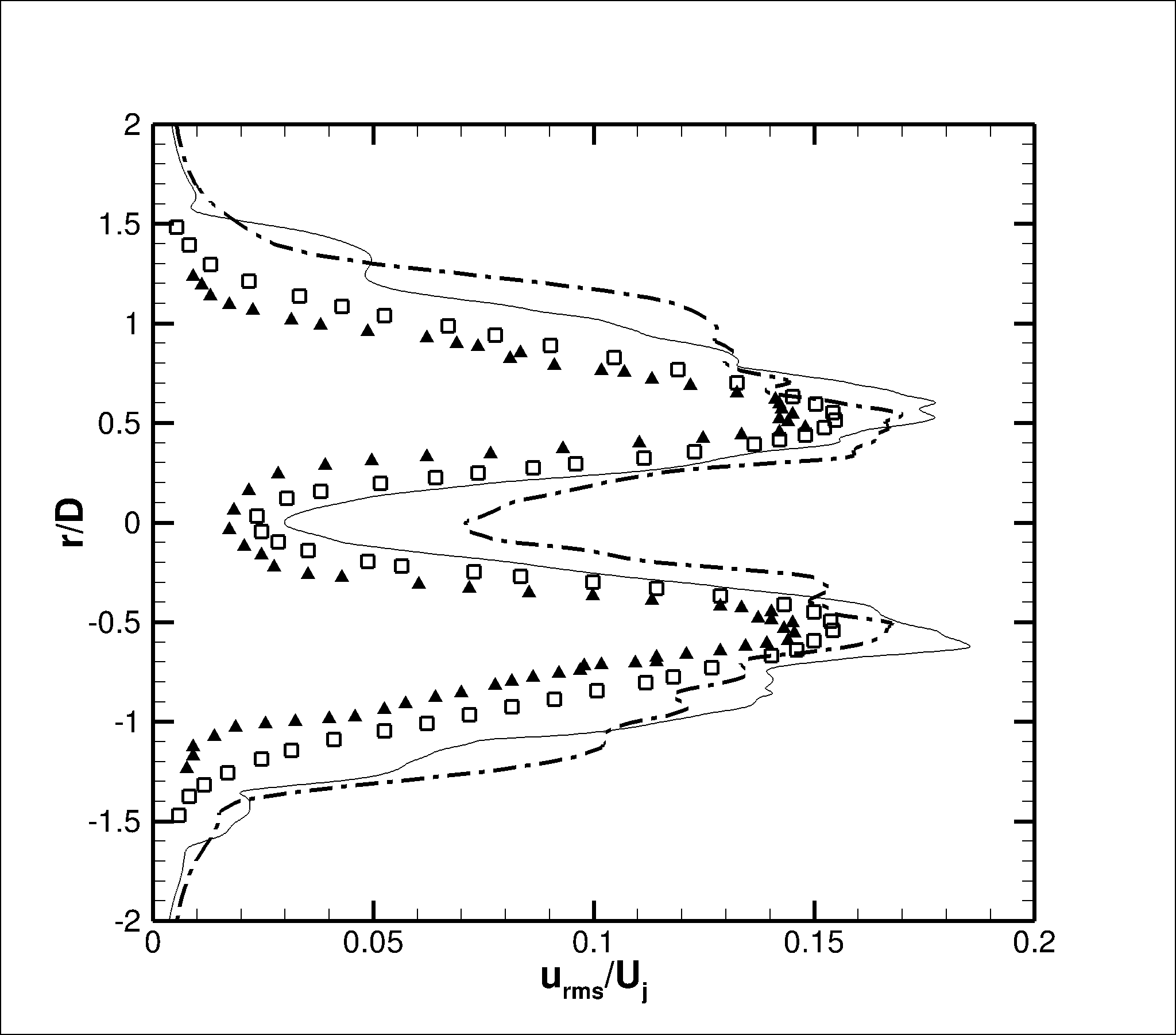}}
  \subfigure[$u^{*}_{RMS}$ - X=10D ; $-1.5D\leq Y\leq 1.5D$]
    {\includegraphics[width=0.45\textwidth]
	{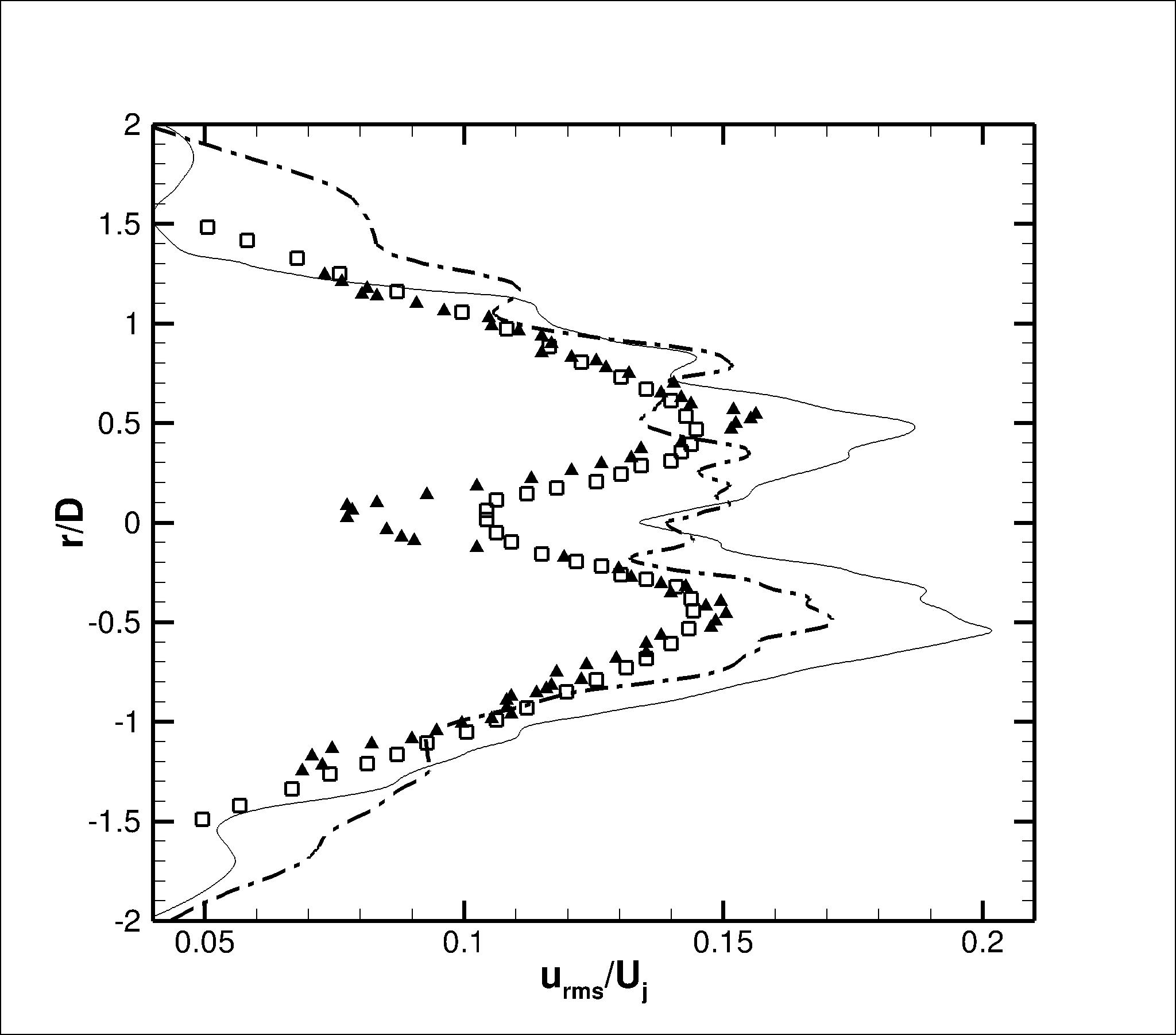}}
  \subfigure[$u^{*}_{RMS}$ - X=15D ; $-1.5D\leq Y\leq 1.5D$]
    {\includegraphics[width=0.45\textwidth]
	{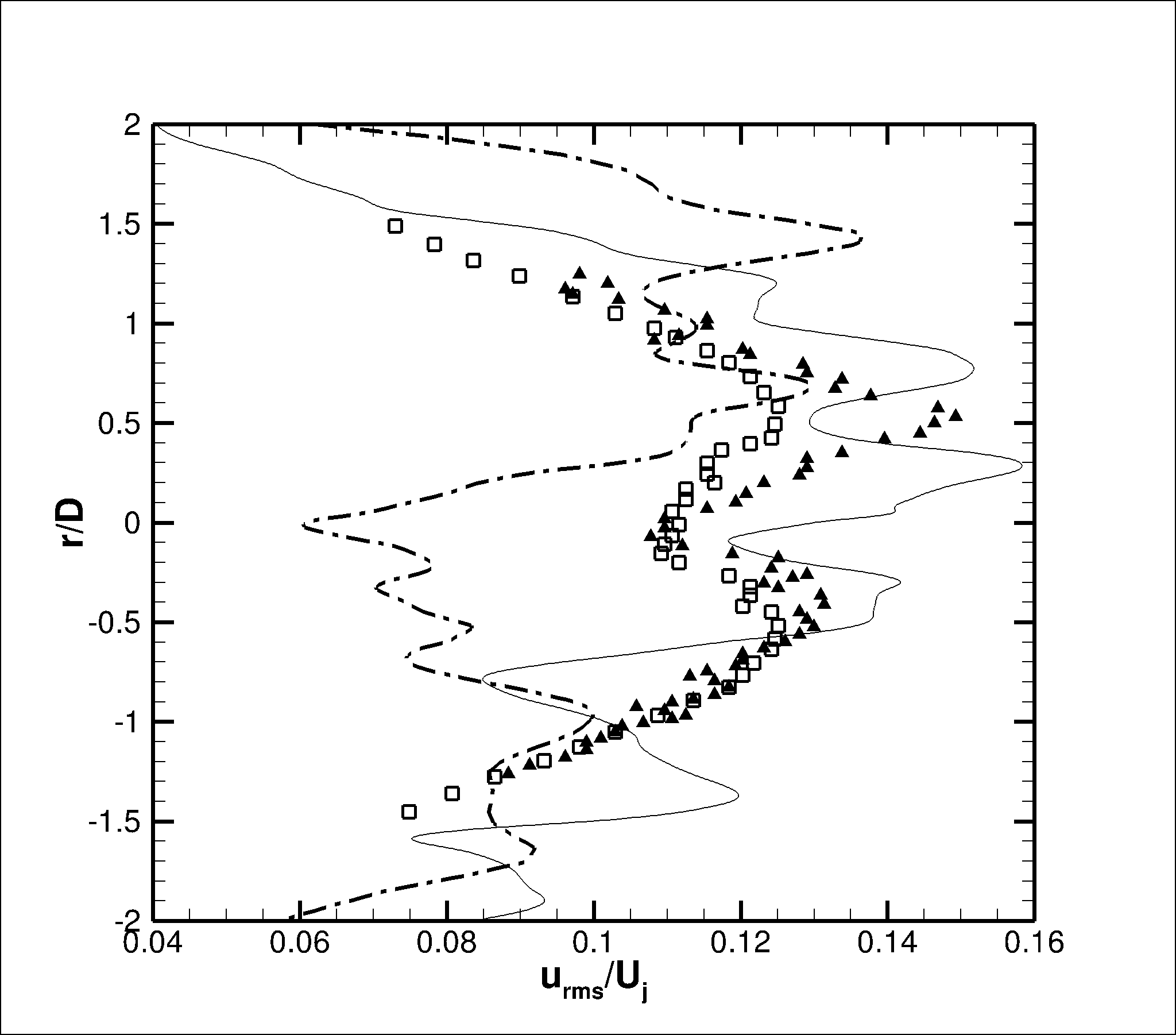}}
  \subfigure[$u^{*}_{RMS}$ - Centerline - Y=0 ; $0\leq X \leq 20D$ ]
    {\includegraphics[width=0.45\textwidth]
	{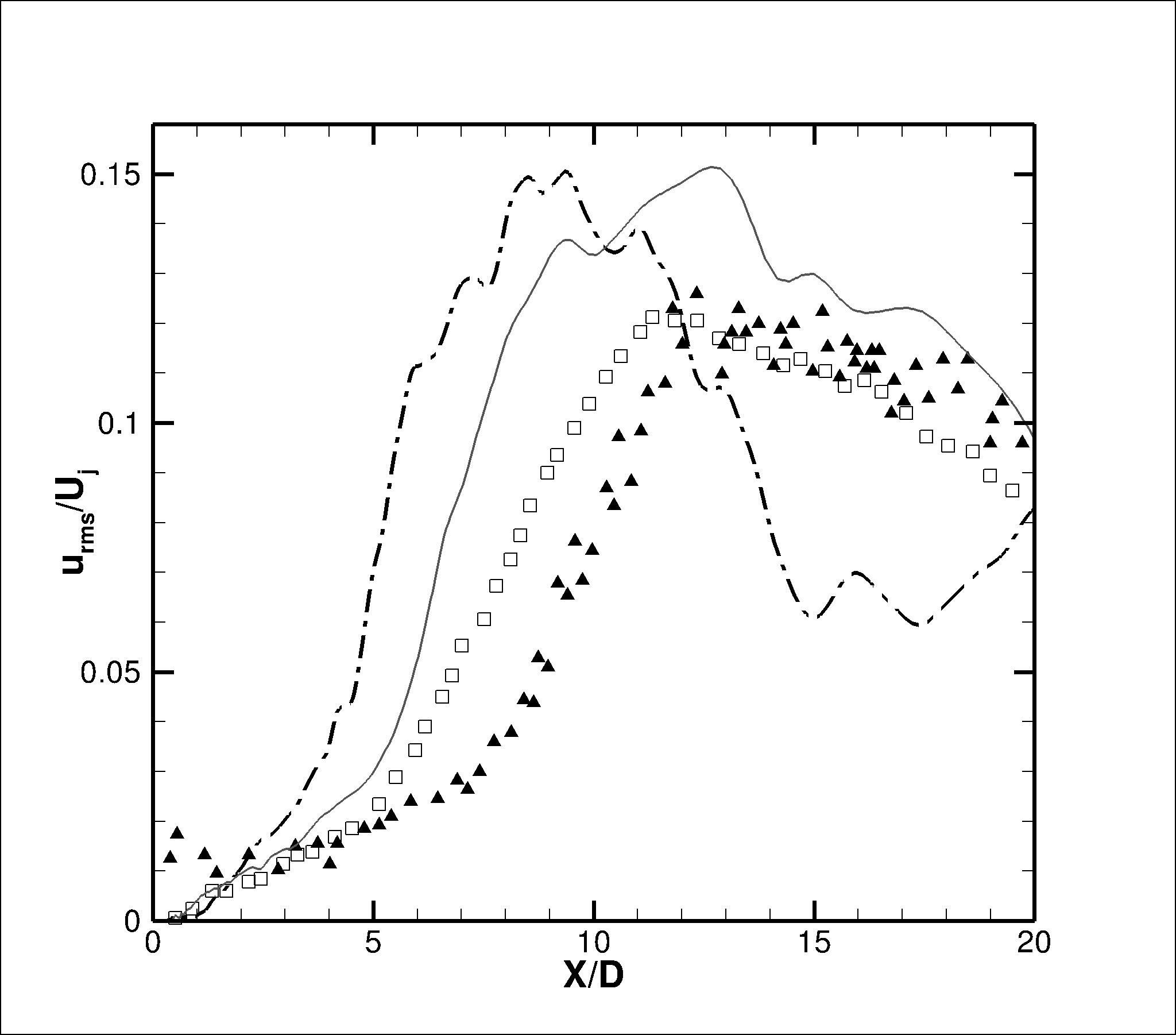}\label{subfig:urms-centerline}}
  \subfigure[$u^{*}_{RMS}$ - Lipline - Y=0.5D ; $0\leq X \leq 20D$ ]
    {\includegraphics[width=0.45\textwidth]
	{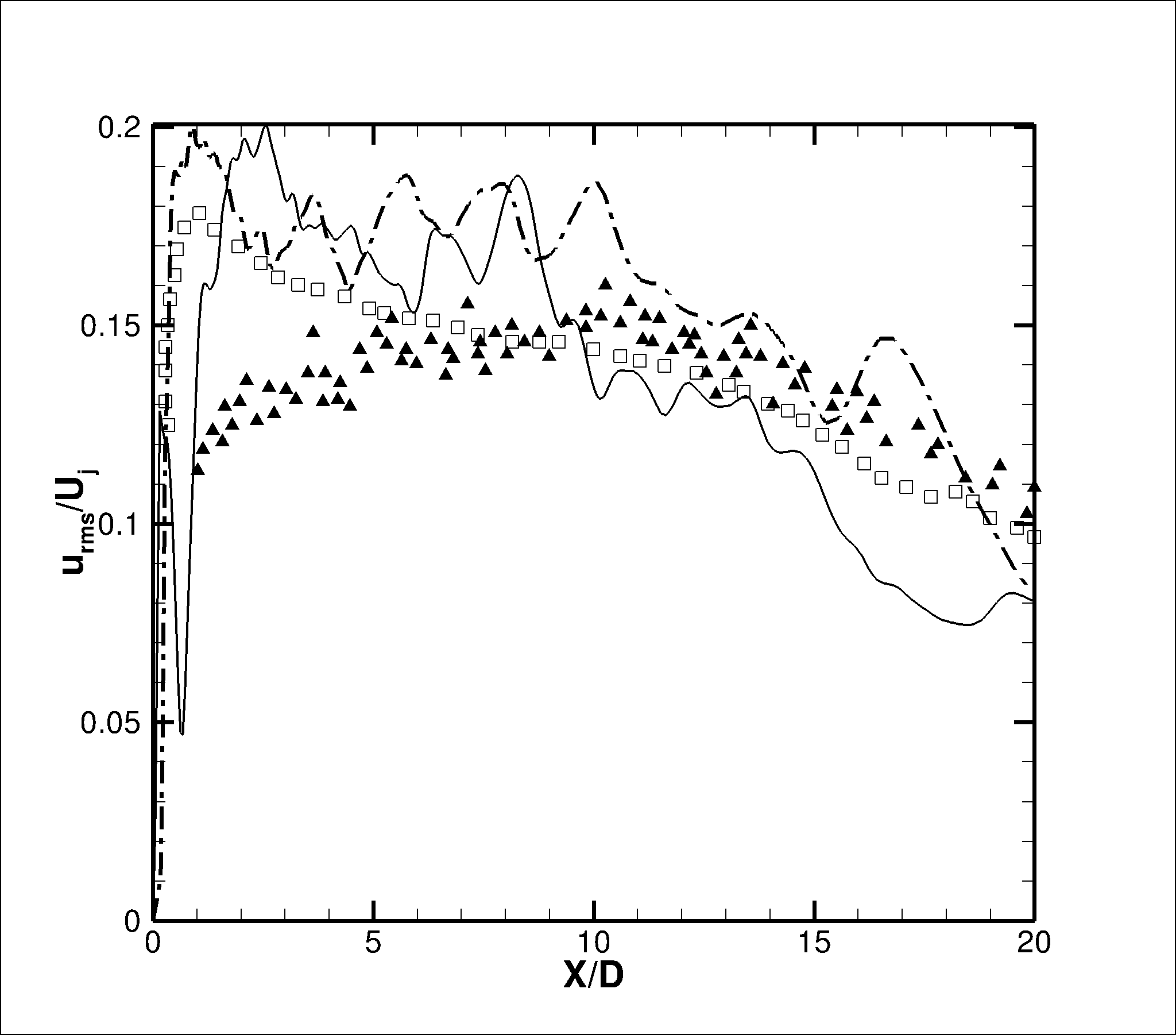}\label{subfig:urms-lipline}}
	\caption{Profiles of RMS of time fluctuation of axial 
	component of velocity, $u_{RMS}^{*}$, for S1 and S2,
	at different positions within the computational domain.
	(\LARGE \textbf{-}$\cdot$\textbf{-}\scriptsize), S1;
	(\LARGE \textbf{--}\scriptsize), S2; 
	($\square$), numerical data; ($\blacktriangle$), experimental data.}
	\label{fig:prof-u-rms-mesh}
\end{figure}

The same strategy used to compare the mean profiles of velocity is 
used here for the study of $u^{*}_{RMS}$. Figure \ref{fig:prof-u-rms-mesh}
presents the comparison of root mean square profiles of $u^{*}$
calculated by S1 and S2 with reference results. The profile of 
$u^{*}_{RMS}$ calculated by S2 fits perfectly the reference profiles 
at $X=2.5D$. The profile calculated by S1, at the same position, 
presents a good correlation with numerical and experimental data. 
However, it does not correctly represent the two peaks of the profile. 
For $X=5.0D$ and $X=10.0D$ the profiles start to diverge from the 
reference results. At $X=15.0D$, the $u^{*}_{RMS}$ profile, calculated 
by S1, present a different shape and different magnitude from the
reference profiles. At the same position the fluctuation profile 
computed by S2 reproduce the same peaks of the reference data. However, 
the shape of the profile is completely different from the shape of
profiles calculated by the references.

\FloatBarrier

Figures \ref{subfig:urms-centerline} and \ref{subfig:urms-lipline}
presents the distribution of $u^{*}_{RMS}$ along the centerline and 
lipline of the jet. The distributions calculated by S1 and S2 are
somewhat different from the results of the references. However,
one can notice an upgrade on the solution when comparing S1 and S2
results. The results achieved using the more refined mesh are closer
to the reference than the results obtained using mesh A.

\subsubsection*{Root Mean Square Distribution of Time Fluctuations 
of Radial Velocity Component}

The time fluctuation of the radial component of velocity is also 
compared with the reference data. Distributions of root mean square 
of $v_{RMS}^{*}$ are presented in the subsection. Figures 
\ref{subfig:vrms-det-s1} and \ref{subfig:vrms-det-s2} illustrate
a lateral view of the distribution of $v_{RMS}^{*}$ computed by S1 and S2,
respectively. A significant divergence between the results can be easily
noticed on the lateral view of the $v_{RMS}^{*}$ distribution. 
From $X=2.5D$ towards the exit boundary the magnitude of fluctuation 
calculated by S1 is much higher than the magnitude of $v_{RMS}^{*}$ 
computed by S2.

\begin{figure}[htb!]
  \centering
  \subfigure[$v^{*}_{RMS}$ - X=2.5D ; $-1.5D\leq Y\leq 1.5D$]
    {\includegraphics[width=0.45\textwidth]
	{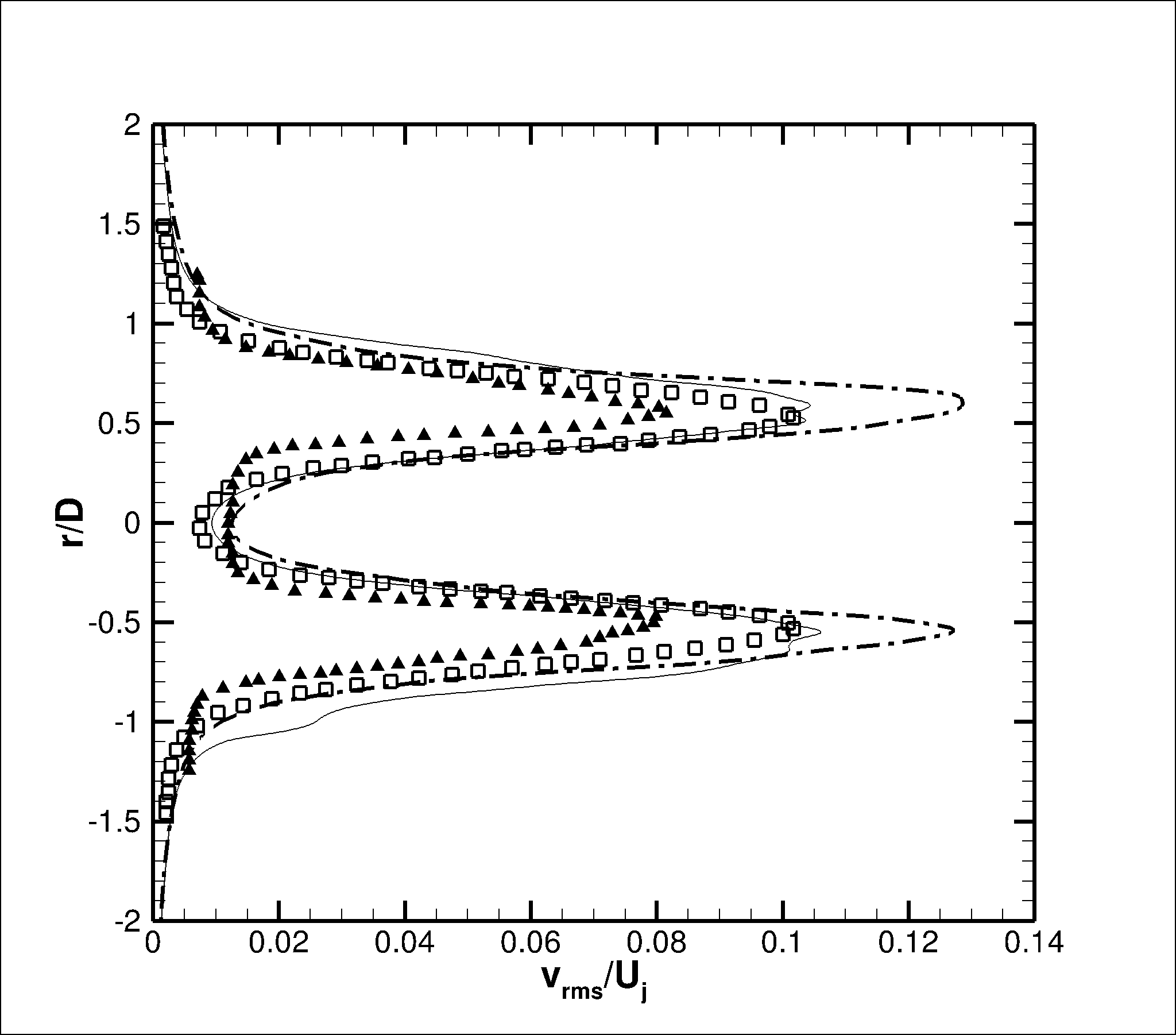}}
  \subfigure[$v^{*}_{RMS}$ - X=5.0D ; $-1.5D\leq Y\leq 1.5D$]
    {\includegraphics[width=0.45\textwidth]
	{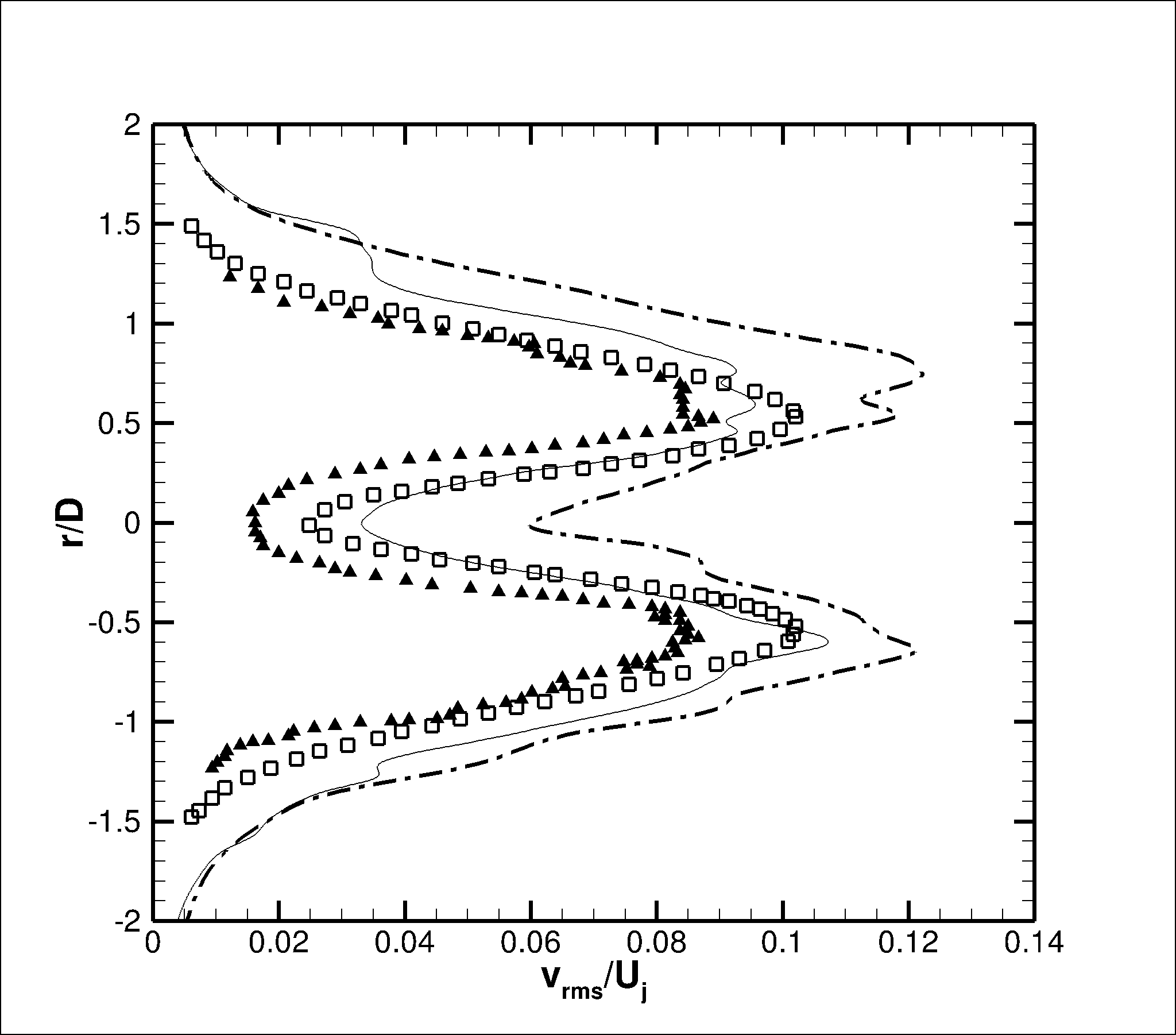}}
  \subfigure[$v^{*}_{RMS}$ - X=10D ; $-1.5D\leq Y\leq 1.5D$]
    {\includegraphics[width=0.45\textwidth]
	{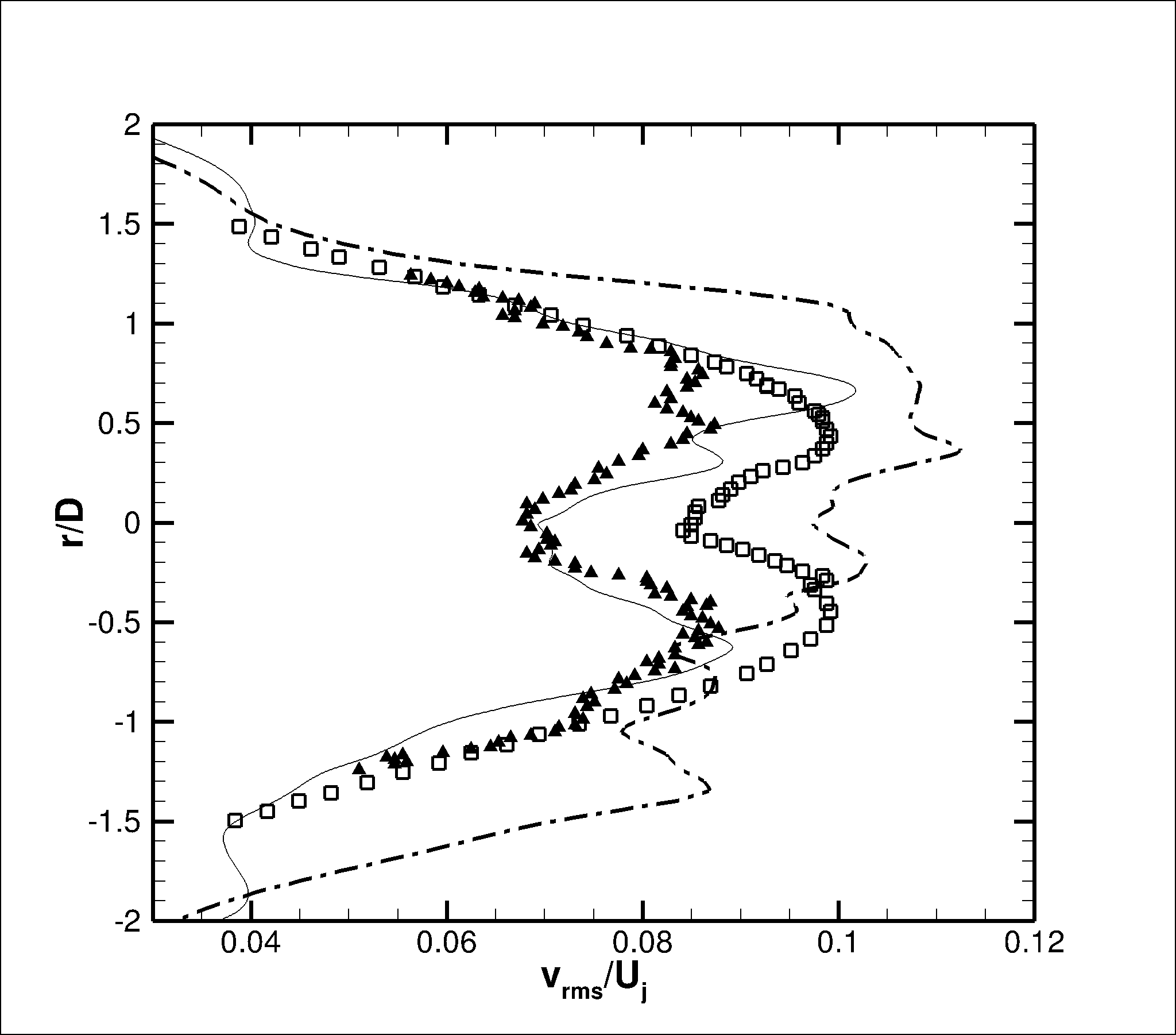}}
  \subfigure[$v^{*}_{RMS}$ - X=15D ; $-1.5D\leq Y\leq 1.5D$]
    {\includegraphics[width=0.45\textwidth]
	{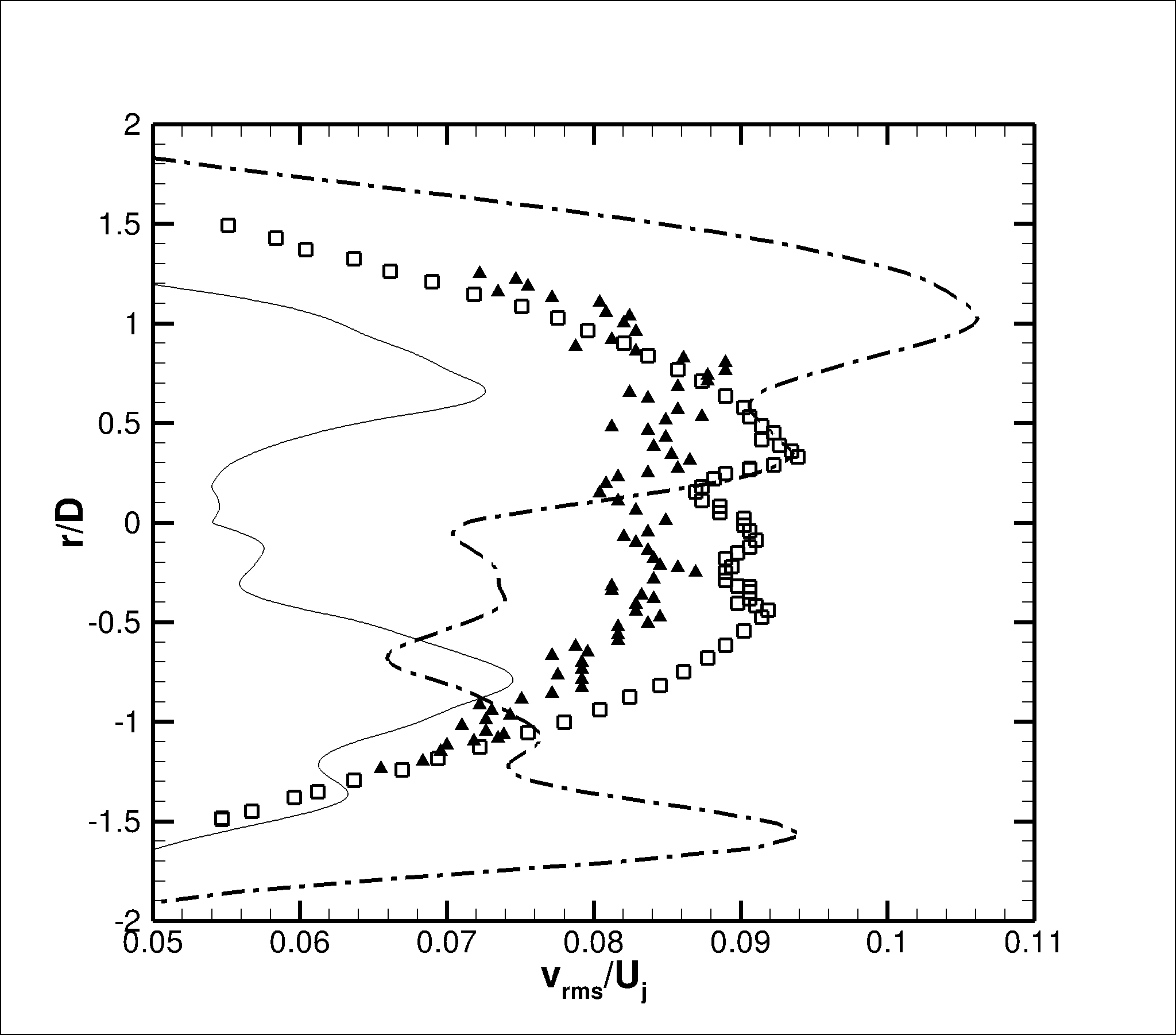}}
	\caption{Profiles of RMS of time fluctuation of radial 
	component of velocity, $v_{RMS}^{*}$, for S1 nd S2,
	at different positions within the computational domain.
	(\LARGE \textbf{-}$\cdot$\textbf{-}\scriptsize), S1;
	(\LARGE \textbf{--}\scriptsize), S2; 
	($\square$), numerical data; ($\triangle$), experimental data.}
	\label{fig:prof-v-rms-mesh}
\end{figure}

Four profiles of $v_{RMS}^{*}$ in the radial direction at $X=2.5D$, 
$X=5.0D$, $X=10.0D$ and $X=15.0D$ are presented in 
Fig.\ \ref{fig:prof-v-rms-mesh}. S1 results presented a good correlation 
with the reference at $X=2.5D$, where only the peaks of the profile 
are not well represented. For all other positions on the axial 
direction studied in the present research the $v_{rms}^{*}$
profiles of S1 are overestimated and poorly correlates with the 
reference. On the other hand, $v_{RMS}^{*}$ profiles calculated 
using a refined grid fits very well with the results of the numerical 
reference at $X=2.5D$ and $X=5.0D$. At $X=10.0D$ the fluctuation 
profile calculated by S2 presents a better correlation with the 
experimental data than the numerical reference. At $X=15.0D$ the S2
does not present a good profile of $v_{RMS}^{*}$.

\subsubsection*{Component of Reynolds Stress tensor}

\begin{figure}[htb!]
  \centering
  \subfigure[$\langle u^{*}v^{*}\rangle$ - X=2.5D ; $-1.5D\leq Y\leq 1.5D$]
    {\includegraphics[width=0.45\textwidth]
	{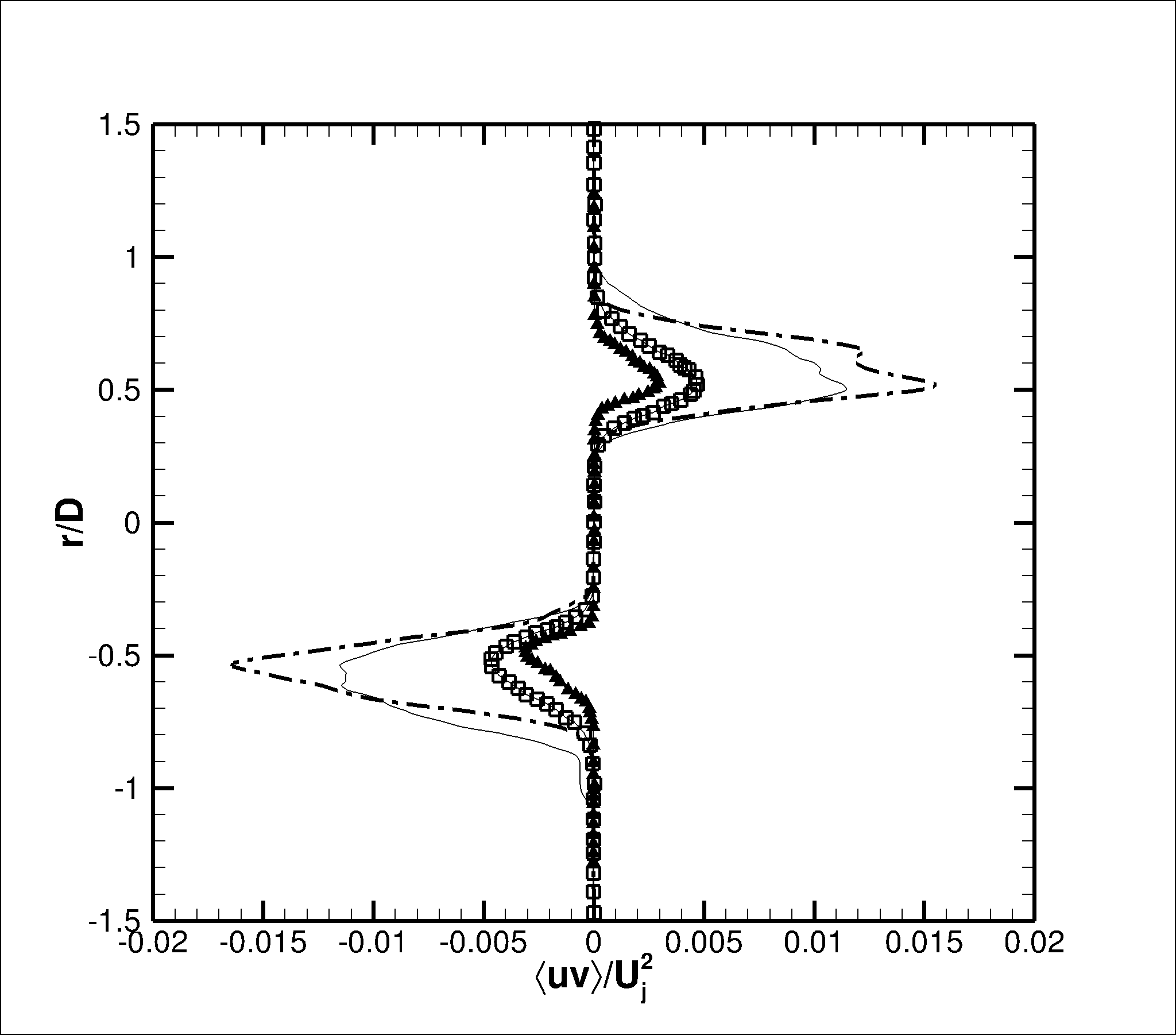}}
  \subfigure[$\langle u^{*}v^{*}\rangle$ - X=5.0D ; $-1.5D\leq Y\leq 1.5D$]
    {\includegraphics[width=0.45\textwidth]
	{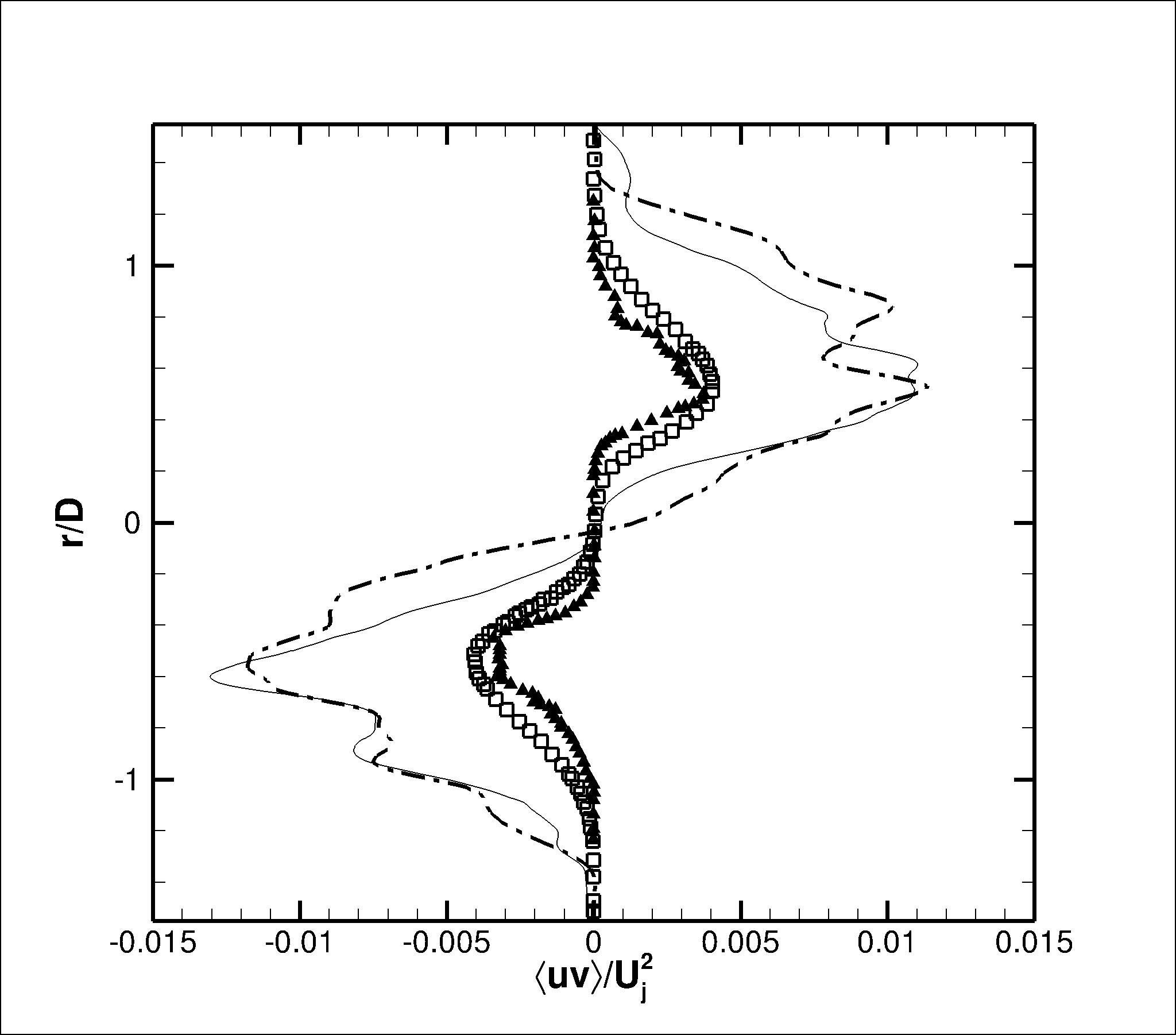}}
  \subfigure[$\langle u^{*}v^{*}\rangle$ - X=10D ; $-1.5D\leq Y\leq 1.5D$]
    {\includegraphics[width=0.45\textwidth]
	{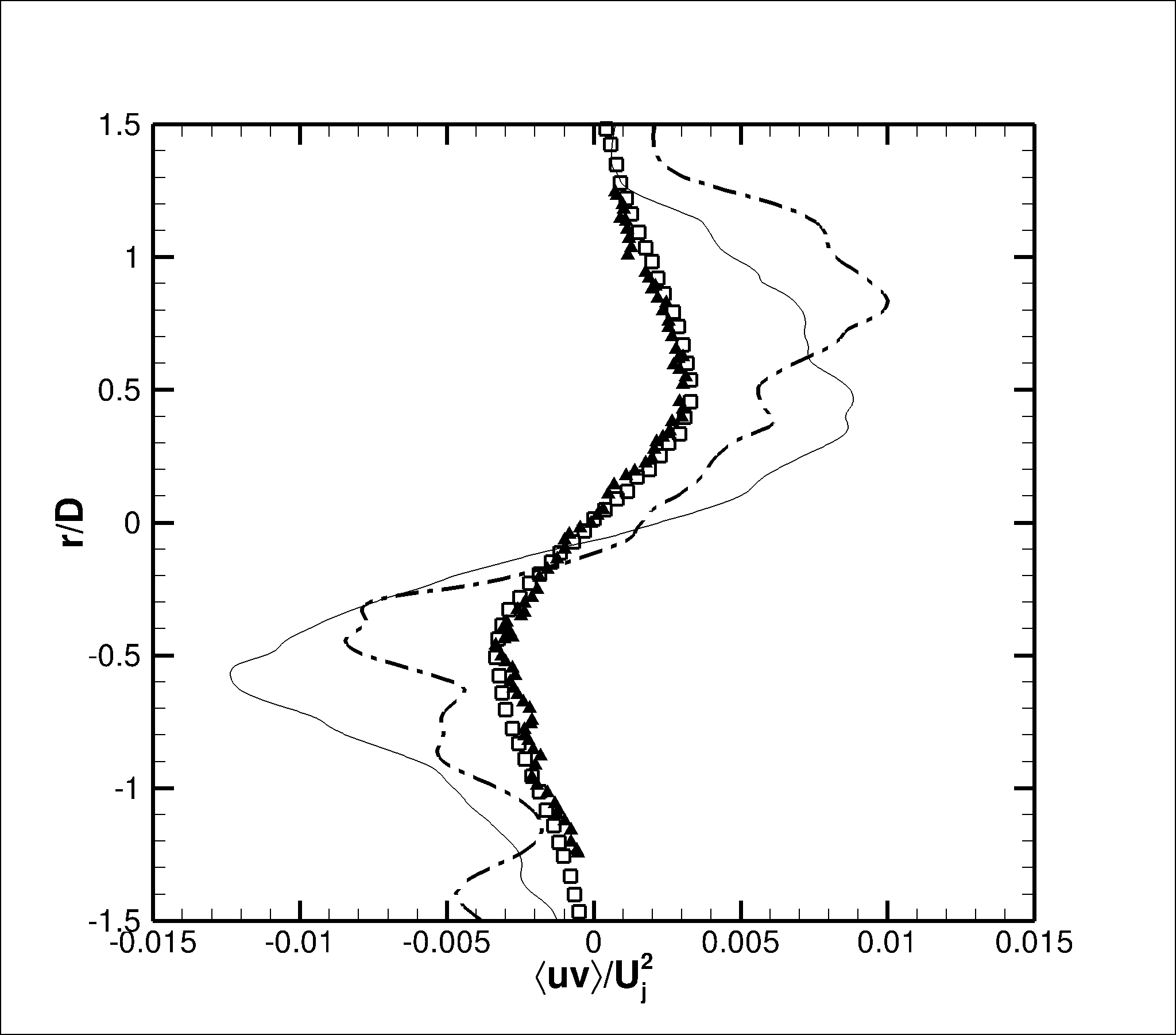}}
  \subfigure[$\langle u^{*}v^{*}\rangle$ - X=15D ; $-1.5D\leq Y\leq 1.5D$]
    {\includegraphics[width=0.45\textwidth]
	{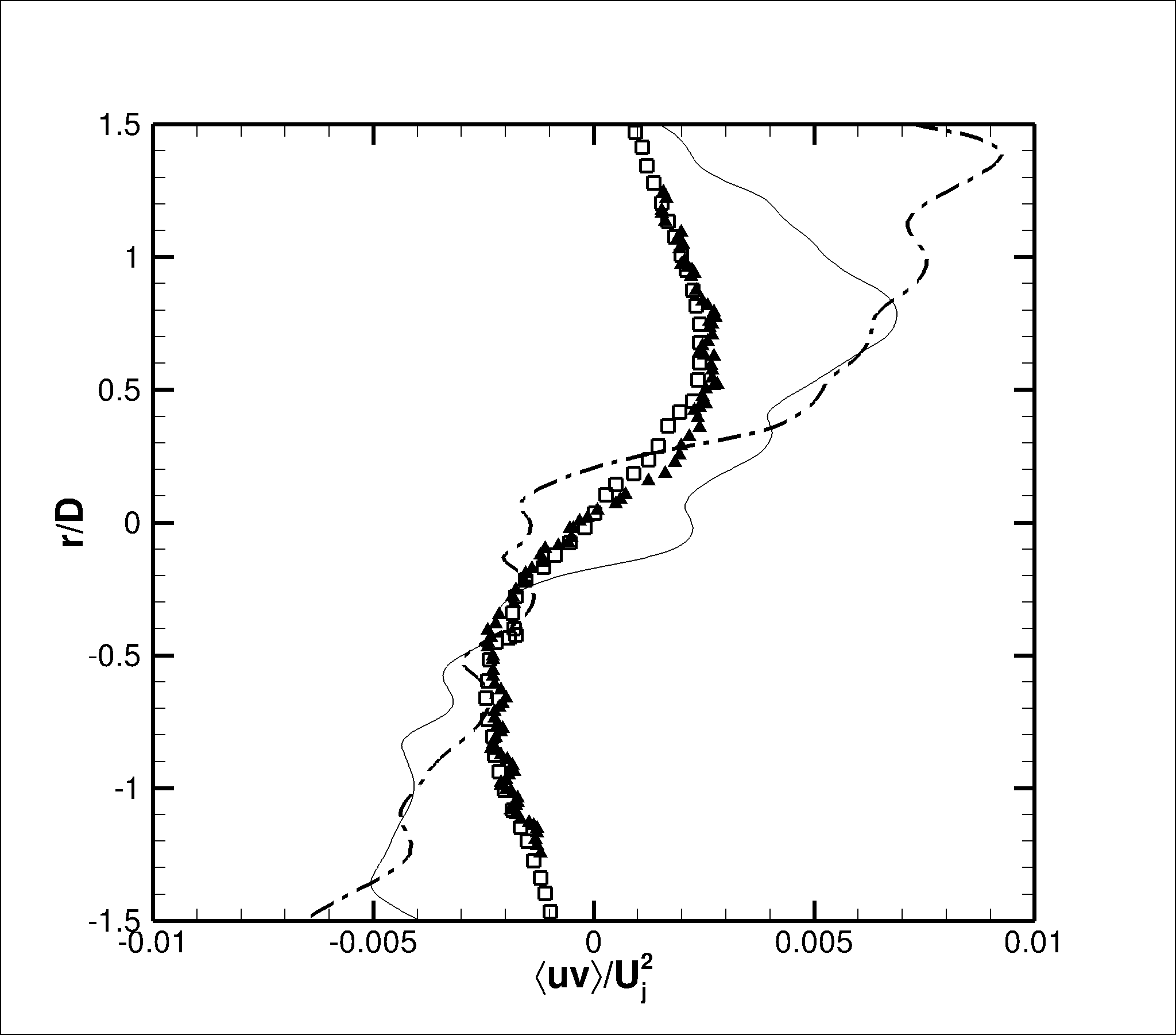}}
	\caption{Profiles of the $\langle u^{*}v^{*} \rangle$ 
	Reynolds shear stress tensor component, for S1 and S2, at 
    different positions within the computational domain.
	(\LARGE \textbf{-}$\cdot$\textbf{-}\scriptsize), S1;
    (\LARGE \textbf{--}\scriptsize), S2;
    ($\square$), numerical data; ($\triangle$), experimental data.}
	\label{fig:prof-uv-av-mesh}
\end{figure}

Figures \ref{subfig:uv-det-s1}, \ref{subfig:uv-det-s2} and 
\ref{fig:prof-uv-av-mesh} present lateral views and profiles of 
$\langle u^{*}v^{*}\rangle$ component of the Reynolds stress tensor.
One can observe that the distributions of the property obtained by S1 
is over dissipated when compared with results collected from S2. Comparing 
the profiles with the reference, one can notice that the profiles achived in 
S1 and S2 are really far from the numerical and ex\-pe\-ri\-men\-tal data. 
The solver has produced with succes the shape of $\langle u^{*}v^{*}\rangle$ 
profile. However, it fails to represent the peak of 
$\langle u^{*}v^{*}\rangle$ for all profiles compared.

\subsubsection*{Time Averaged Eddy Viscosity}

\begin{figure}[htb!]
  \centering
  \subfigure[Lateral view of $\langle\mu_{t}\rangle$ for S1.]
    {\includegraphics[width=0.495\textwidth]
	{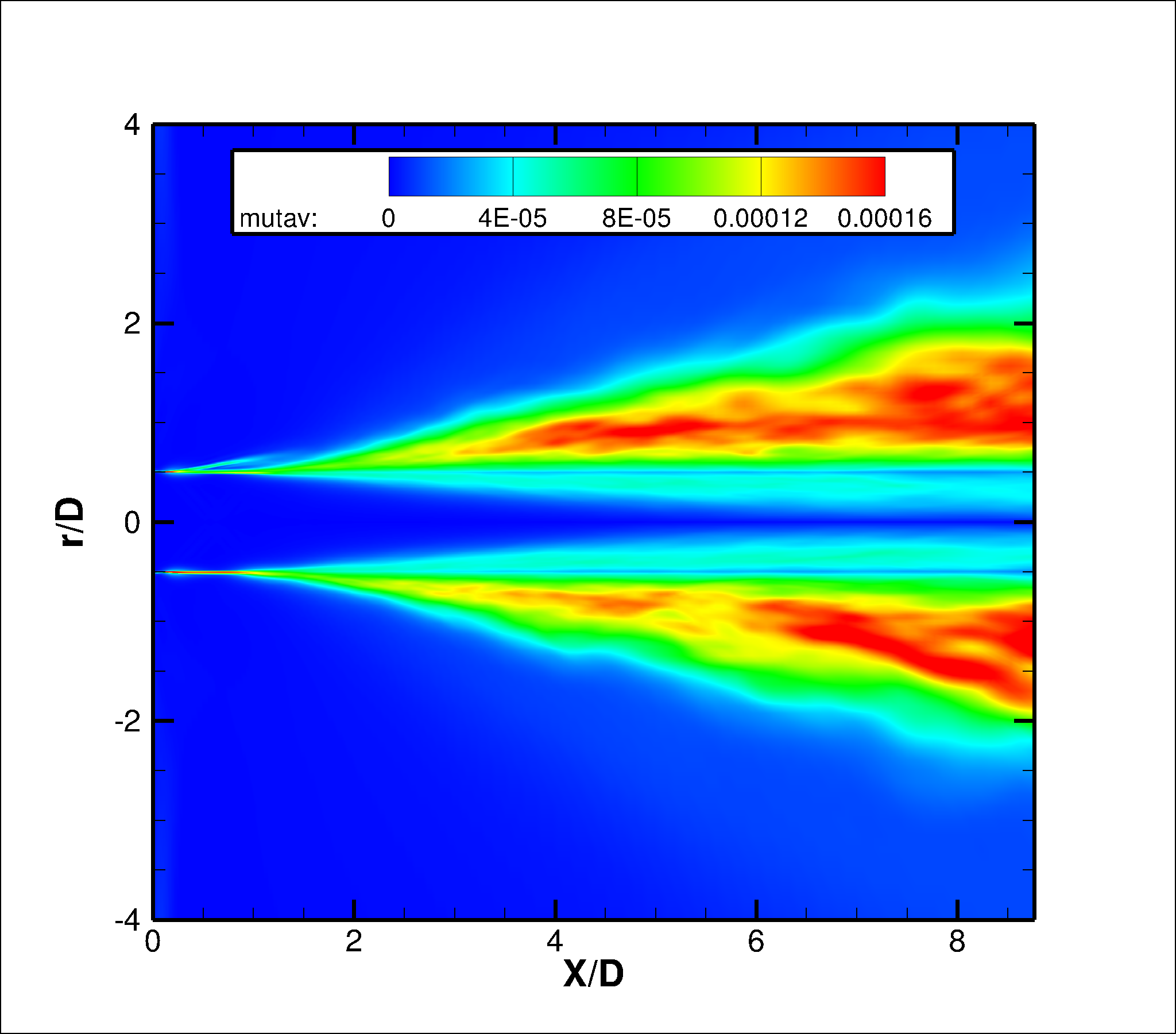}}
  \subfigure[Lateral view of $\langle\mu_{t}\rangle$ for S2.]
    {\includegraphics[width=0.495\textwidth]
	{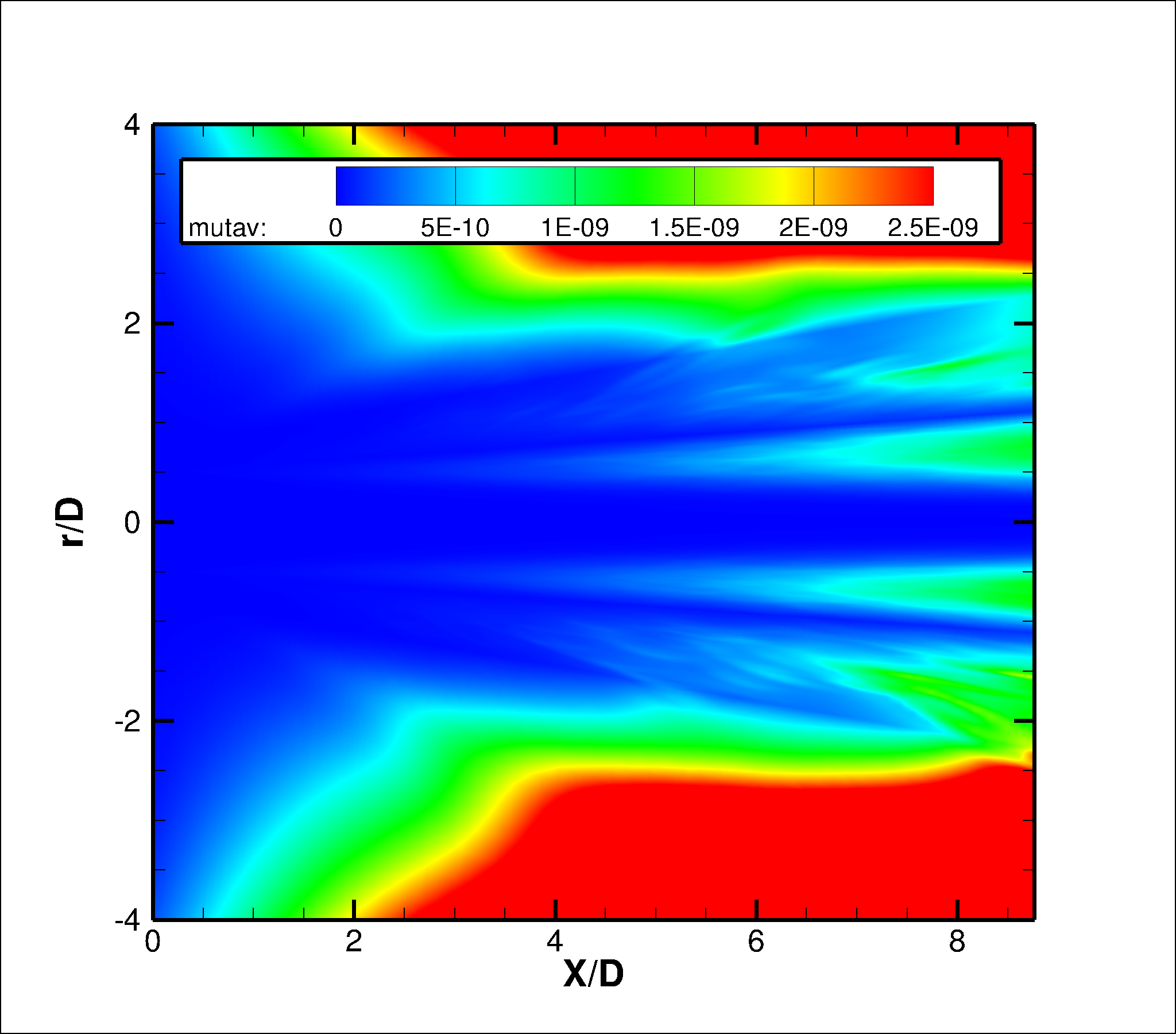}}
	\caption{Lateral view and detailed view of time averaged
	eddy viscosity, $\langle\mu_{t}\rangle$, for S1 and S2.} 
	\label{fig:lat-mut-mesh}
\end{figure}

The effects of the mesh on the SGS modeling are also studied in 
the present subsection. Figure \ref{fig:lat-mut-mesh}
presents distributions of time averaged eddy viscosity, 
$\langle\mu_{t}\rangle$, calculated on S1 and S2. The Smagorinsky 
model \cite{Smagorinsky63,Lilly65,Lilly67} is used on both simulations. 
This SGS closure is highly dependent on the local mesh size. One can 
notice that $\langle\mu_{t}\rangle$ presents higher values on the 
distributions obtained by S1. On the other hand, the eddy viscosity is only 
acting on the regions where the mesh is no longer very refined for the 
S2 study. The $\langle\mu_{t}\rangle$ is very low in the region where 
the grid spacing is small.
%
%

The eddy viscosity can contribute to the dissipative 
characteristic of the simulations. Specially for meshes with
low point resolution. The divergence observed on the 
distribution of $\langle\mu_{t}\rangle$ calculated by S1 and S2
is an example of such effect. However, it is important to 
notice that, even in regions where the mesh is not very refined, 
yet, not coarse, and where the eddy viscosity can be neglected, 
some distributions of properties, calculated by S2, have shown to 
be very dissipative when compared with the LES reference and with
the experimental data. Therefore, one can state that the 
truncation errors originated from the second order spatial 
discretization, used on the simulations here performed, can easily 
overcome the effects of SGS modeling if the grid spacing is not 
small enough. The issue is very important for the structured mesh
approach. Increasing mesh resolution in the region of interest
expressively rises up the number of points all over the computational
domain. Local refinement for structured mesh is not straight 
forward and the code used in the current work does not have such 
approach available.

%


\subsubsection*{Power Spectral Density}

\begin{figure}[htb!]
  \centering
    {\includegraphics[width=0.7\textwidth]
	{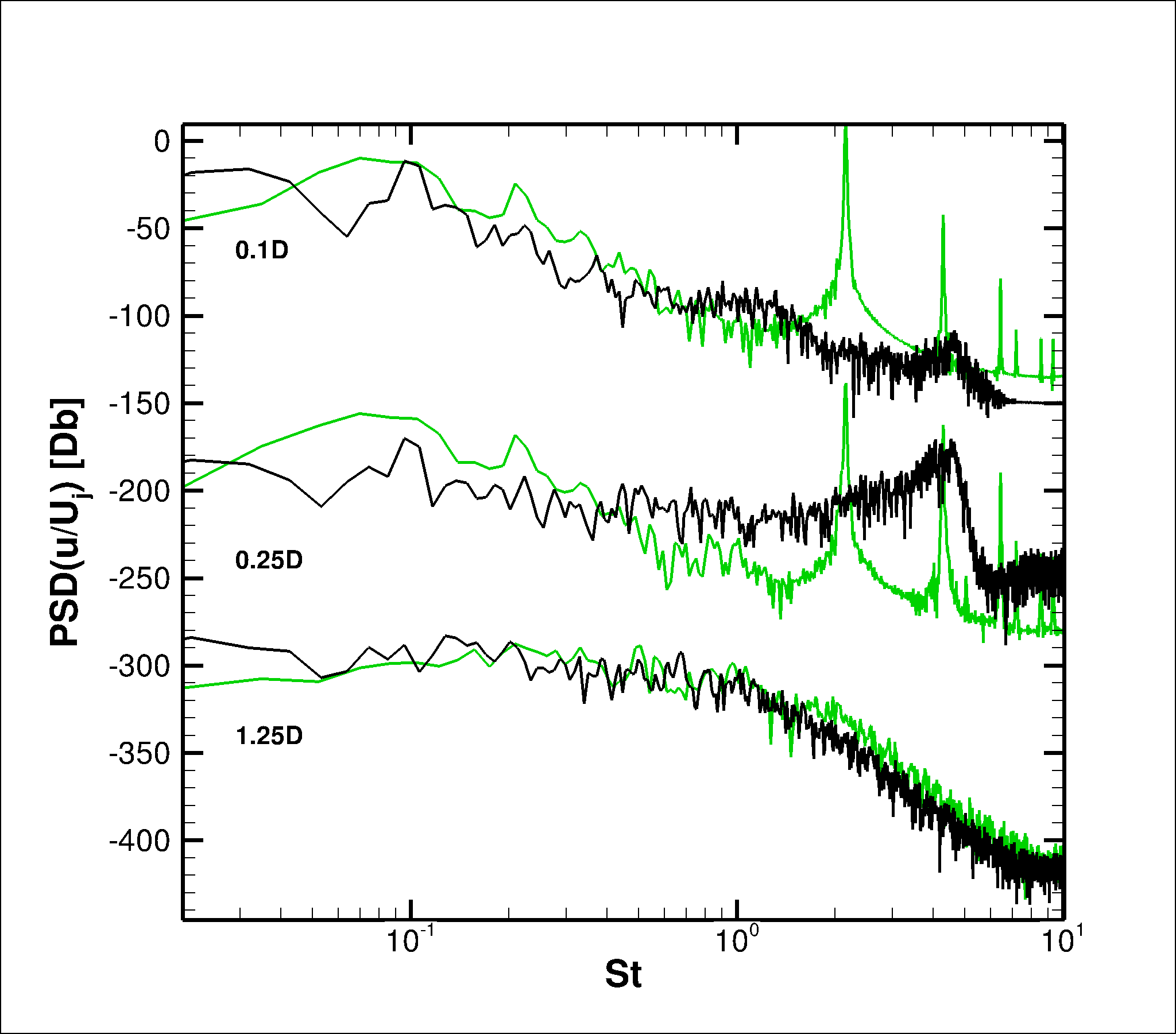}}
	\caption{Power spectral density of $u^{*}$ as function of the Strouhal
	number along the lipline of the jet. 
	(\LARGE {\color{green}\textbf{--}}\scriptsize), S1; 
	(\LARGE \textbf{--}\scriptsize), S2. A shift of -150 {\bf dB} and 
	-300 {\bf dB} has been added to the PSD in order to separate plots 
	for $X=0.25D$ and $X=1.25D$, respectively.}
	\label{fig:psd-mesh}
\end{figure}

The power spectral density of time fluctuation of the axial component 
of velocity is studied in the present work in order to better 
understand the transient portion of the solution. Figure 
\ref{fig:psd-mesh} presents the PSD of $u^{*}$, in $dB$, as function 
of the Strouhal number for S1 and S2. The signals are collected from 
the sensors allocated at the positions presented in Tab.\ 
\ref{tab:sensor}. The PSD of $u^{*}$ are shifted of -150{\bf dB} 
and -300{\bf dB} for $X=0.25D$ and $X=1.25$, respectively, in order 
to separate plots. 

One can observe that PSD signals obtained using S1 and S2 present a 
similar behavior at $X=1.25D$ on the lipline. On the other hand, it 
is possible to notice significant differences on the shape and on the 
peaks positioning for $St>1.0$ at $X=0.1D$ and $X=0.25D$. The divergence 
indicates that the dissipative characteristic of S1 have changed the
positioning of the turbulent transition when compared with the S2 study.

\subsection{Subgrid Scale Modeling Study}

After the mesh refinement study the three SGS models added
to the solver are compared. S2, S3 and S4 simulations
are performed using the static Smagorinsky model 
\cite{Smagorinsky63,Lilly65,Lilly67}, the dynamic Smagorinsky 
model \cite{germano90,moin91} and the Vreman model 
\cite{vreman2004}, respectively. The same mesh with 50 
million points is used for all three simulations.
The stagnated flow condition is used as intial condition
for S2 and S3. A restart of S2 is used as initial condition
for the S4 simulation. The configuration of the numerical 
studies is presented at Tab.\ \ref{tab:simu}. The same comparisons 
performed on the study of mesh refinement effects, Sec.\ 
\ref{sec:mesh-study}, are performed for the SGS modeling study.

\subsubsection*{Time Averaged Axial Component of Velocity}

\begin{figure}[htb!]
  \centering
  \subfigure[Lateral view of $\langle U \rangle$ for S2.]
    {\includegraphics[width=0.32\textwidth]
	{sources/results/pictures/stat-cs/XY-zoom/u-av-zoom-XY}}
  \subfigure[Lateral view of $\langle U \rangle$ for S3.]
    {\includegraphics[width=0.32\textwidth]
	{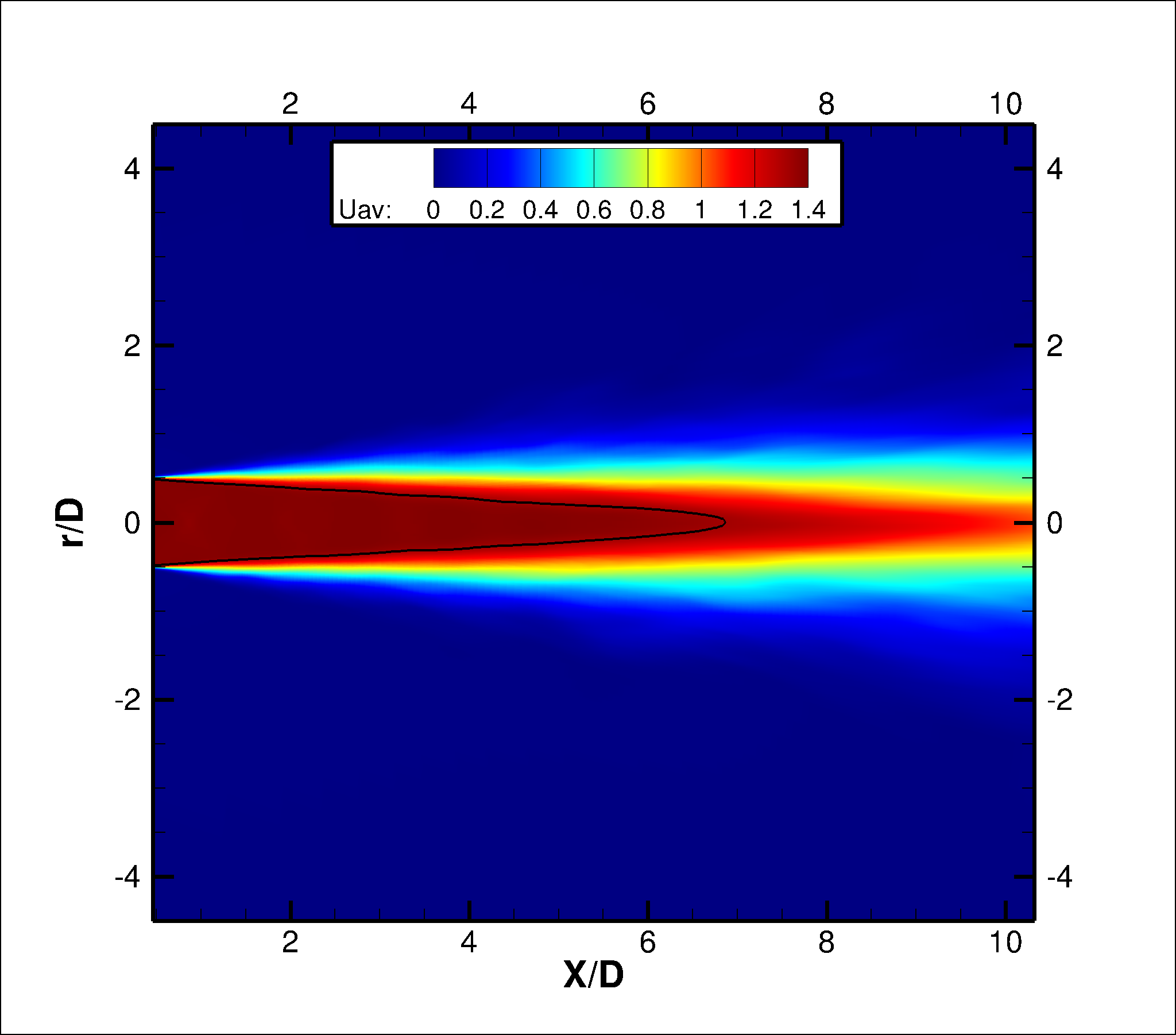}}
  \subfigure[Lateral view of $\langle U \rangle$ for S4.]
    {\includegraphics[width=0.32\textwidth]
	{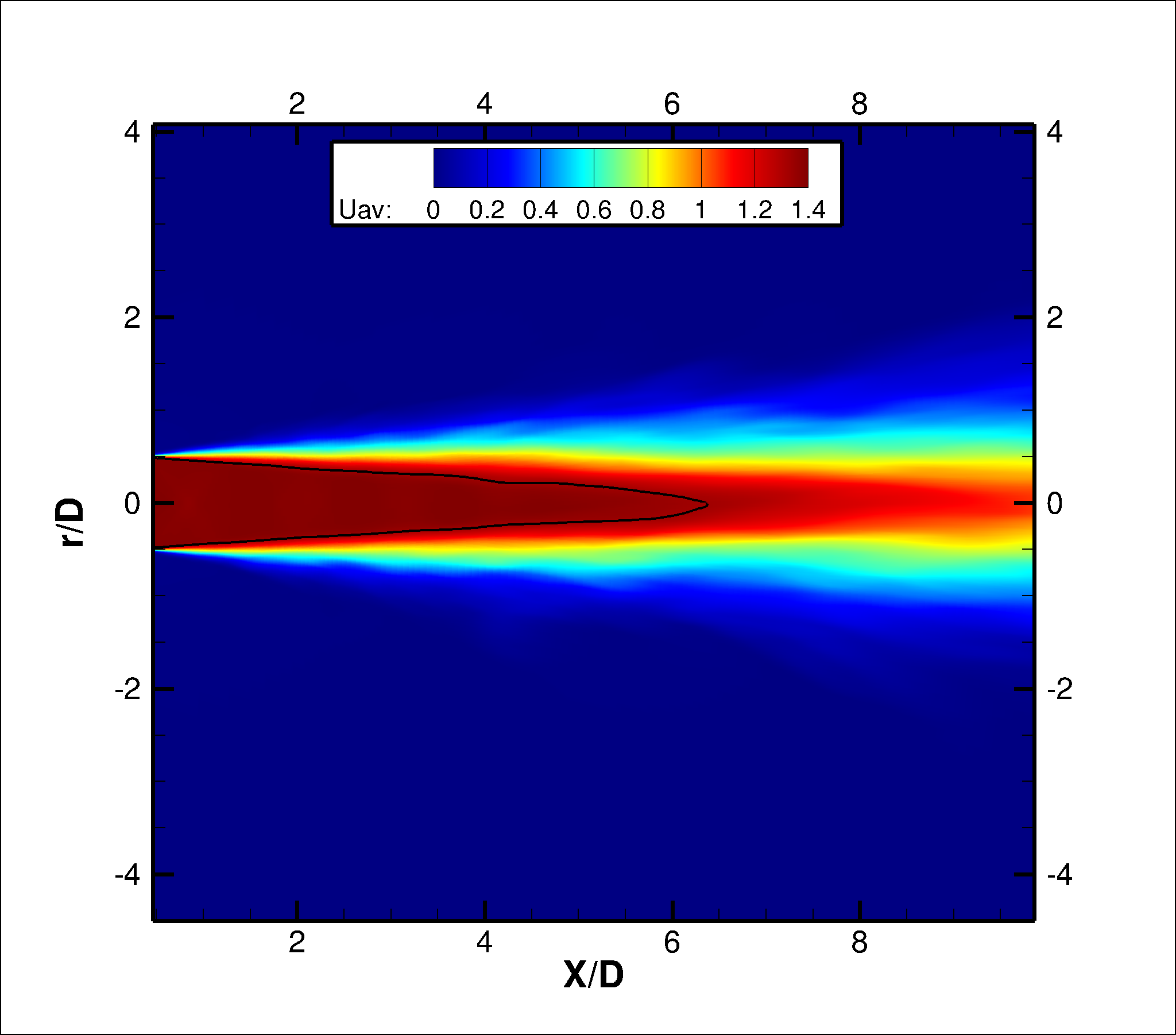}}
	\caption{Lateral view of the averaged axial component 
	of velocity, $\langle U \rangle$, for S2, S3 and S4.
	(\LARGE \textbf{--}\scriptsize) indicates the 
	potential core of the jet, $U_{j}^{95\%}$.} 
	\label{fig:lat-u-av-sgs}
\end{figure}

Effects of the SGS modeling on the time averaged results of the 
axial component of velocity are presented in the subsection. A 
lateral view of $\langle U \rangle$ for S2, S3 and S4, side by 
side, are presented in Fig.\ \ref{fig:lat-u-av-sgs}, where 
$U_{j}^{95\%}$ is indicated by the solid line. 
%
Table \ref{tab:core-sgs} presents the size of 
the potetial core of S2, S3 and S4 and the numerical reference 
\cite{Mendez10,Mendez12} along with the relative error compared 
with the experimental data \cite{bridges2008turbulence}.
\begin{table}[htb!]
\begin{center}
  \caption{Potential core length and relative error of S2, S3 and S4.}
  \label{tab:core-sgs}
  \begin{tabular}{|c|c|c|}
  \hline
  Simulation & $\delta_{j}^{95\%}$ & Relative error\\
  \hline
  S2 & 6.84 & 26\%\\
  S3 & 6.84 & 26\%\\
  S4 & 6.28 & 32\%\\
  Mendez {\it et al.} & 8.35 & 8\%\\
  \hline
  \end{tabular}
\end{center}
\end{table}

Comparing the results, one cannot observe significant differences 
on the potential core length between S2, S3 and S4. The distribution
of $\langle U \rangle$ calculated using the dynamic Smagorinky model
has shown to be slightly more concentrated at the centerline region.
S2 and S4 time averaged distribution of $U$ are, on some small scale,
more spread than the distribution obtained by S3.

\begin{figure}[htb!]
  \centering
  \subfigure[$\langle U \rangle$ - X=2.5D ; $-1.5D\leq Y\leq 1.5D$]
    {\includegraphics[width=0.45\textwidth]
	{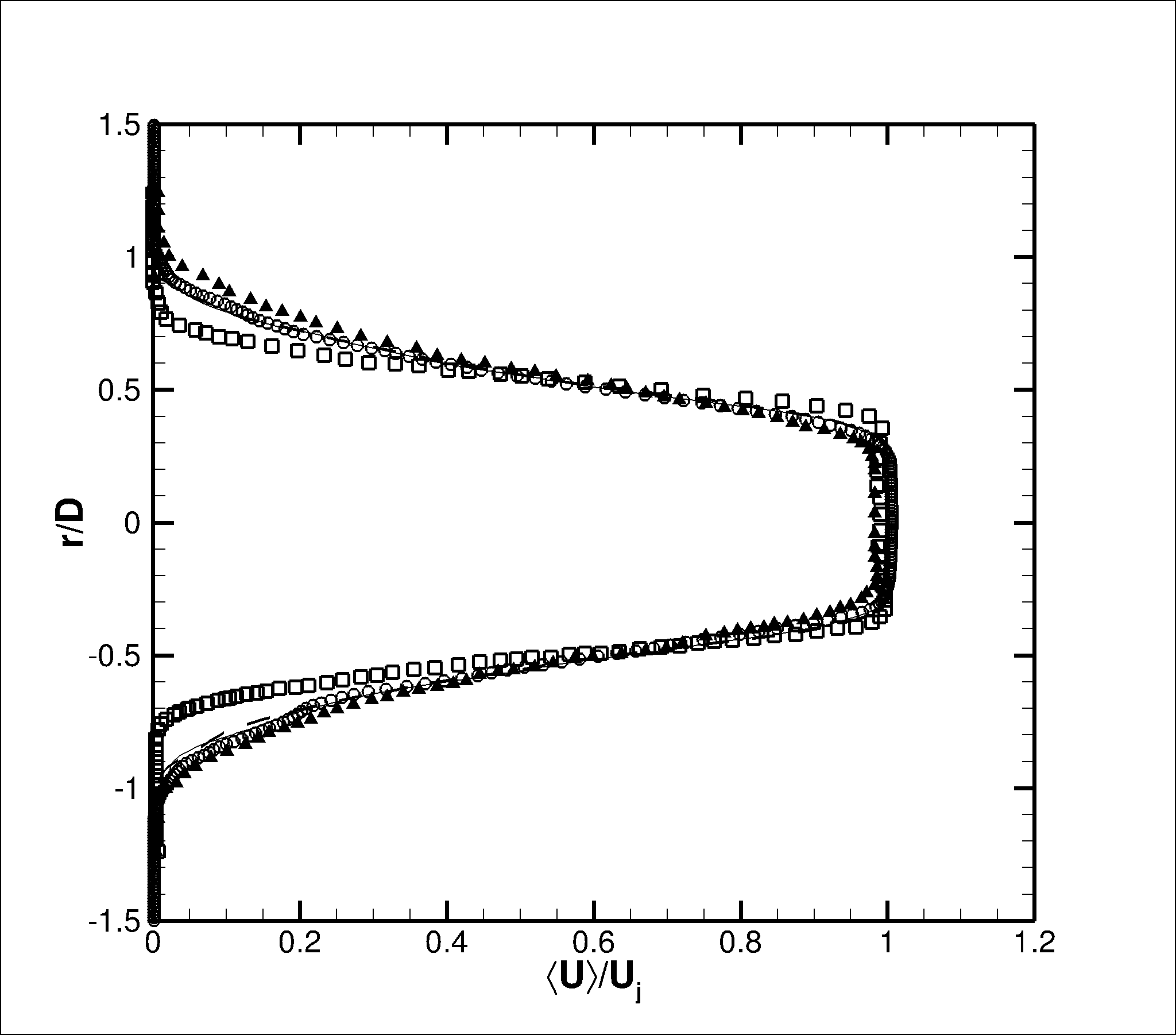}}
  \subfigure[$\langle U \rangle$ - X=5.0D ; $-1.5D\leq Y\leq 1.5D$]
    {\includegraphics[width=0.45\textwidth]
	{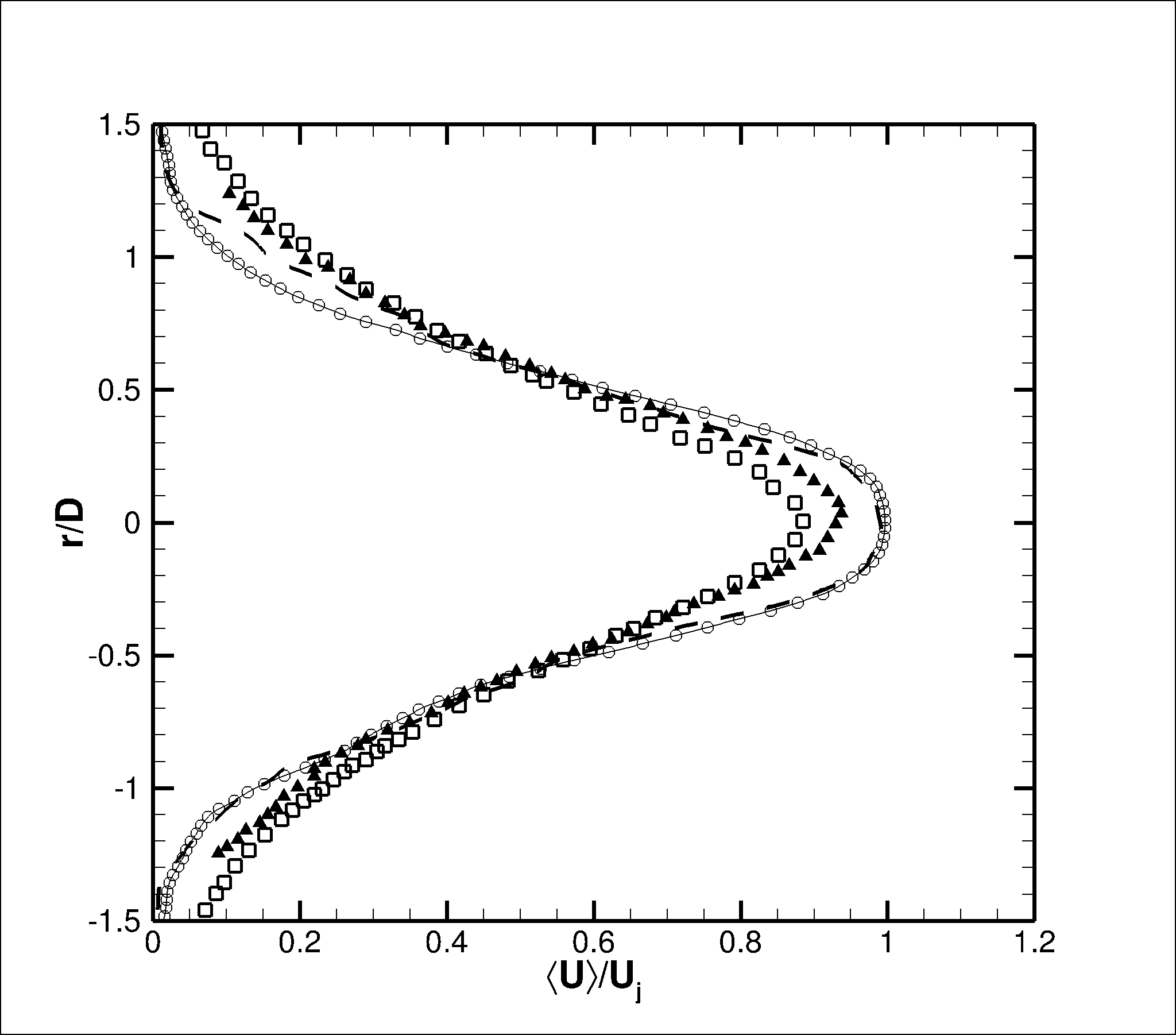}}
  \subfigure[$\langle U \rangle$ - X=10D ; $-1.5D\leq Y\leq 1.5D$]
    {\includegraphics[width=0.45\textwidth]
	{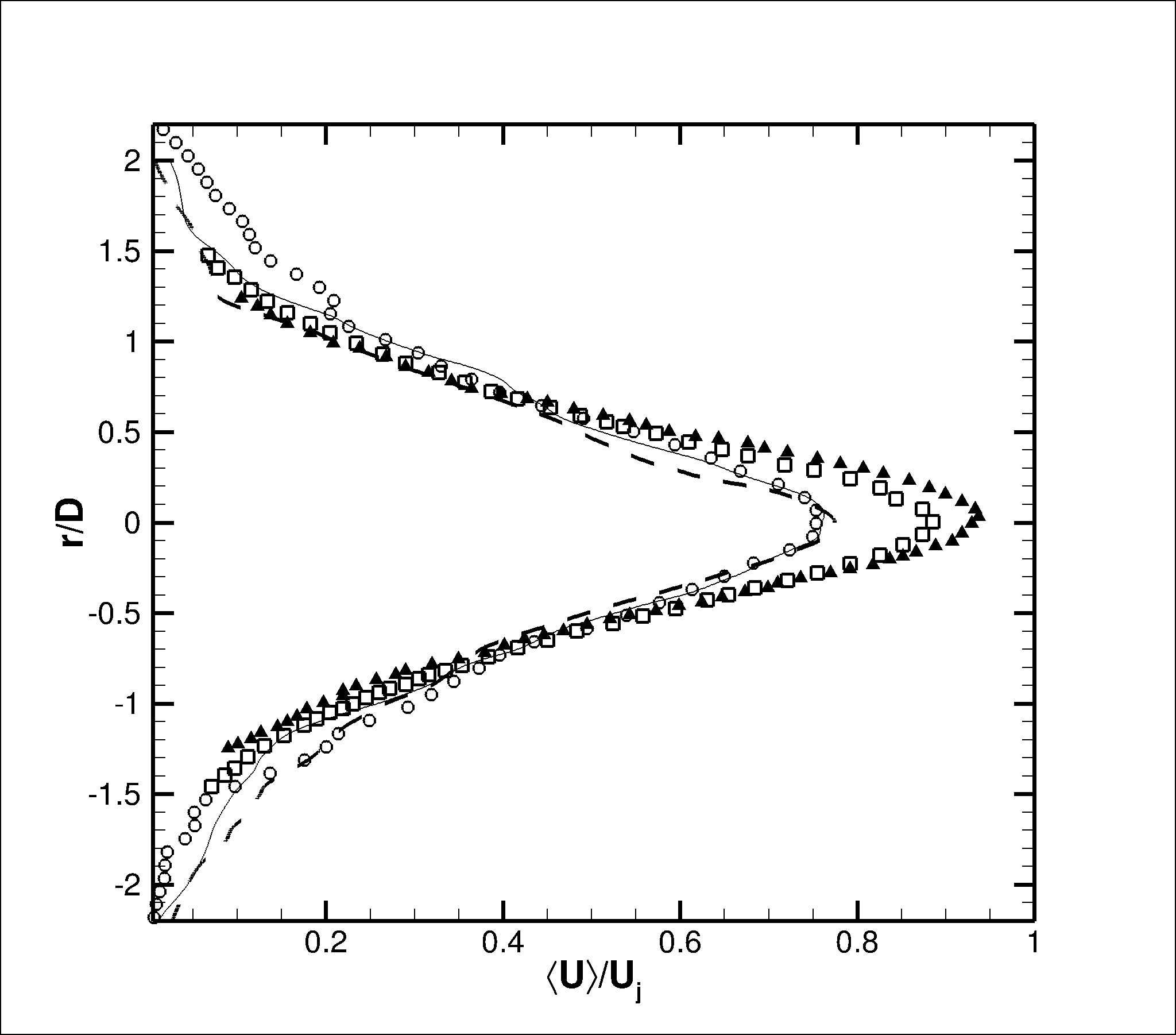}}
  \subfigure[$\langle U \rangle$ - X=15D ; $-1.5D\leq Y\leq 1.5D$]
    {\includegraphics[width=0.45\textwidth]
	{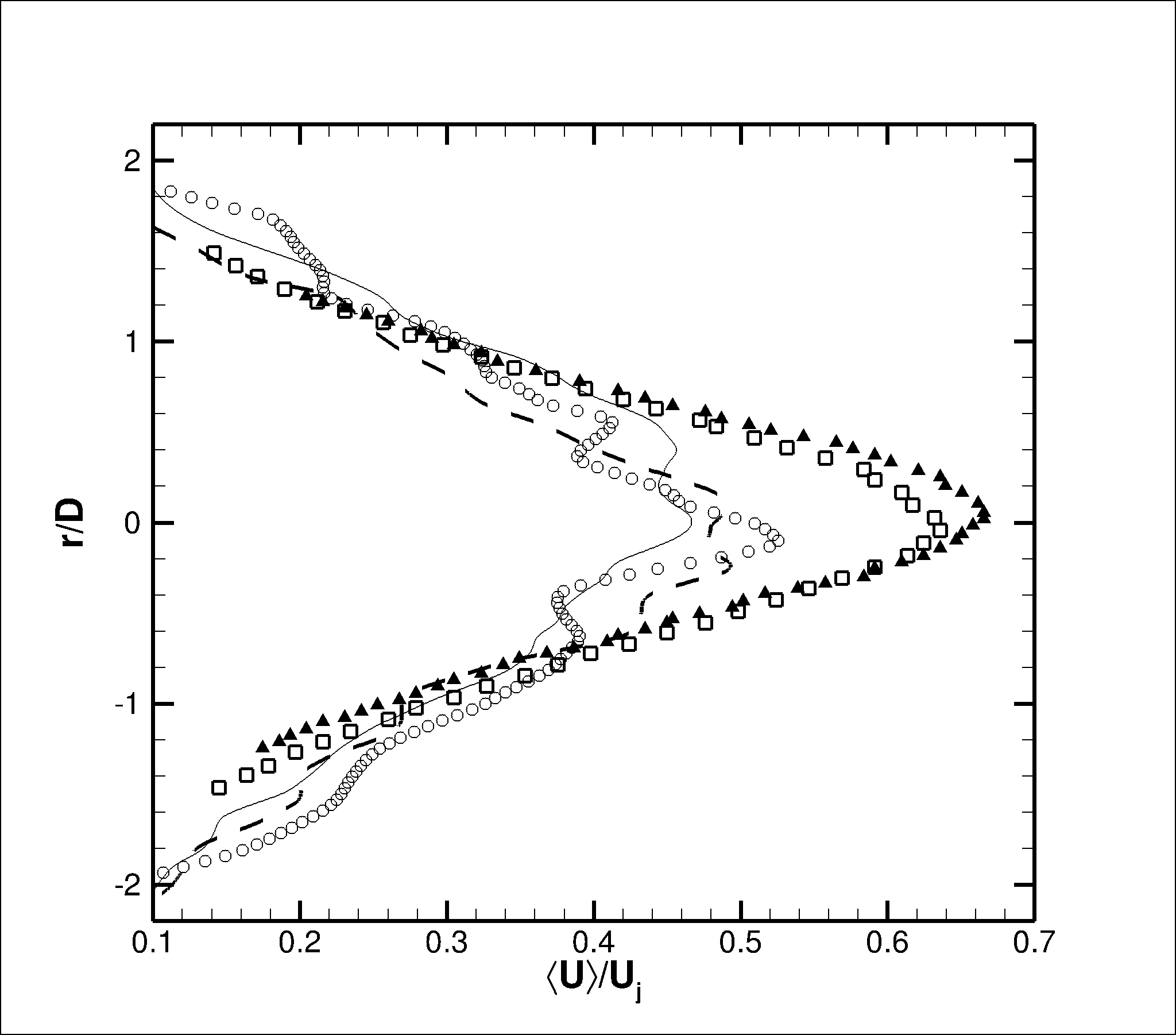}}
  \subfigure[$\langle U \rangle$ - Centerline - Y=0 ; $0\leq X \leq 20D$ ]
    {\includegraphics[width=0.45\textwidth]
	{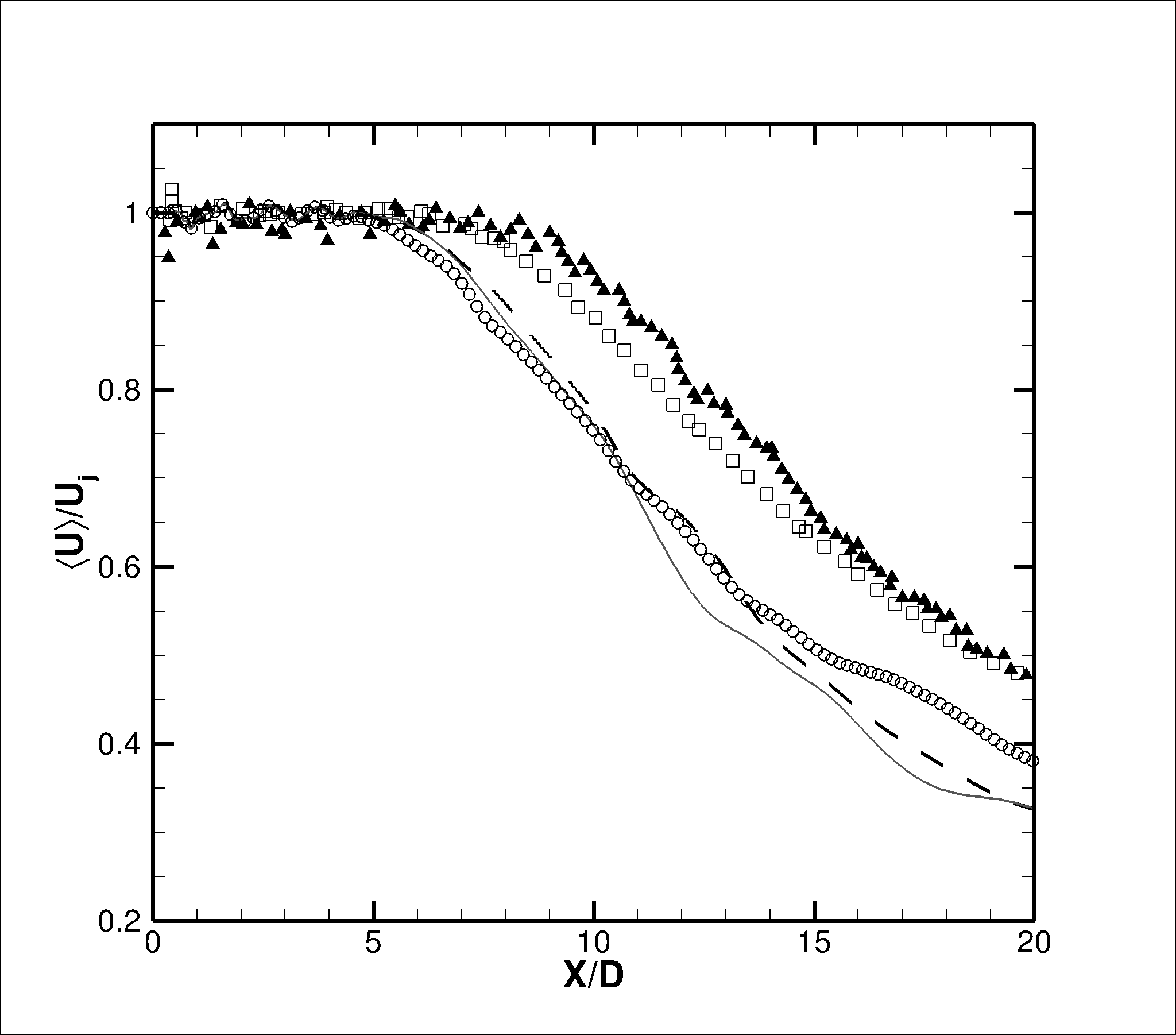}}
  \subfigure[$\langle U \rangle$ - Lipline - Y=0.5D ; $0\leq X \leq 20D$ ]
    {\includegraphics[width=0.45\textwidth]
	{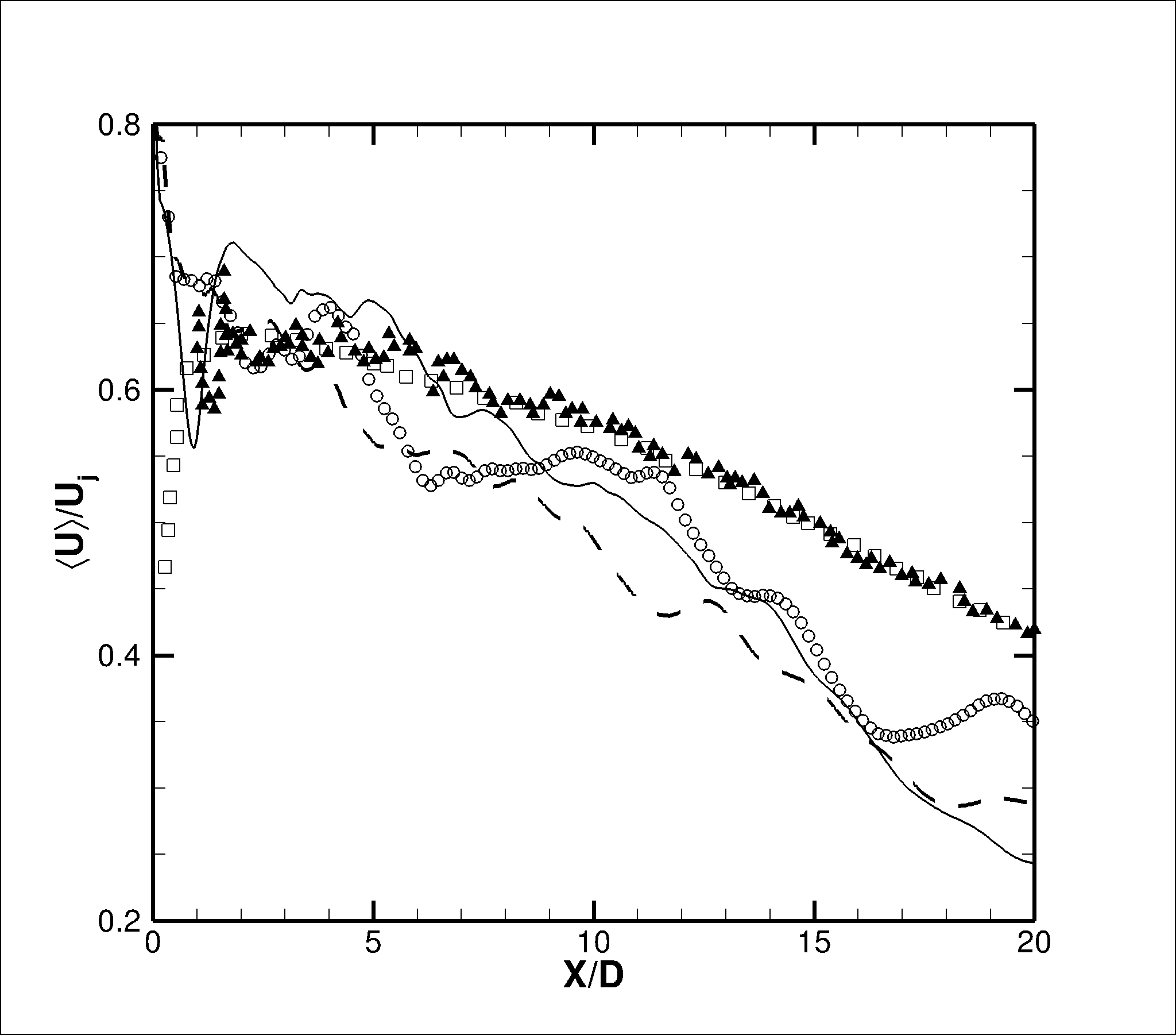}}
	\caption{Profiles of averaged axial component of velocity at 
	different positions within the computational domain.
	(\LARGE \textbf{--}\scriptsize), S2; 
	(\LARGE \textbf{-}\textbf{-}\scriptsize), S3;
	($\bigcirc$), S4; ($\square$), numerical data; 
	($\blacktriangle$), experimental data.}
	\label{fig:prof-u-av-sgs}
\end{figure}

Profiles of $\langle U \rangle$ from S2, S3 and S4, along the
mainstream direction, and the evolution of $\langle U \rangle$
along the centerline and, also along the lipline, are compared 
with numerical and experimental results in Fig.\ \ref{fig:prof-u-av-sgs}.  
The solid line, the dashed line and 
the circular symbol stand for the profiles of $\langle U \rangle$
computed by S2, S3 and S4, respectively. The reference data are 
represented by the same symbols presented in the mesh refinement
study. 

The comparison of profiles indicates that distributions of 
$\langle U \rangle$ calculated on S2, S3 and S4 correlates 
well with the references until $X=5.0D$. For $X>10.0D$ 
all SGS models fail to predict the correct profile.
One can notice that the evolution of $\langle U\rangle$
along the centerline, calculated by all three simulations,
are in good agreement with the numerical and experimental 
reference data at the region where the mesh presents a good 
resolution. Moreover, the three distributions calculated using
different SGS closures have presented the very similar behavior.
The dynamic Smagorinsky model and the Vreman model correlates
better with the experimental data for $X<5.0D$ than the classic 
Smagorinsky closure does. However, all simulations tend to not 
predict well the magnitude of $\langle U \rangle$ on the lipline
when the mesh size increases, $X>5.0D$. 



\subsubsection*{Root Mean Square Distribution of Time Fluctuations 
of Axial Velocity Component}

\begin{figure}[htb!]
  \centering
  \subfigure[Lateral view of $u^{*}_{RMS}$ for S2.]
    {\includegraphics[width=0.32\textwidth]
	{sources/results/pictures/stat-cs/XY-zoom/u-rms-zoom-XY}\label{subfig:urms-sgs-s2}}
  \subfigure[Lateral view of $u^{*}_{RMS}$ for S3.]
    {\includegraphics[width=0.32\textwidth]
	{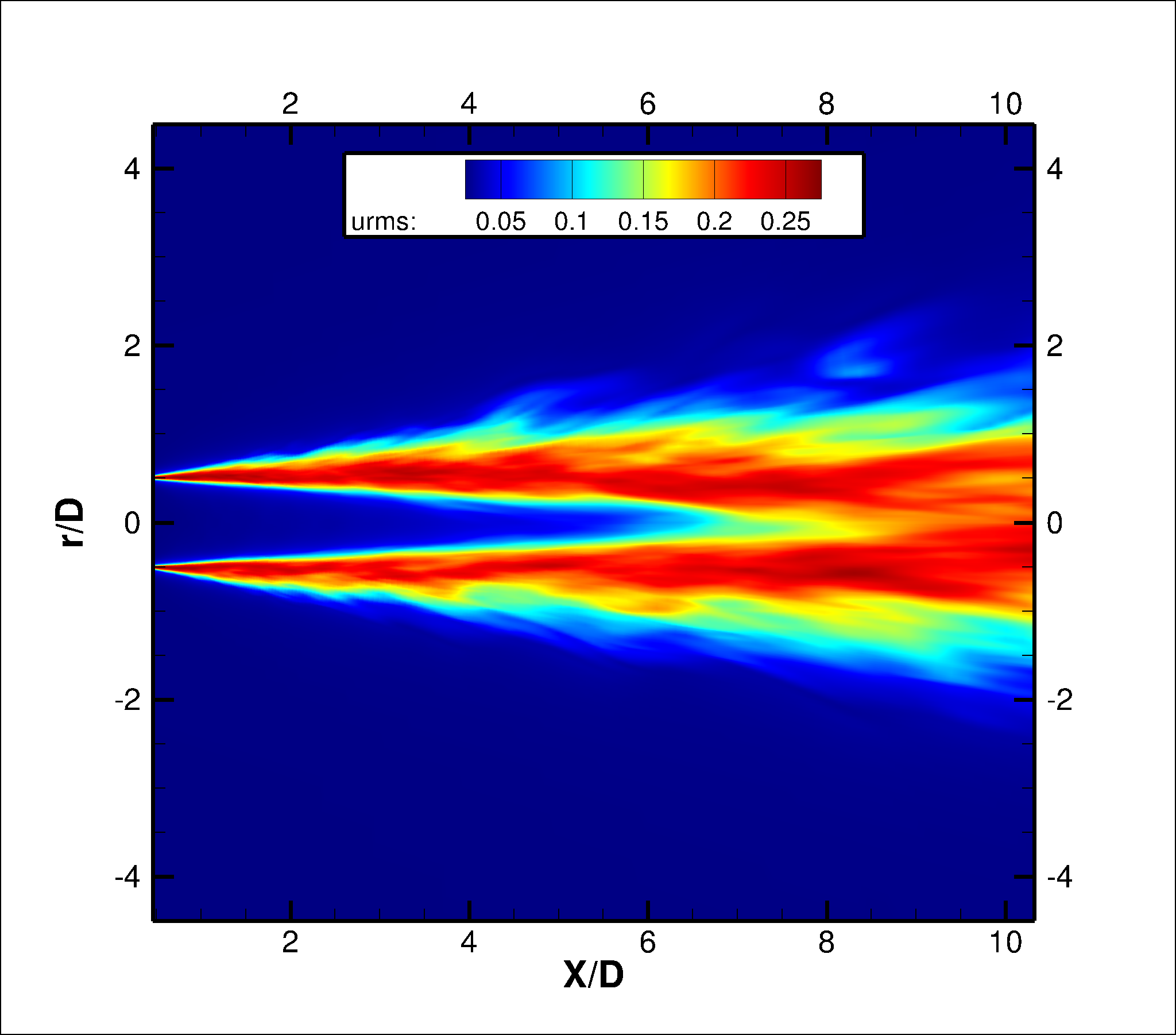}\label{subfig:urms-sgs-s3}}
  \subfigure[Lateral view of $u^{*}_{RMS}$ for S4.]
    {\includegraphics[width=0.32\textwidth]
	{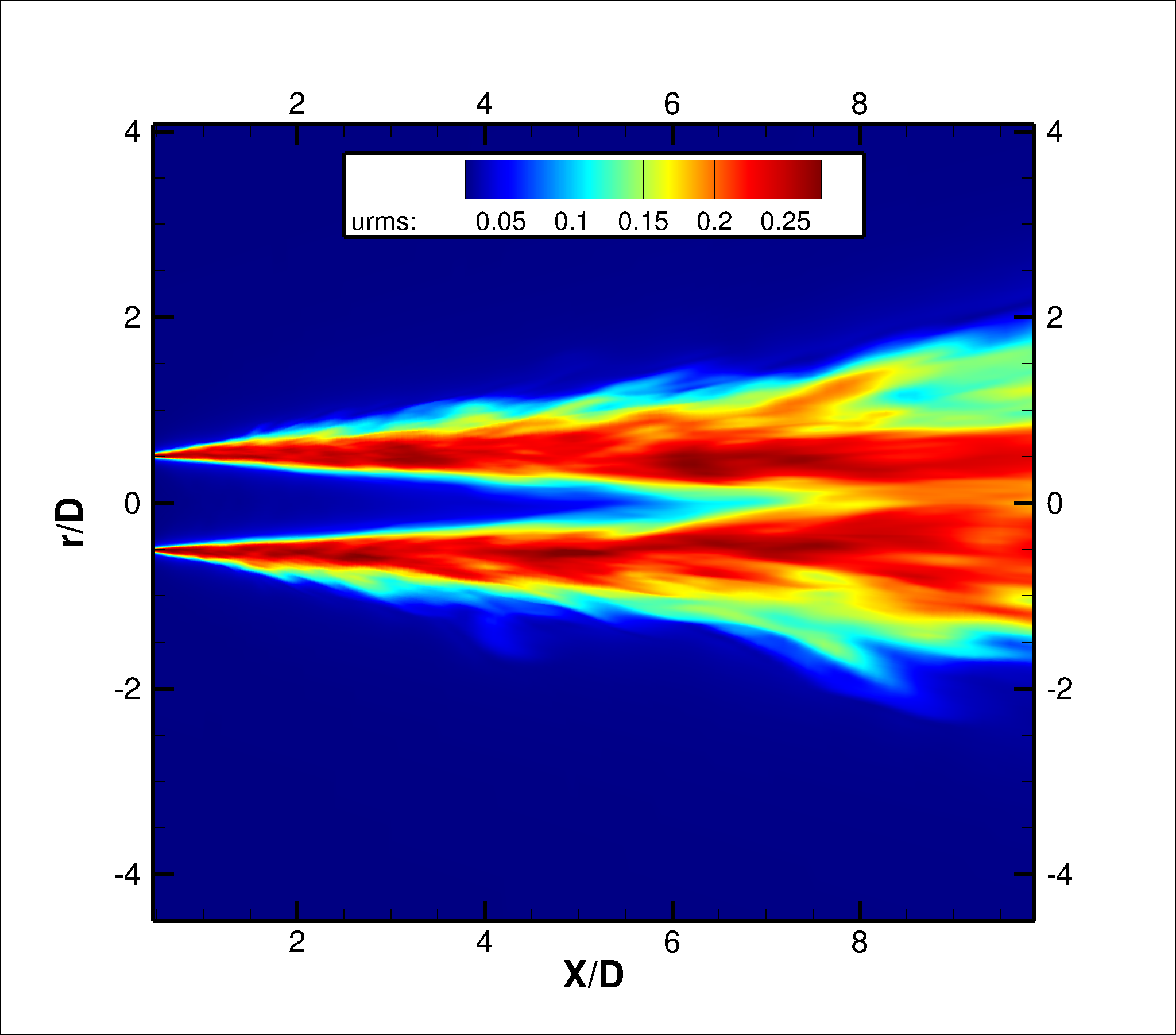}\label{subfig:urms-sgs-s4}}
  \subfigure[Lateral view of $v^{*}_{RMS}$ for S2.]
    {\includegraphics[width=0.32\textwidth]
	{sources/results/pictures/stat-cs/XY-zoom/v-rms-zoom-XY}\label{subfig:vrms-sgs-s2}}
  \subfigure[Lateral view of $v^{*}_{RMS}$ for S3.]
    {\includegraphics[width=0.32\textwidth]
	{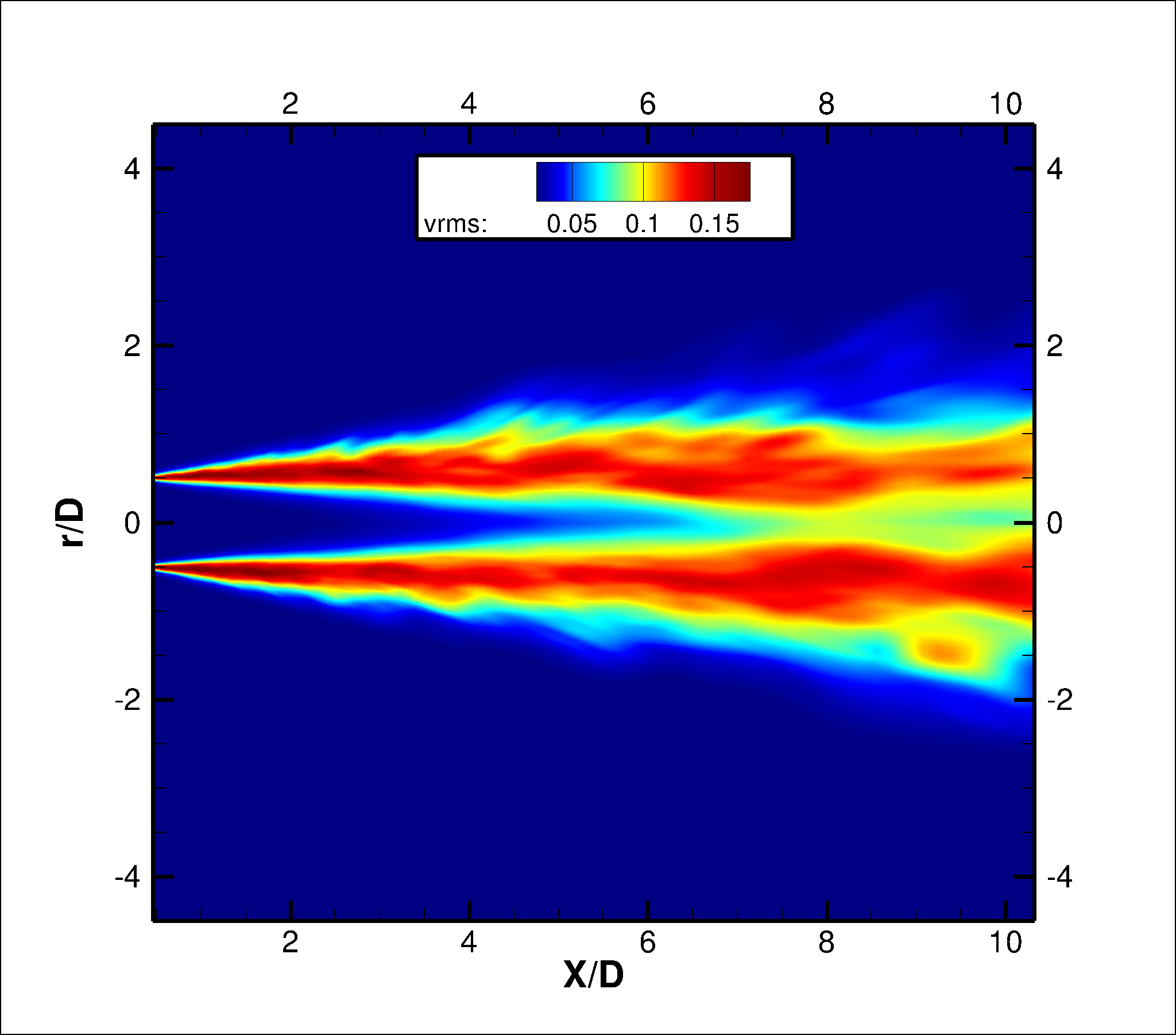}\label{subfig:vrms-sgs-s3}}
  \subfigure[Lateral view of $v^{*}_{RMS}$ for S4.]
    {\includegraphics[width=0.32\textwidth]
	{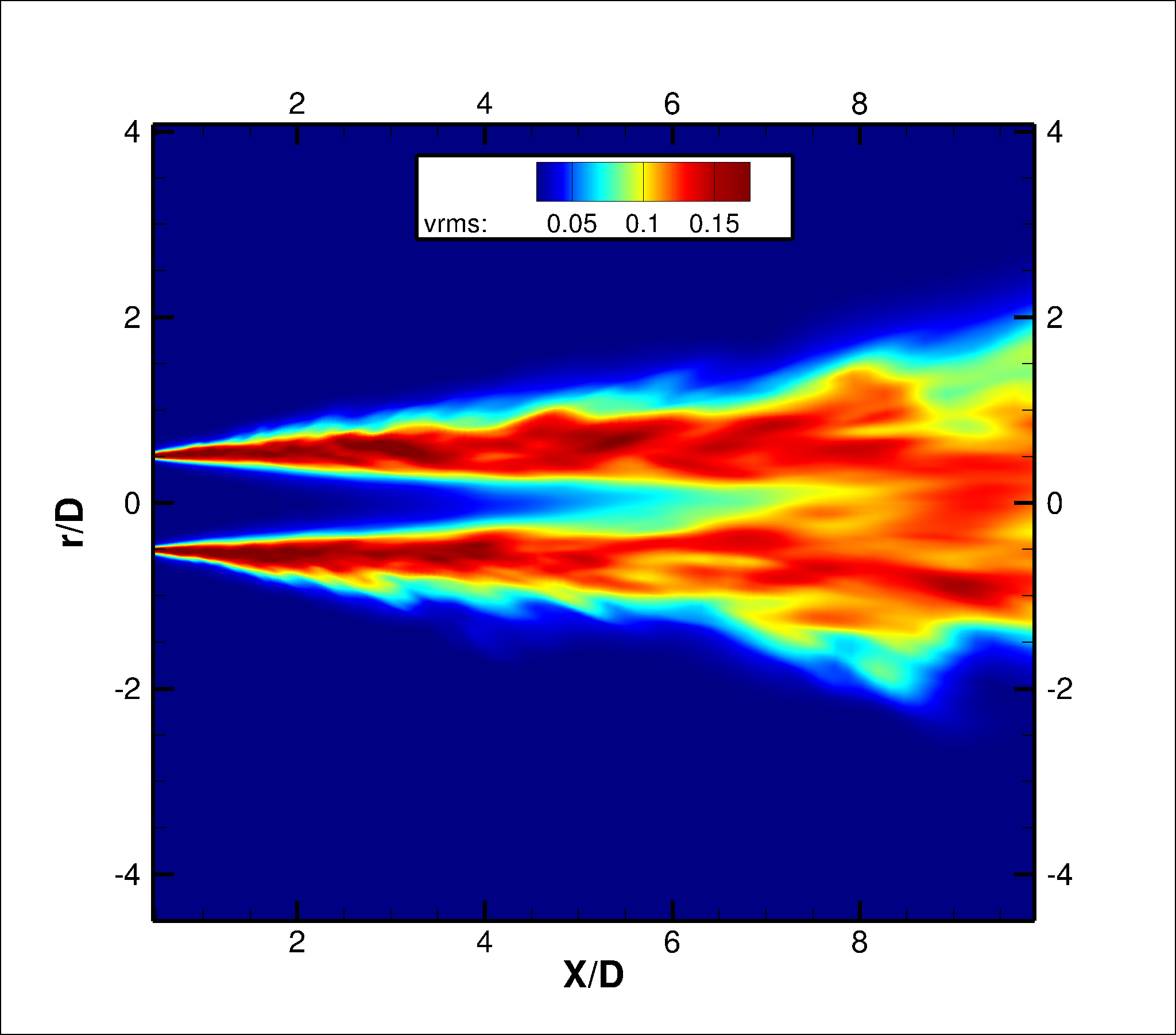}\label{subfig:vrms-sgs-s4}}
  \subfigure[Lateral view of $\langle u^{*}v^{*}\rangle$ for S2.]
    {\includegraphics[width=0.32\textwidth]
	{sources/results/pictures/stat-cs/XY-zoom/uv-av-zoom-XY}\label{subfig:uv-sgs-s2}}
  \subfigure[Lateral view of $\langle u^{*}v^{*}\rangle$ for S3.]
    {\includegraphics[width=0.32\textwidth]
	{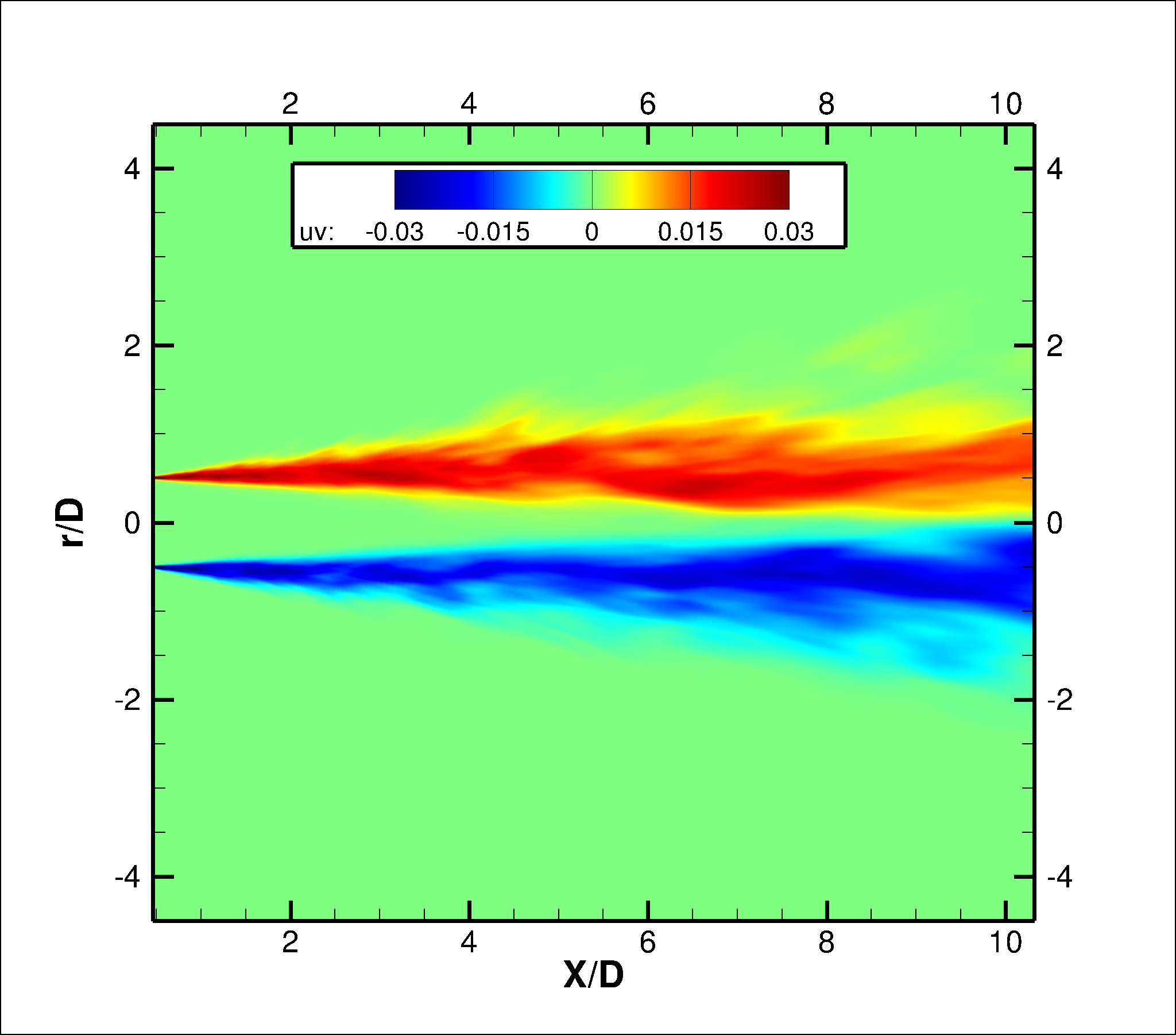}\label{subfig:uv-sgs-s3}}
  \subfigure[Lateral view of $\langle u^{*}v^{*}\rangle$ for S4.]
    {\includegraphics[width=0.32\textwidth]
	{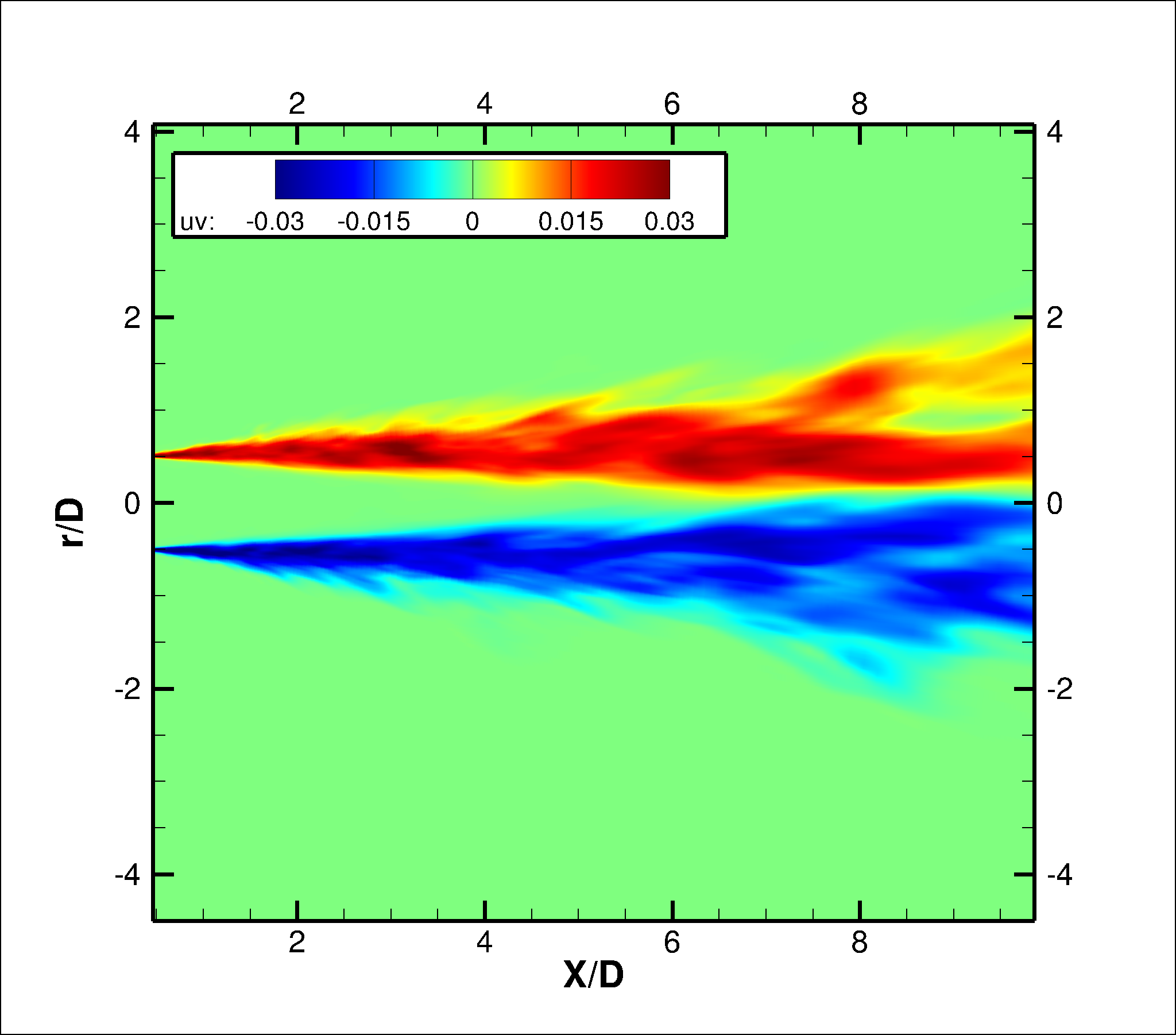}\label{subfig:uv-sgs-s4}}
	\caption{Lateral view of RMS of time fluctuation
	of axial component of velocity, $u_{RMS}^{*}$, RMS 
	of time fluctuation of radial component of velocity, 
	$v_{RMS}^{*}$ and $\langle u^{*}v^{*}\rangle$ Reynolds 
	shear stress tensor component, for S2, S3 and S4.} 
	\label{fig:lat-u-v-uv-sgs}
\end{figure}

\begin{figure}[htb!]
  \centering
  \subfigure[$u^{*}_{RMS}$ - X=2.5D ; $-1.5D\leq Y\leq 1.5D$]
    {\includegraphics[width=0.45\textwidth]
	{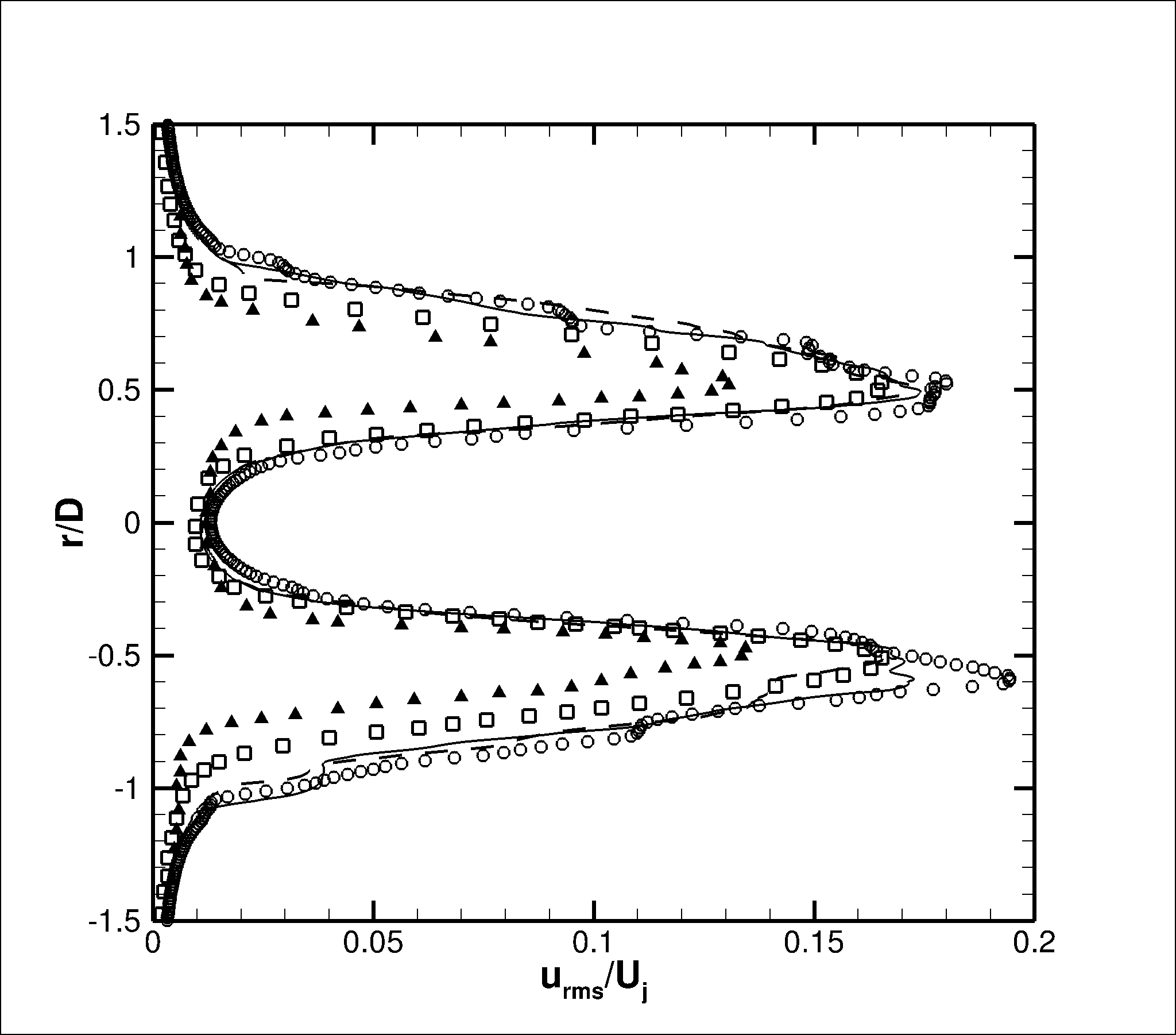}}
  \subfigure[$u^{*}_{RMS}$ - X=5.0D ; $-1.5D\leq Y\leq 1.5D$]
    {\includegraphics[width=0.45\textwidth]
	{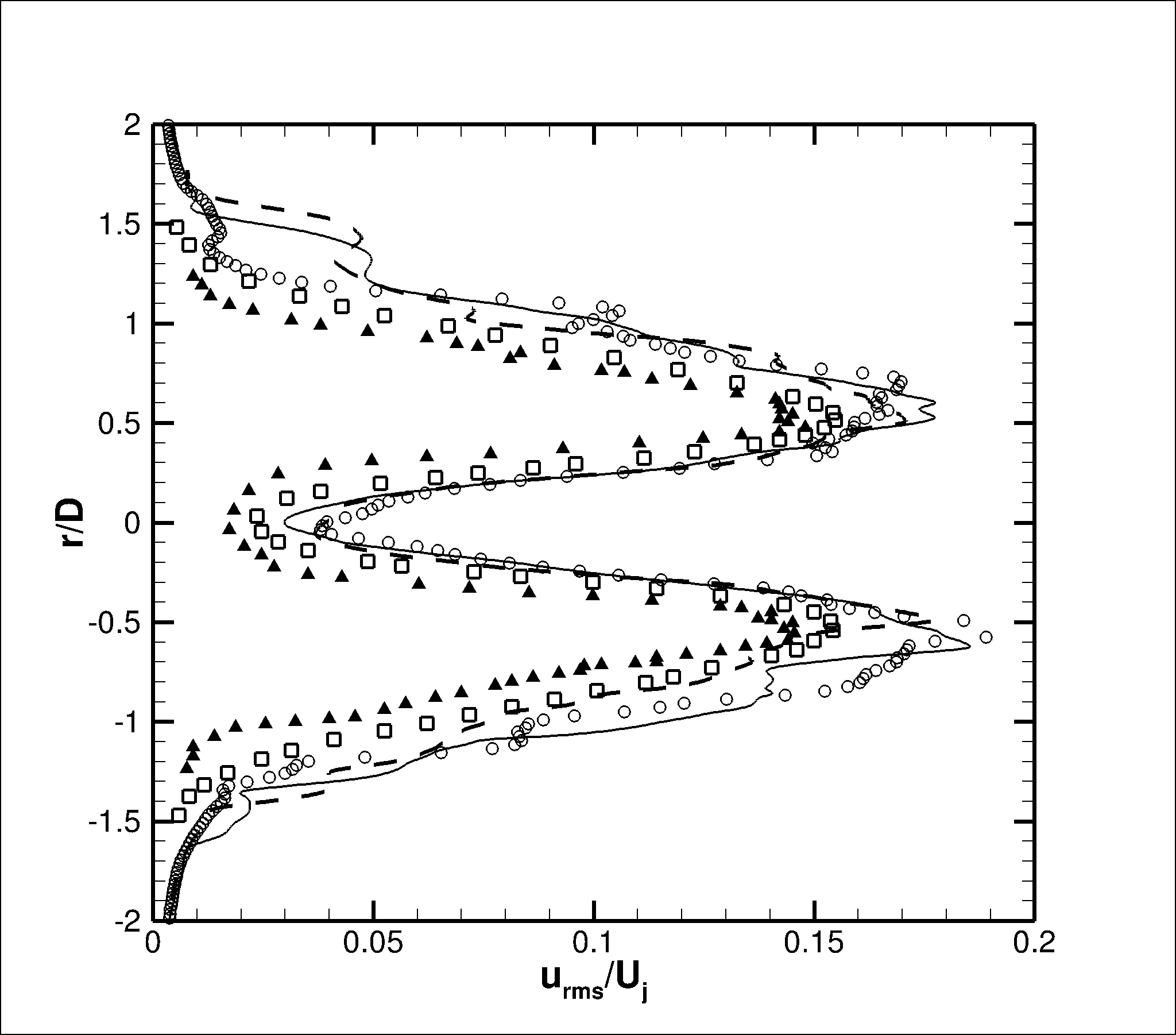}}
  \subfigure[$u^{*}_{RMS}$ - X=10D ; $-1.5D\leq Y\leq 1.5D$]
    {\includegraphics[width=0.45\textwidth]
	{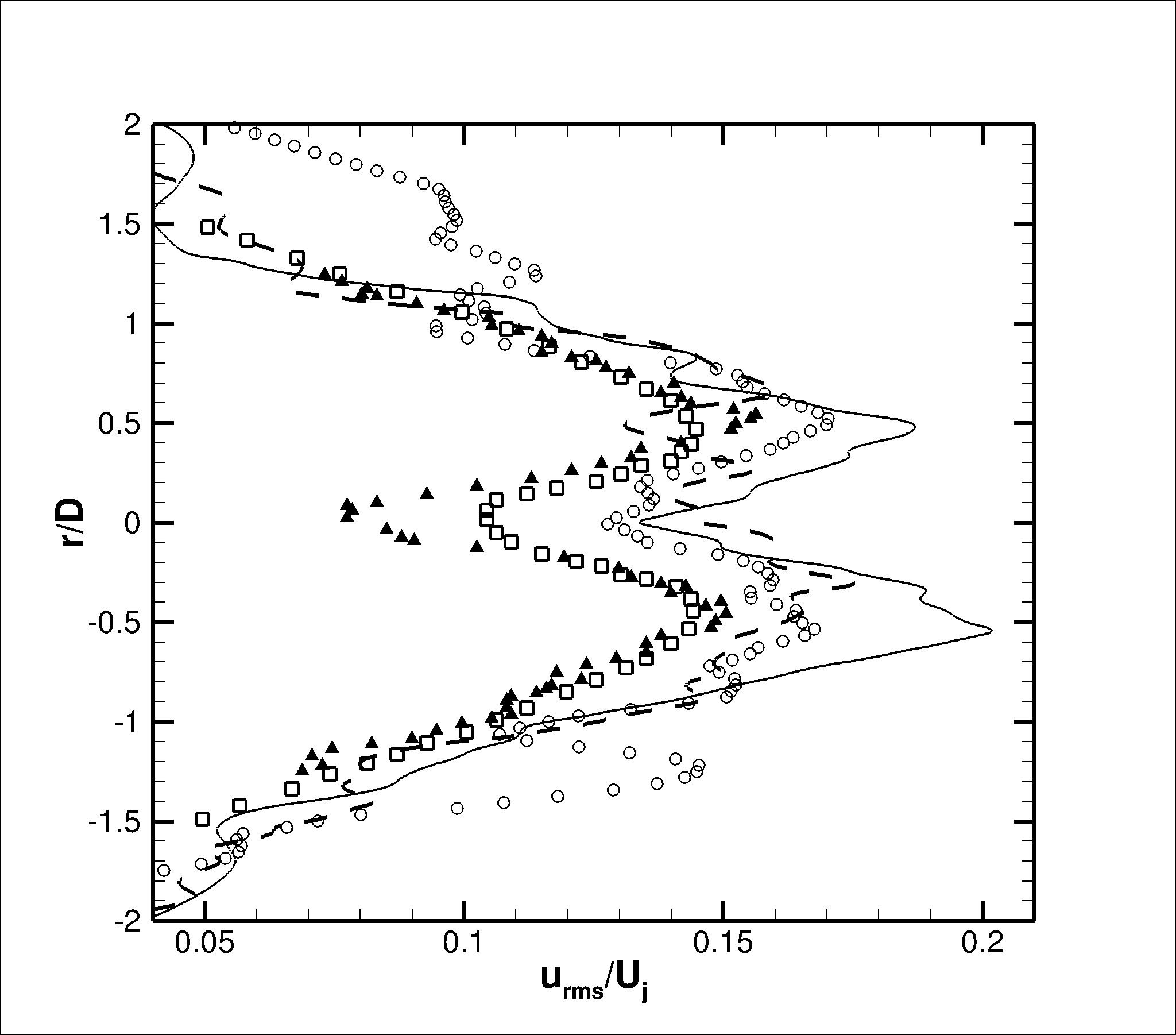}}
  \subfigure[$u^{*}_{RMS}$ - X=15D ; $-1.5D\leq Y\leq 1.5D$]
    {\includegraphics[width=0.45\textwidth]
	{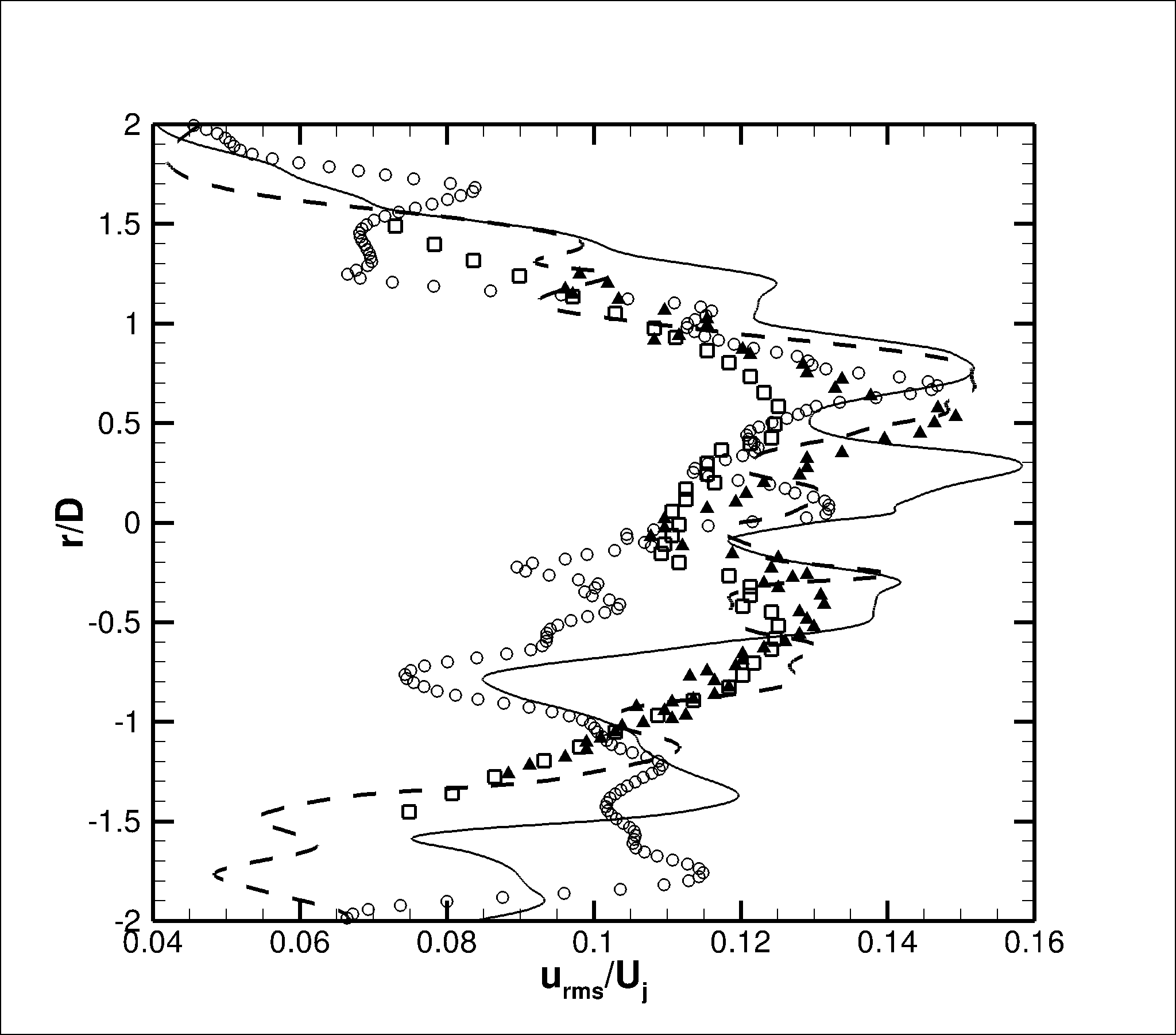}}
  \subfigure[$u^{*}_{RMS}$ - Centerline - Y=0 ; $0\leq X \leq 20D$ ]
    {\includegraphics[width=0.45\textwidth]
	{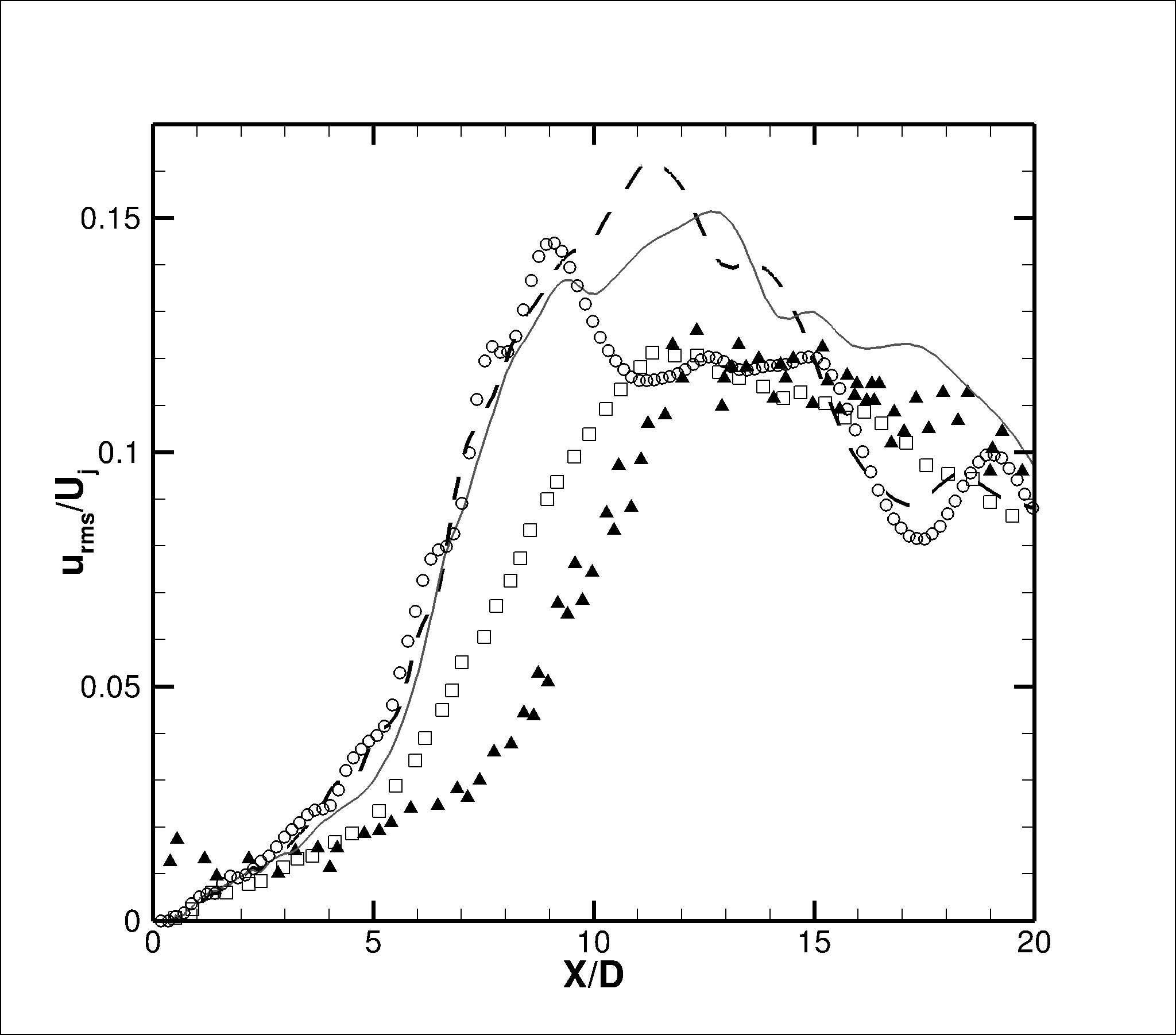}}
  \subfigure[$u^{*}_{RMS}$ - Lipline - Y=0.5D ; $0\leq X \leq 20D$ ]
    {\includegraphics[width=0.45\textwidth]
	{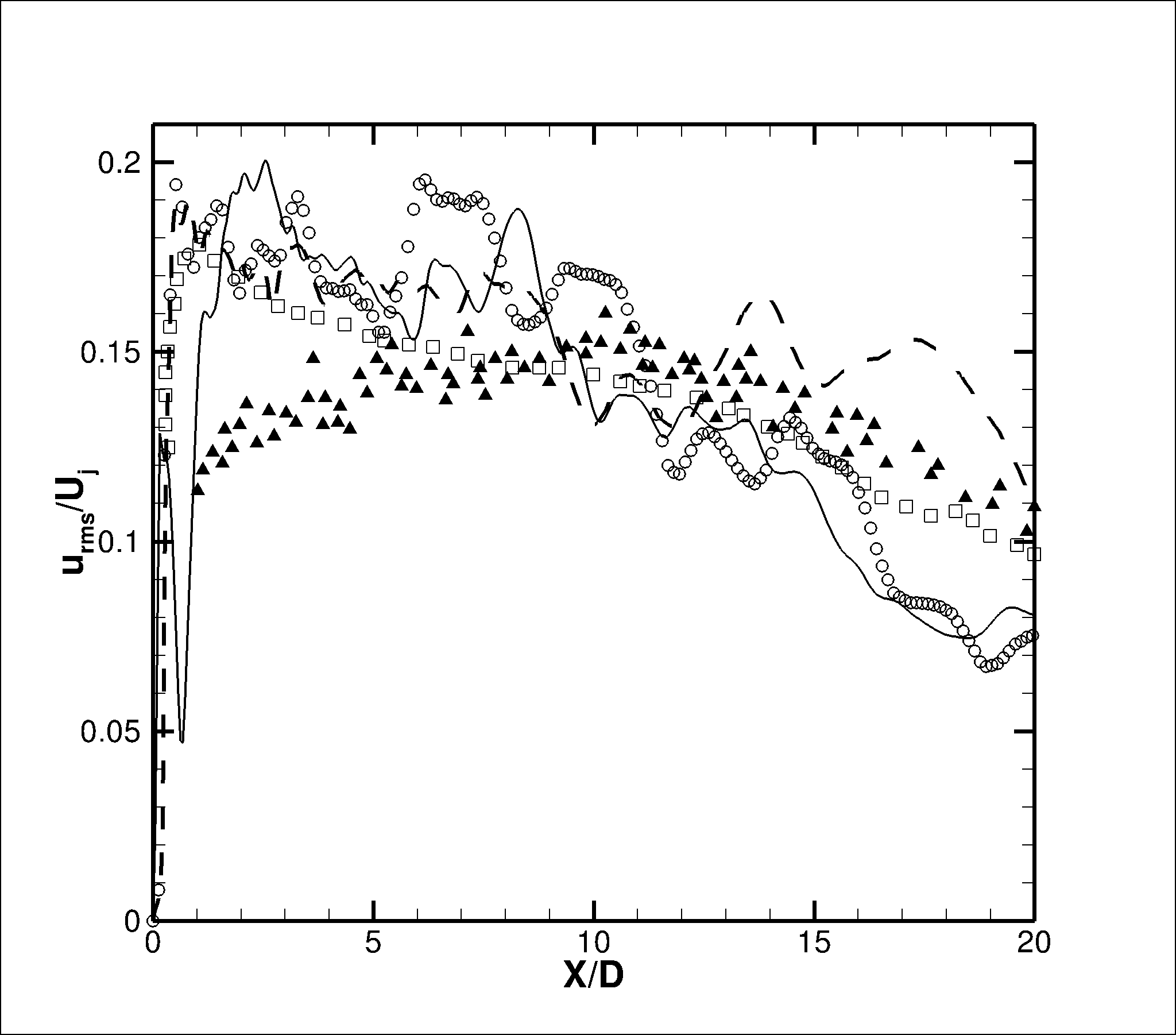}}
	\caption{Profiles of RMS of time fluctuation of axial 
	component of velocity, $u_{RMS}^{*}$, for S2, S3 and S4,
	at different positions within the computational domain.
	(\LARGE \textbf{--}\scriptsize), S2; 
	(\LARGE \textbf{-}\textbf{-}\scriptsize), S3;
	($\bigcirc$), S4; ($\square$), numerical data; 
	($\blacktriangle$), experimental data.}
	\label{fig:prof-u-rms-sgs}
\end{figure}

A lateral view of $u^{*}_{RMS}$ computed by S2, S3 and S4 simulations 
are presented in Figs.\ \ref{subfig:urms-sgs-s2}, \ref{subfig:urms-sgs-s3}
and \ref{subfig:urms-sgs-s4}, respectively. 
The profiles of $u^{*}_{RMS}$ at $X=2.5D$ obtained by S2, S3 and S4 are
in good agreement with the numerical reference, as one can observe in 
Fig.\ \ref{fig:prof-u-rms-sgs}. However, all simulations,
including the LES reference, fail to predict the peaks of $u^{*}_{RMS}$.
At $X=5.0D$ all three simulations have difficulties to predict the peaks 
of the profile. Nonetheless, the results are still in good agreement with
the literature. In the sequence, the profile of $u^{*}_{RMS}$ at $X=10.0D$
calculated by S2, S3 and S4 starts to diverge from the reference results.
Finally, at $X=15.0D$, all SGS closures, but the dynamic Smagorinsky model,
fail the predict the correct profile. S3 simulation have produced a profile 
of $u_{RMS}^{*}$ at $X=15.0D$ that is closer to the experimental data than
the numerical reference.

All three simulations have presented overestimated distributions of 
$u_{RMS}^{*}$ along the centerline. However, for $10D<X<15D$, the 
Vreman model correctly reproduces the magnitude of $u_{RMS}^{*}$.
All simulations performed in the present work have produced noisy 
distributions that diverge from the experimental data along the lipline.
One can notice that the numerical reference has also produced an 
overestimated distribution of $u_{RMS}^{*}$ at $X<10D$.


\subsubsection*{Root Mean Square Distribution of Time Fluctuations 
of Radial Velocity Component}

Effects of SGS modeling on the time fluctuation of the radial 
component of velocity are also compared with the reference data. 
Figures \ref{subfig:vrms-sgs-s2}, \ref{subfig:vrms-sgs-s3} and
\ref{subfig:vrms-sgs-s4} illustrate a lateral view of the 
distribution of $v_{RMS}^{*}$ computed by S2, S3 and S4, respectively. 
The SGS models does not significantly affect the distribution 
of $v_{RMS}^{*}$. All distributions calculated by S2, S3, S4 have 
shown similar behavior.

\begin{figure}[htb!]
  \centering
  \subfigure[$v^{*}_{RMS}$ - X=2.5D ; $-1.5D\leq Y\leq 1.5D$]
    {\includegraphics[width=0.45\textwidth]
	{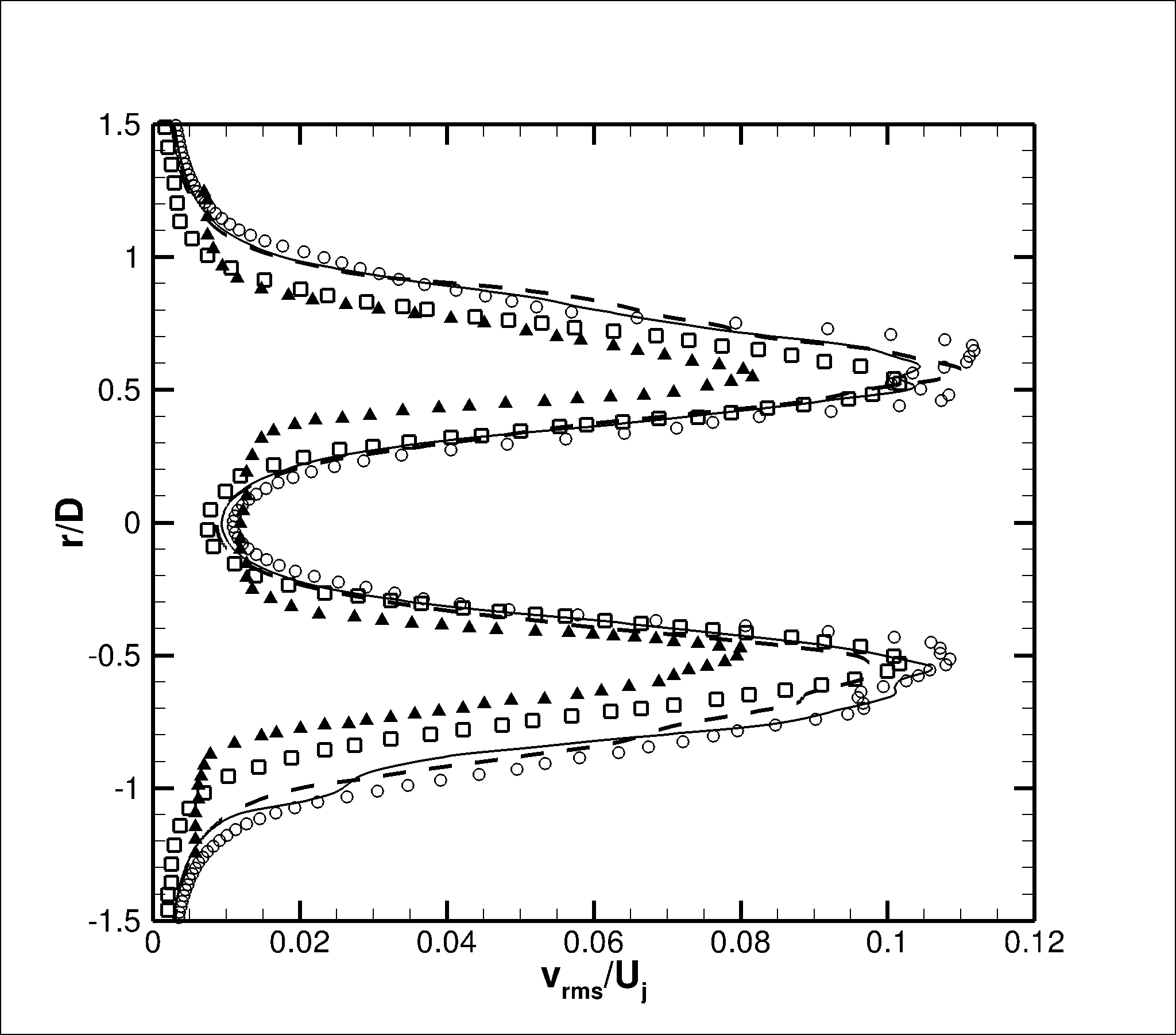}}
  \subfigure[$v^{*}_{RMS}$ - X=5.0D ; $-1.5D\leq Y\leq 1.5D$]
    {\includegraphics[width=0.45\textwidth]
	{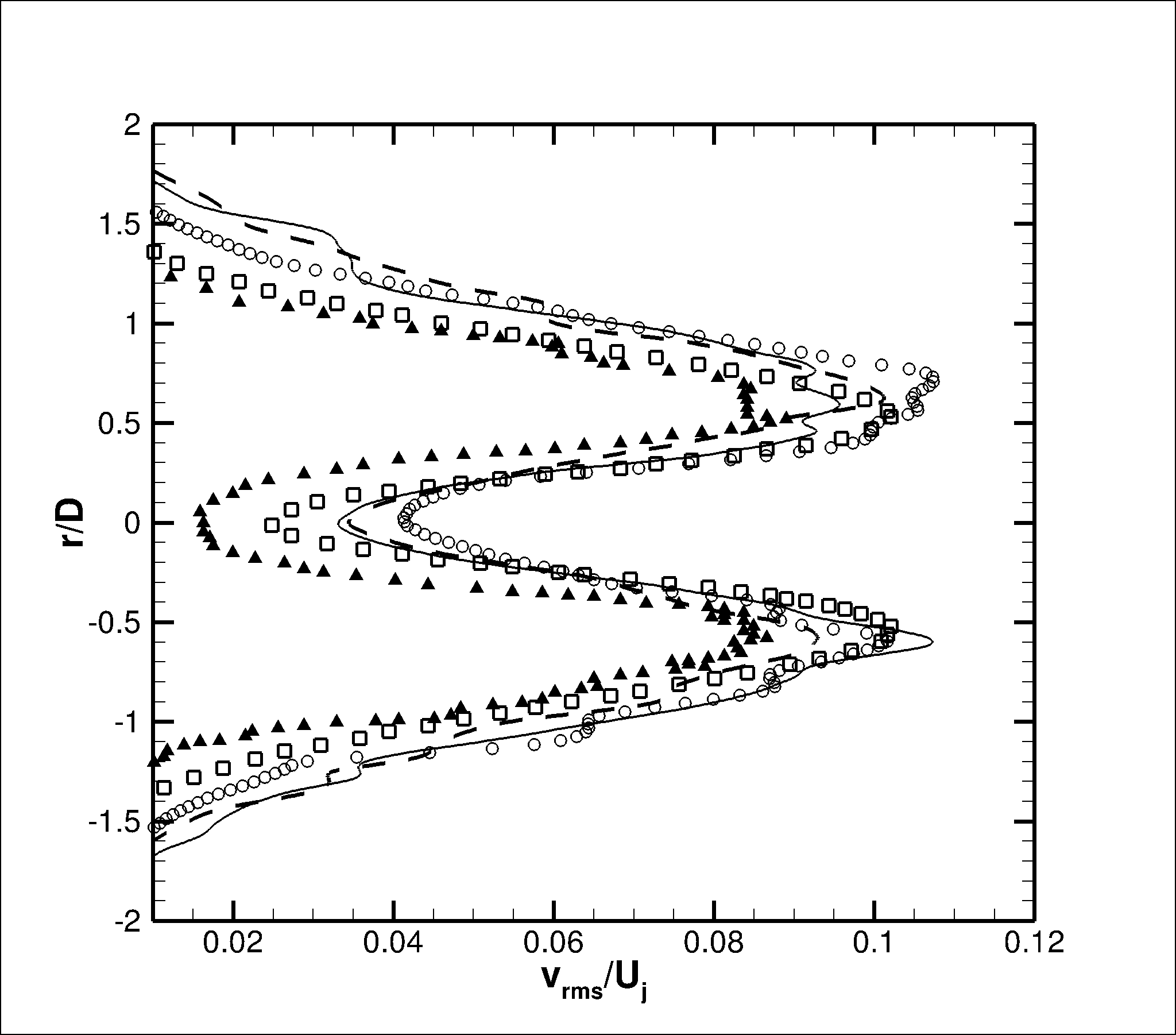}}
  \subfigure[$v^{*}_{RMS}$ - X=10D ; $-1.5D\leq Y\leq 1.5D$]
    {\includegraphics[width=0.45\textwidth]
	{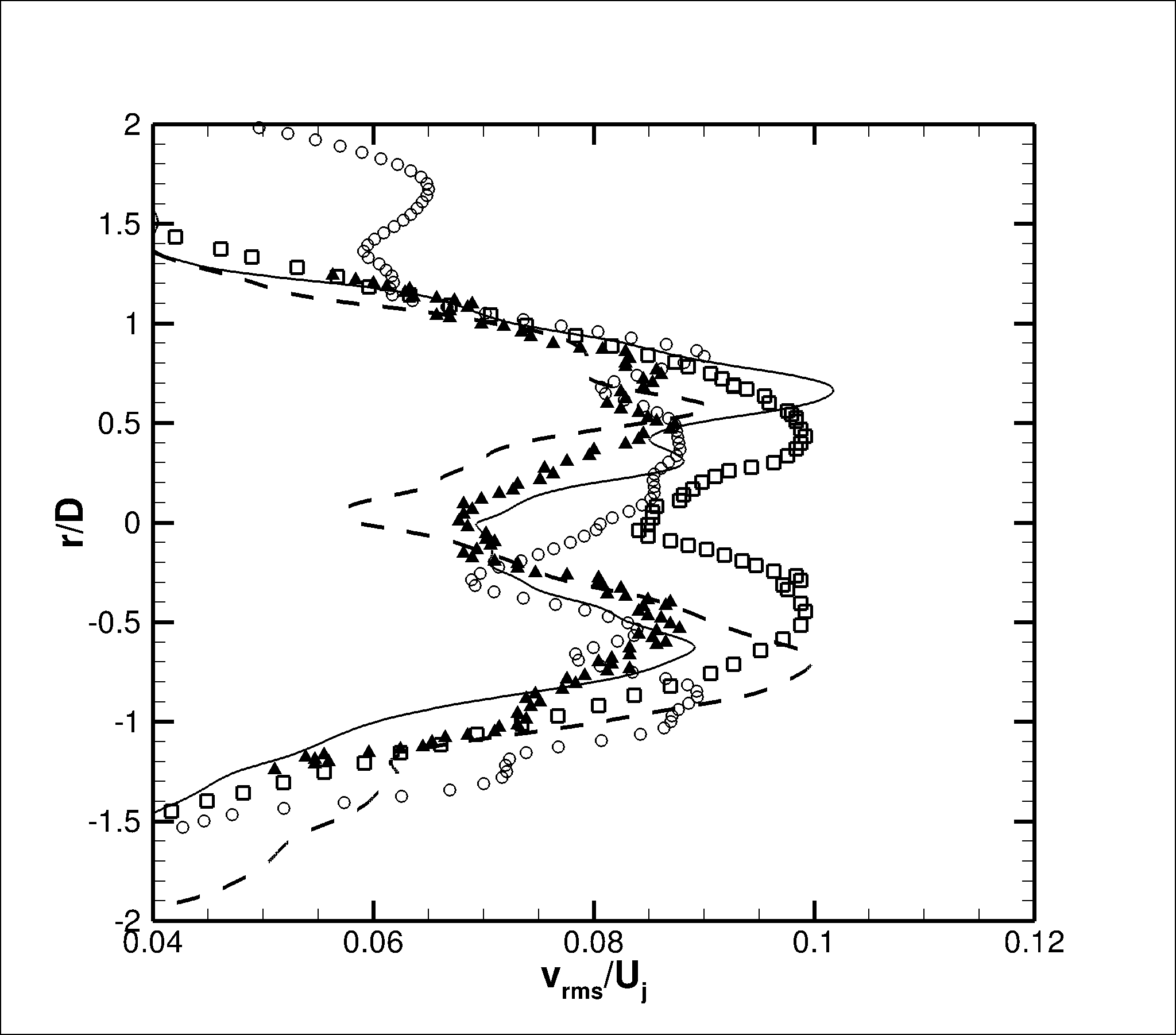}}
  \subfigure[$v^{*}_{RMS}$ - X=15D ; $-1.5D\leq Y\leq 1.5D$]
    {\includegraphics[width=0.45\textwidth]
	{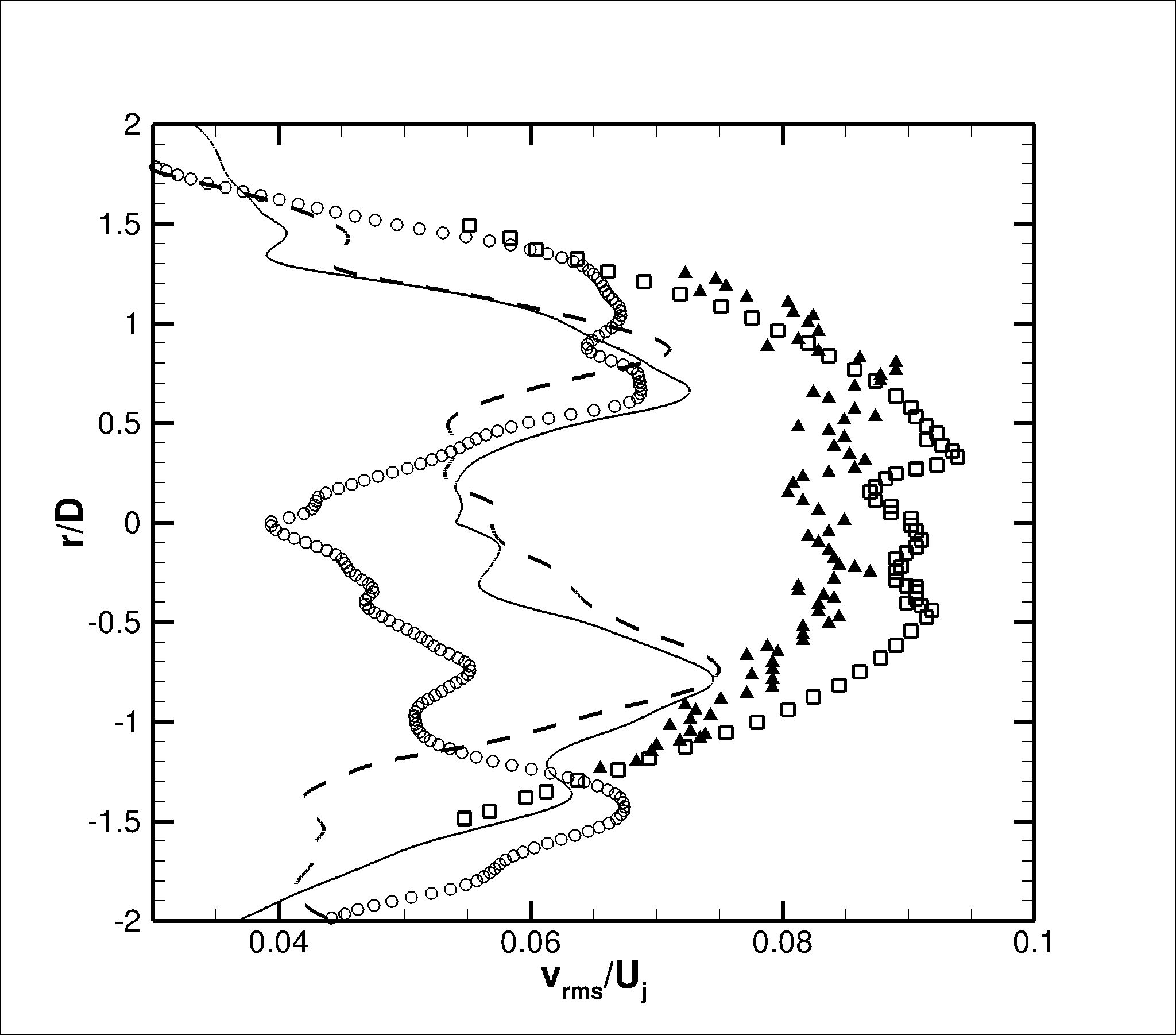}}
	\caption{Profiles of RMS of time fluctuation of radial 
	component of velocity, $v_{RMS}^{*}$, for S1 nd S2,
	at different positions within the computational domain.
	(\LARGE \textbf{--}\scriptsize), S2; 
	(\LARGE \textbf{-}\textbf{-}\scriptsize), S3;
	($\bigcirc$), S4; ($\square$), numerical data; 
	($\blacktriangle$), experimental data.}
	\label{fig:prof-v-rms-sgs}
\end{figure}

Four profiles of $v_{RMS}^{*}$ in the radial direction at $X=2.5D$, 
$X=5.0D$, $X=10.0D$ and $X=15.0D$ are presented in Fig.\ \ref{fig:prof-v-rms-sgs}. 
One can observe that, for $X\leq10.0D$, all the profiles calculated on S2, 
S3 and S4 are close to the reference. Moreover, the results of the static and the 
dynamic Smagorinsky models are in better agreement with experimental data than the LES
reference. At $X=15.0D$, all simulations performed in the current
work fail to predict the correct $v_{RMS}^{*}$ profile.

\FloatBarrier

%
%

\subsubsection*{Component of Reynolds Stress Tensor}

Figures \ref{subfig:uv-sgs-s2}, \ref{subfig:uv-sgs-s3} and 
\ref{subfig:uv-sgs-s4} present lateral view and profiles of 
$\langle u^{*}v^{*}\rangle$ component of the Reynolds stress tensor 
computed using three different SGS models, respectively. One can 
observe that the simulation performed using different SGS models have 
produced very similar distributions of $\langle u^{*}v^{*} \rangle$ 
for the region where the mehs is refined. However, for $X>8.0D$ the 
properties calculated by the different SGS closures present different 
behavior. The spreading is not the same for S2, S3 and S4 where 
$X>8.0D$. Therefore, one can state that the static Smagorinsky, 
the dynamic Smagorinsky and the Vreman models react differently to 
the coarsening of the grid.

\begin{figure}[htb!]
  \centering
  \subfigure[$\langle u^{*}v^{*}\rangle$ - X=2.5D ; $-1.5D\leq Y\leq 1.5D$]
    {\includegraphics[width=0.45\textwidth]
	{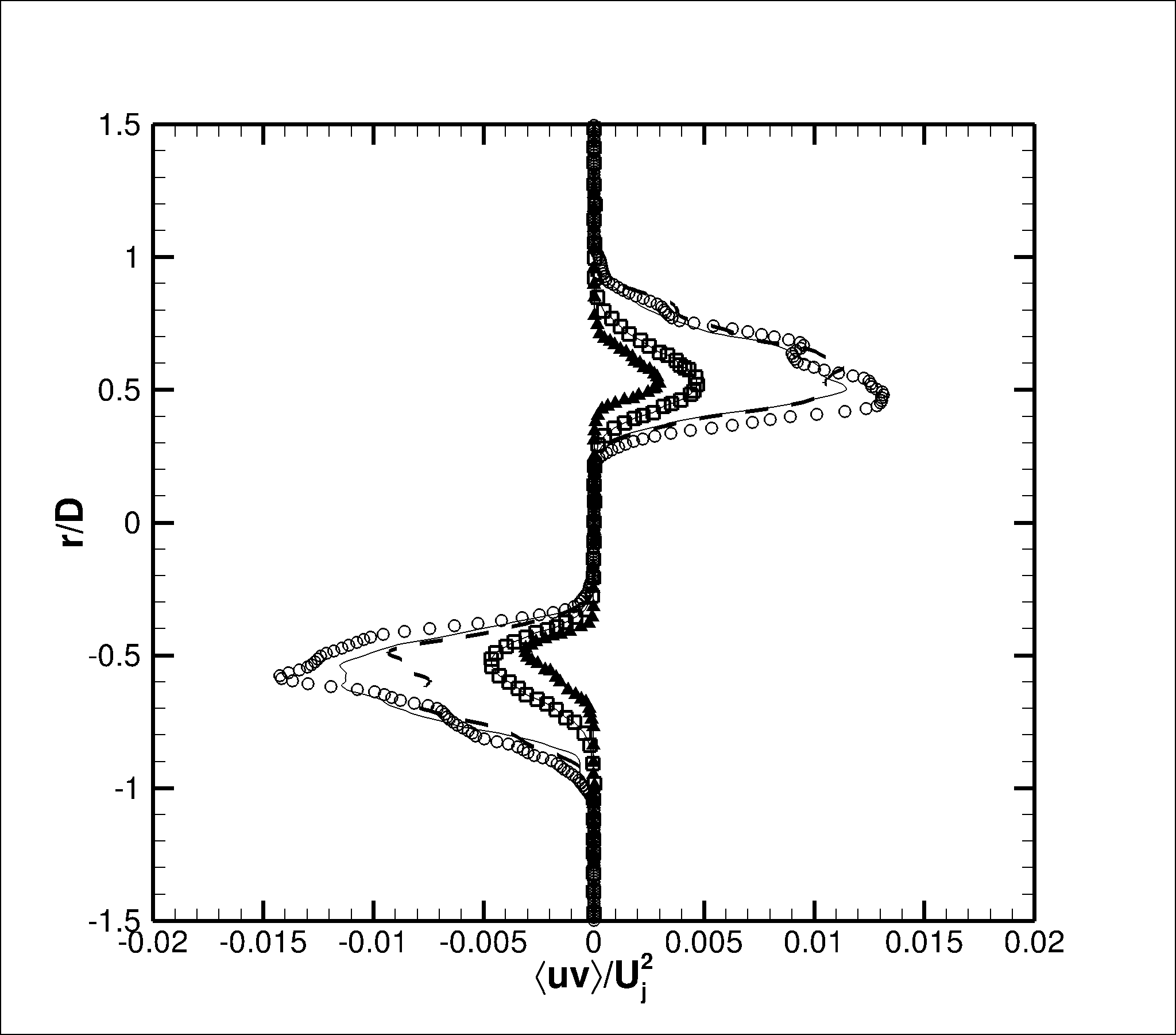}}
  \subfigure[$\langle u^{*}v^{*}\rangle$ - X=5.0D ; $-1.5D\leq Y\leq 1.5D$]
    {\includegraphics[width=0.45\textwidth]
	{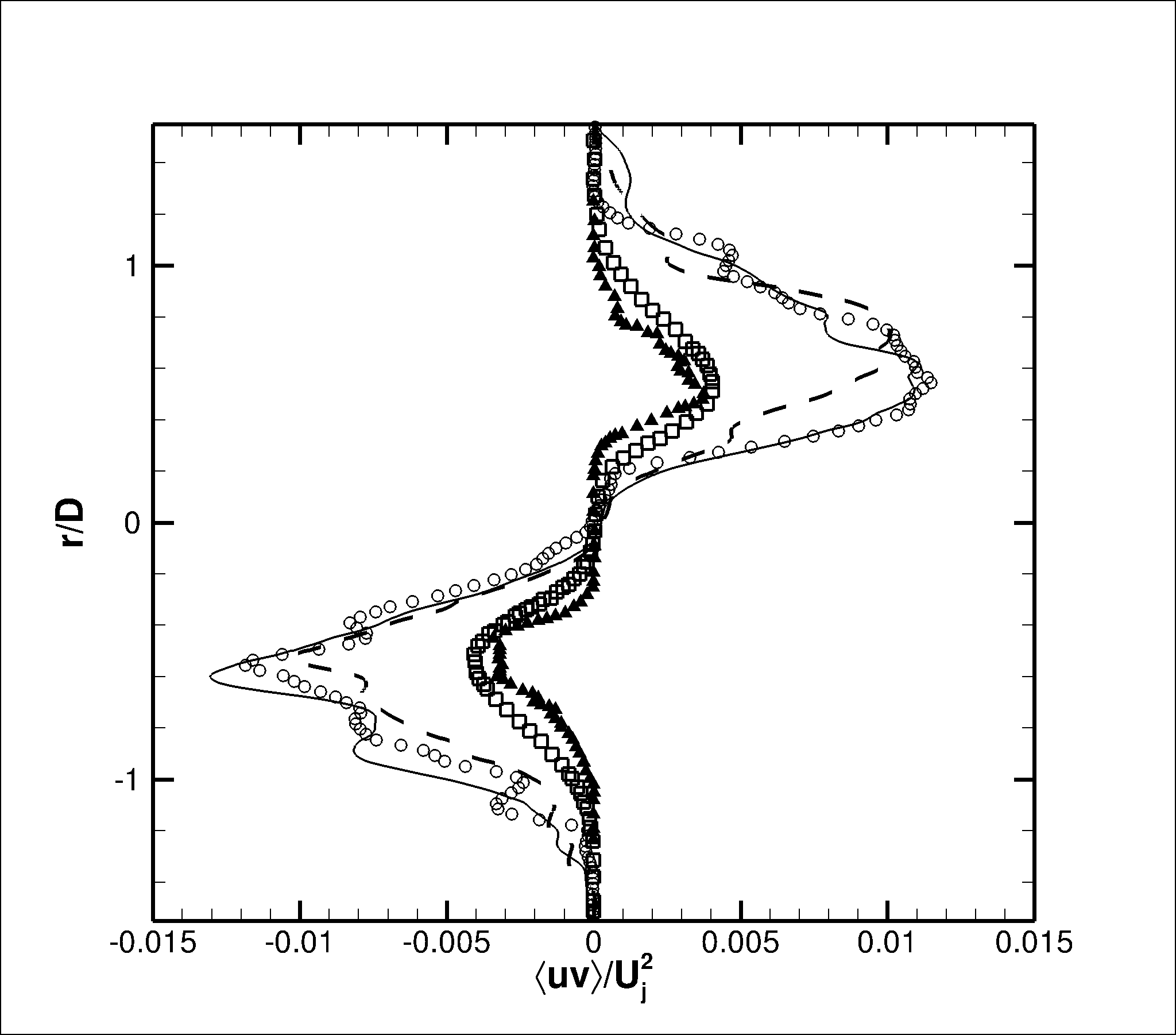}}
  \subfigure[$\langle u^{*}v^{*}\rangle$ - X=10D ; $-1.5D\leq Y\leq 1.5D$]
    {\includegraphics[width=0.45\textwidth]
	{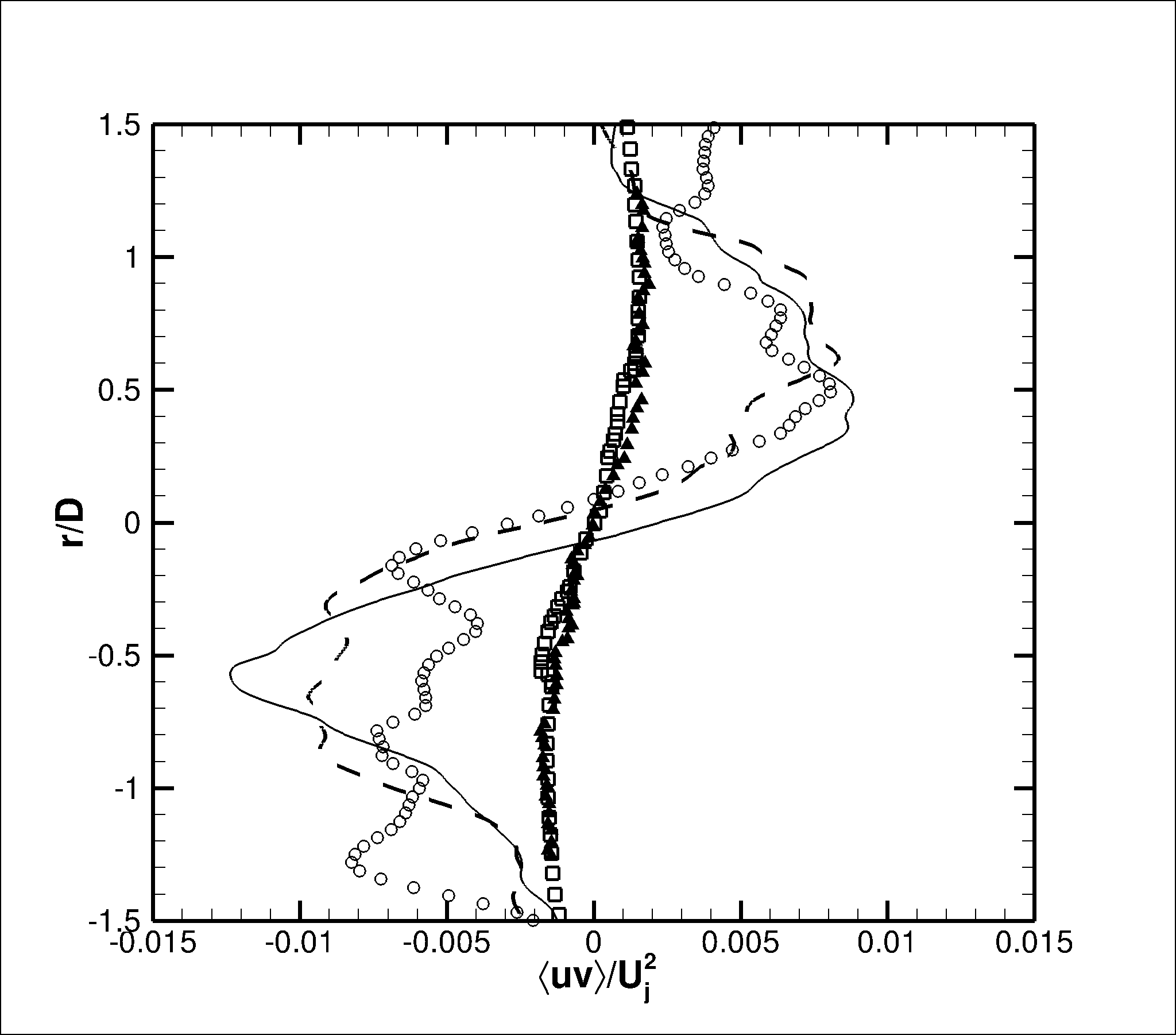}}
  \subfigure[$\langle u^{*}v^{*}\rangle$ - X=15D ; $-1.5D\leq Y\leq 1.5D$]
    {\includegraphics[width=0.45\textwidth]
	{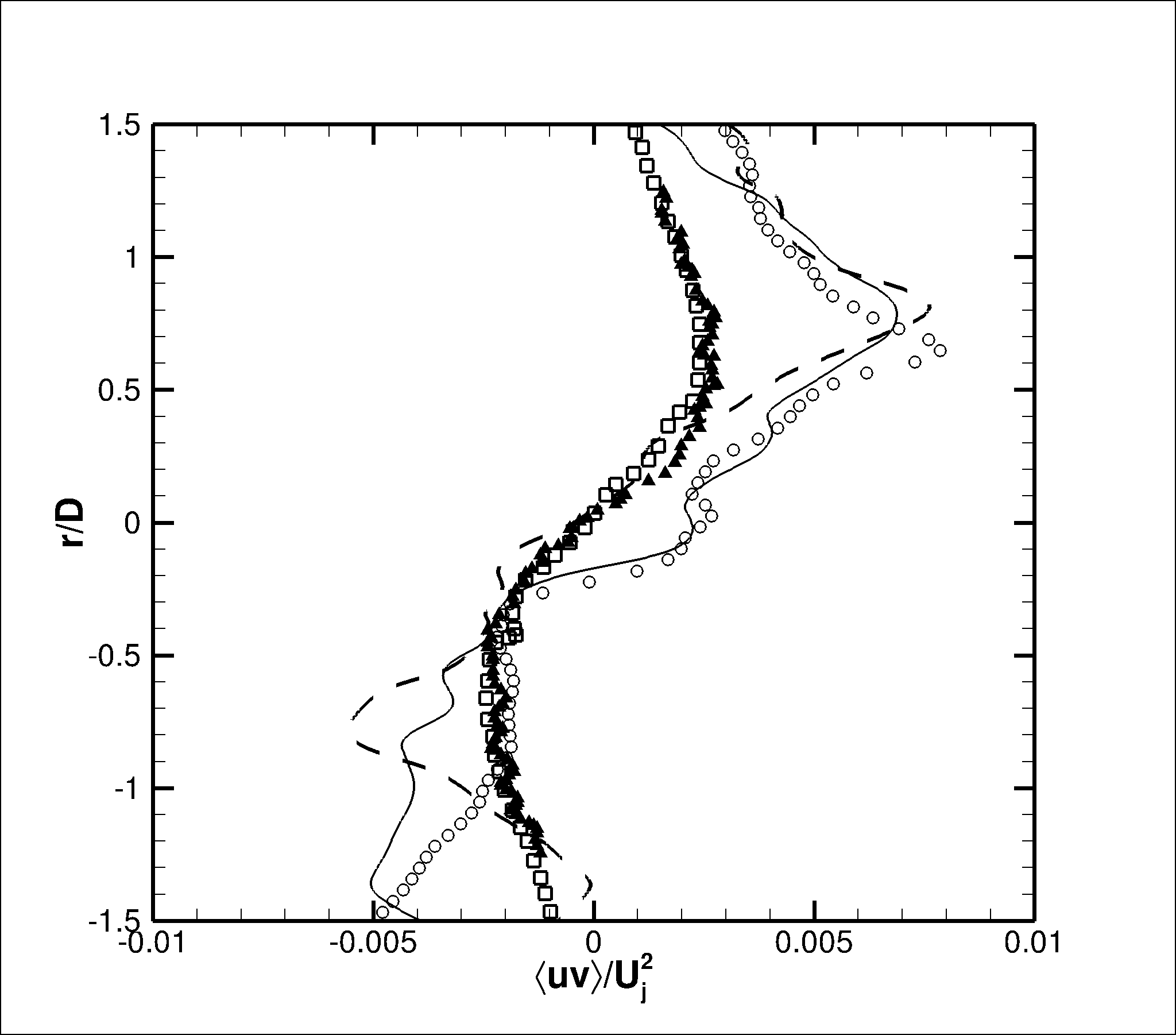}}
	\caption{Profiles of the $\langle u^{*}v^{*} \rangle$ 
	Reynolds shear stress tensor component, for S2, S3 and S4, at 
    different positions within the computational domain.
	(\LARGE \textbf{--}\scriptsize), S2; 
	(\LARGE \textbf{-}\textbf{-}\scriptsize), S3;
	($\bigcirc$), S4; ($\square$), numerical data; 
	($\blacktriangle$), experimental data.}
	\label{fig:prof-uv-av-sgs}
\end{figure}

All numerical simulations performed in the present work have failed
to correct predict the profiles of $\langle u^{*}v^{*} \rangle$ 
presented in Fig.\ \ref{fig:prof-uv-av-sgs}. The peaks of the component
of the Reynolds stress tensor does not correlate with the reference 
results. However, one should notice that the LES performed by the 
reference has also presented difficulties to calculate the same peaks.
The cause of the issue could be related to an eventual lack of grid
points in the radial direction. In spite of that, more studies on the
subject are necessary in order to understand such behavior.

\subsubsection*{Time Averaged Eddy Viscosity}

\begin{figure}[htb!]
  \centering
  \subfigure[$\langle\mu_{t}\rangle$ - S2]
    {\includegraphics[width=0.32\textwidth]
	{sources/results/pictures/stat-cs/XY-zoom/mut-av-zoom-XY}}
  \subfigure[$\langle\mu_{t}\rangle$ - S3]
    {\includegraphics[width=0.32\textwidth]
	{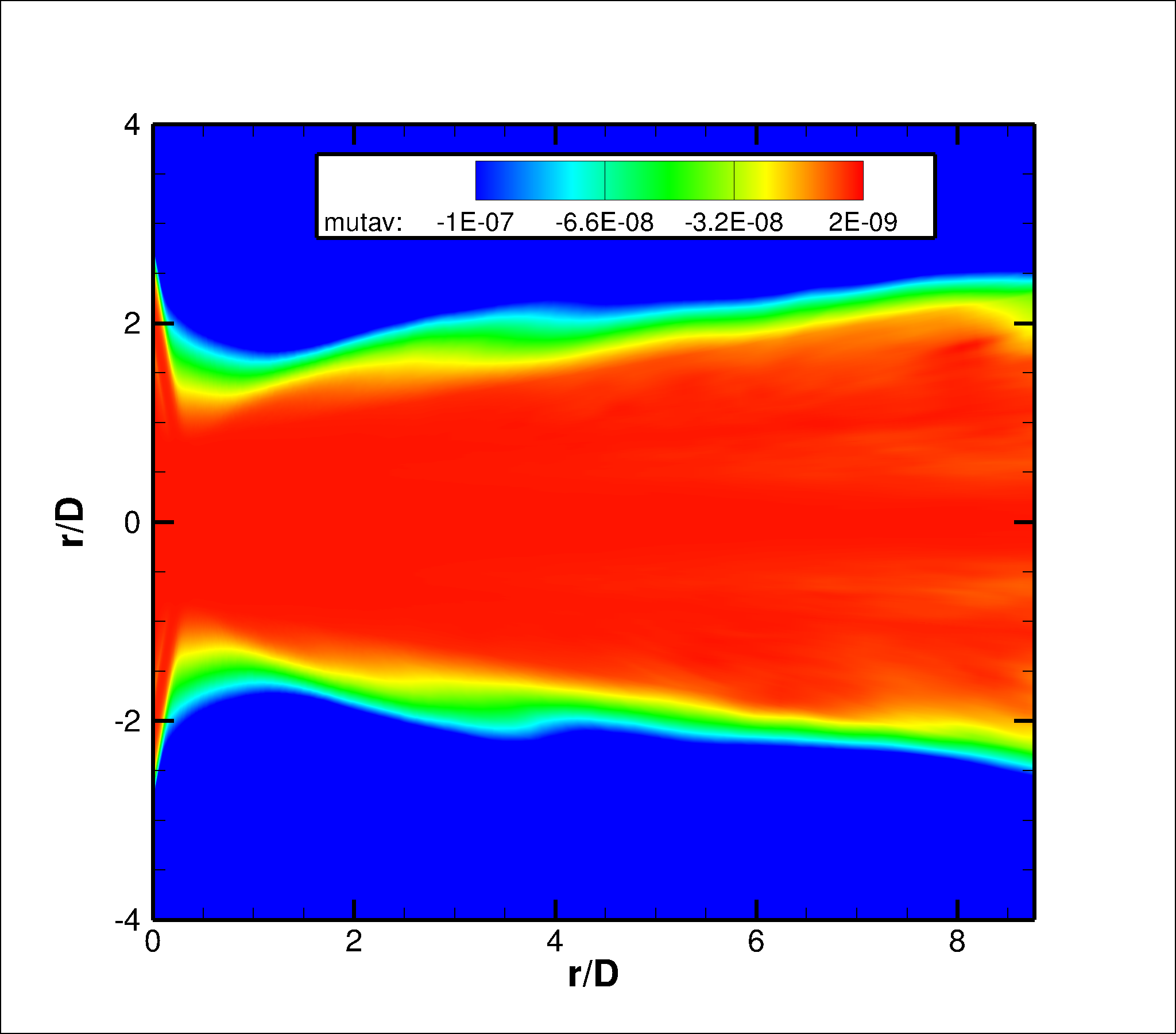}}
  \subfigure[$\langle\mu_{t}\rangle$ - S4]
    {\includegraphics[width=0.32\textwidth]
	{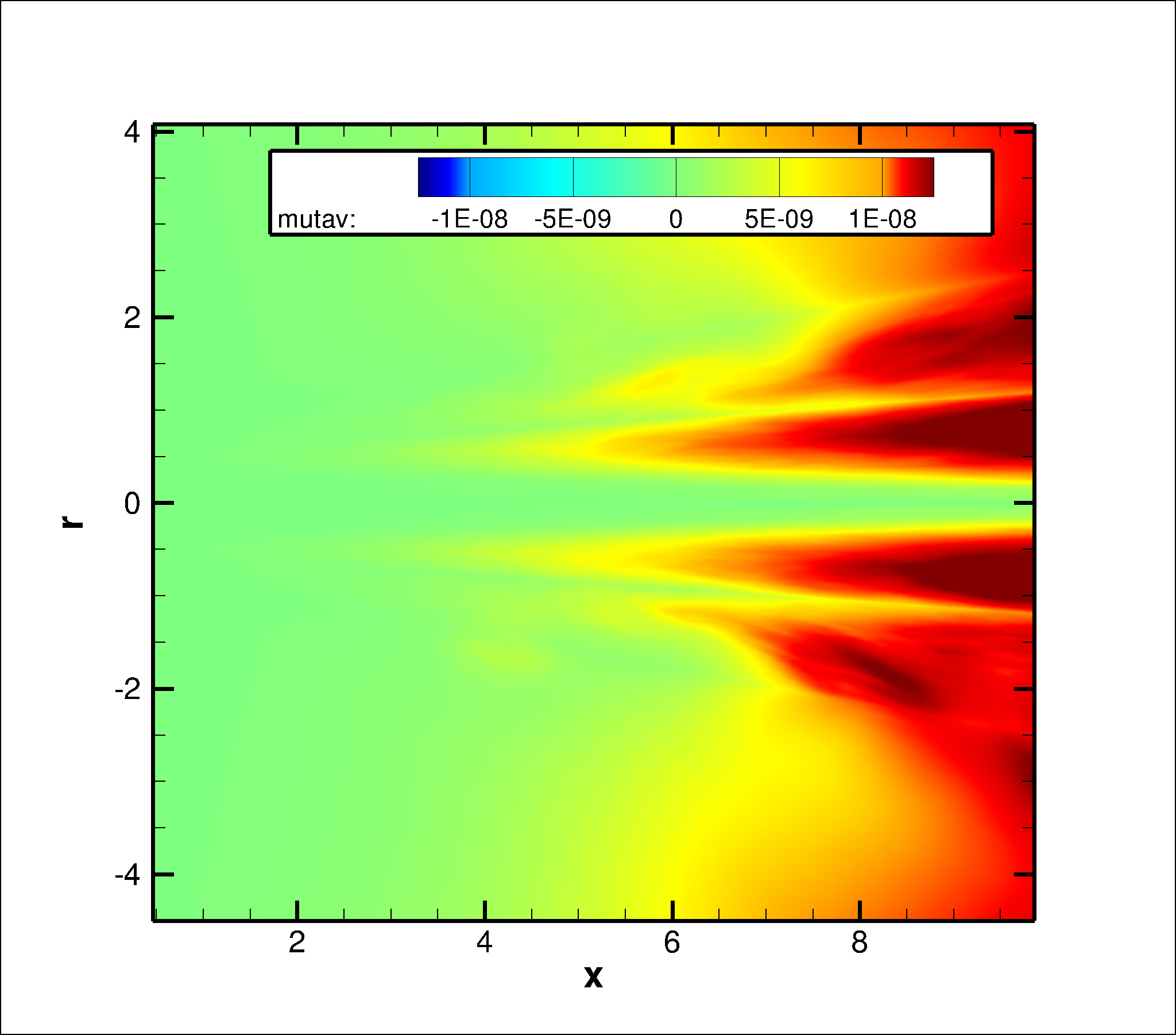}}
	\caption{Lateral view of time averaged eddy viscosity, 
	$\langle\mu_{t}\rangle$, for S2, S3 and S4.} 
	\label{fig:lat-mut-sgs}
\end{figure}

The distribution of the eddy viscosity, $\mu_{t}$, is discussed 
in the current subsection. Figure \ref{fig:lat-mut-sgs} presents 
distributions of time averaged eddy viscosity calculated using 
different SGS models. All subgrid scale closures used in the present 
work, the static Smagorinsky \cite{Smagorinsky63,Lilly65,Lilly67}, 
the dynamic Smagorinsky \cite{Germano91,moin91} and the Vreman 
\cite{vreman2004} models, are dependent of the local mesh size 
by design. This characteristic is exposed on the lateral view 
of the flow presented in Fig.\ \ref{fig:lat-mut-sgs}. The SGS 
models are only acting in the region where mesh presents a low 
resolution. Near the entrance domain, where the computational 
grid is very refined, the eddy viscosity can be neglected. 

The remark goes in the same direction of the work of Li and Wang
\cite{Li15}, which indicates that SGS closures introduce numerical 
dissipation that can be used as a stabilizing mechanism. However, 
this numerical dissipation does not necessarily add more physics 
of the turbulence to the LES solution. Therefore, in the present 
work, the numerical truncation, which generates the dissipative 
characteristic of JAZzY solutions, have show to overcome the 
effects of the SGS modeling. The mesh need to be very fine in order
to achieve good results with second order spatial discretizations.
The grid refinement generates very small grid spacing. Consequently, 
the SGS models, which are strongly dependent on the filter width, does
not affect much the solution. A LES of compressible flow configurations
without the use of SGS closure would be welcome in order to complete 
such discussion.

%

\subsubsection*{Power Spectral Density}

\begin{figure}[htb!]
  \centering
    {\includegraphics[width=0.56\textwidth]
	{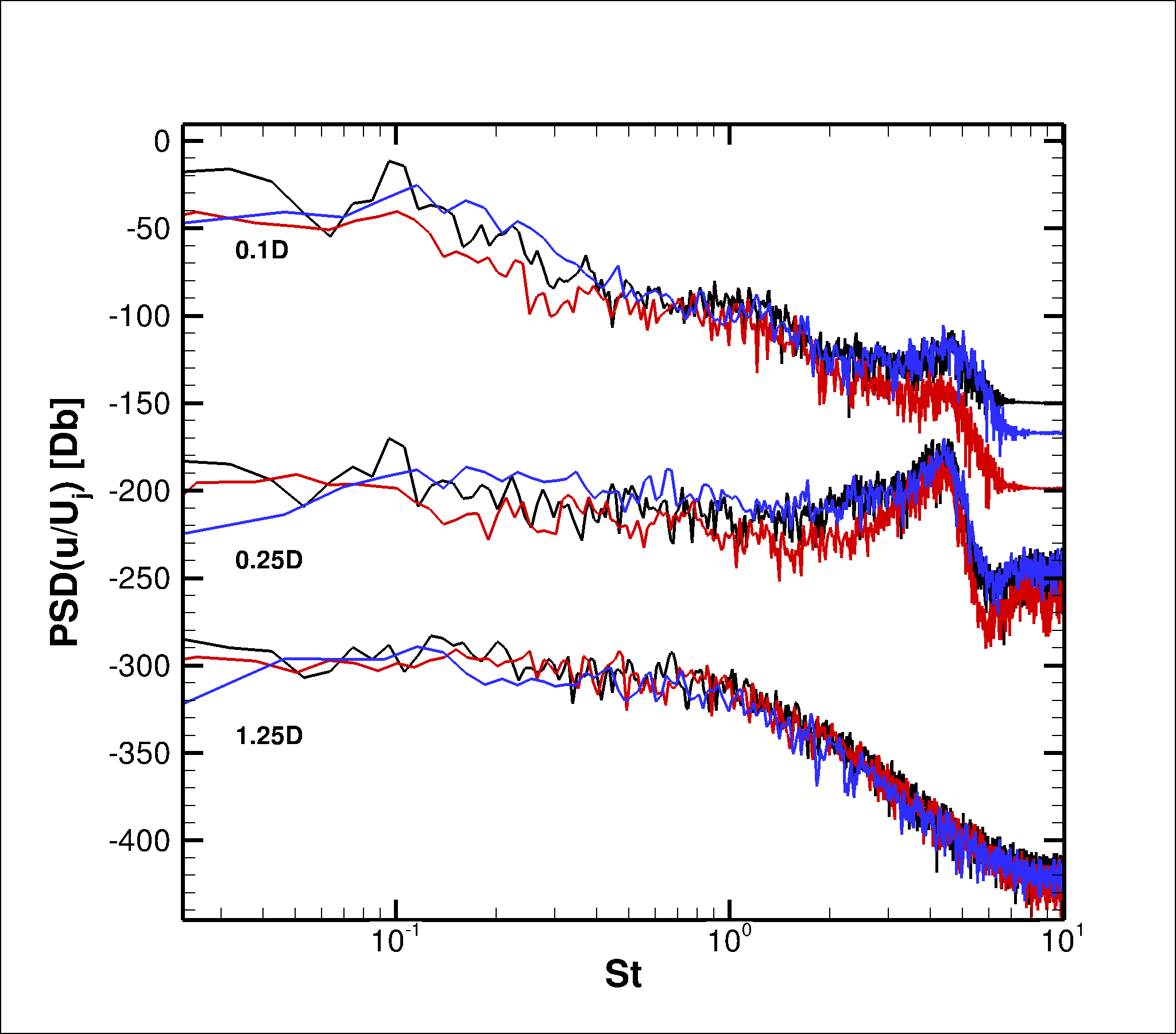}}
	\caption{Power spectral density of $u^{*}$ as function of the Strouhal
	number along the lipline of the jet.
    (\LARGE \textbf{--}\scriptsize), S2;
	(\LARGE {\color{red}\textbf{--}}\scriptsize), S3; 
	(\LARGE {\color{blue}\textbf{--}}\scriptsize), S4.
    A shift of -150 {\bf dB} and -300 {\bf dB} has been added to 
    the PSD in order to separate plots for $X=0.25D$ and $X=1.25D$, 
	respectively.}
	\label{fig:psd-sgs}
\end{figure}

The power spectral density of $u^{*}$ is studied in the comparison 
of SGS modeling. Figure \ref{fig:psd-sgs} presents the PSD of $u^{*}$, 
in $dB$, as function of the Strouhal number for S2, S3 and S4. The same 
methodology used on the mesh refinement study is performed here. The 
signals of $u^{*}$ are collected from the sensors allocated in the 
computational domain. The signals of Fig.\ \ref{fig:psd-sgs} 
are shifted of -150{\bf dB} and -300{\bf dB} for $X=0.25D$ and $X=1.25$, 
respectively, in order to separate plots. 

One can observe that PSD signals along the lipline obtained using S2, S3 
and S4 have shown the same behavior. A small difference can be noticed for
higher Strouhal number for the first two sensors, located at $X=0.1D$ and
$X=0.25D$. Such remark is aligned to the same discussion performed about 
the eddy viscosity for different SGS models. The sensors are located 
in the region where the mesh present excellent resolution. Therefore,
the effects of the static Smagorinsky, the dynamic Smagorinsky and the 
Vreman models, which are strongly dependent on the filter width, can be 
neglected on $u^{*}$ for $X<1.25D$.


\section{Concluding Remarks}

The current work is the study on effects of different 
subgrid scales models on perfectly expanded supersonic jet 
flow configurations using centered second-order spatial 
discretization. A formulation based on the the System I set 
of equations is used in the present work. The time integration 
is performed using a five-steps second order Runge-Kutta scheme.
Four large eddy simulations of compressible jet flows are performed 
in the present research using two different mesh configurations 
an three different subgrid scale models. Their effects
on the large eddy simulation solution are compared and discussed.

The mesh refinement study has indicated that in the region 
where the grid presents high resolution, the simulations
are in good agreement with experimental and numerical 
references. For the mesh with 14 million points the simulation 
has produced good results for $X<2.5D$ and $-1.5D<Y<0.5D$.
For the other mesh, with 50 million points, the simulations
provided good agreement with the literature for $X<5.0D$ and 
$-1.5D<Y<0.5D$. The eddy viscosity, calculated by the static
Smagorinsky model, presents very low levels in the region 
where the results have good correlation with the results of
the literature.

The refined grid used on the mesh refinement study, mesh B,
is used for the comparison of SGS models effects on the 
results of large eddy simulations. Three compressible jet
flow simulations are performed using the classic Smagorinsky
model \cite{Smagorinsky63,Lilly65,Lilly67}, the dynamic 
Smagorinsky model \cite{Germano91,moin91} and the Vreman model
\cite{vreman2004}. All three simulations presented similar 
behavior. Results presented good agreement with the reference 
for $X<5.0D$. In the region where the grid is very fine and 
the results correlates well with the literature, the eddy 
viscosity, provided by the SGS model, is very low values. 
The reason is related to the fact that the SGS closures used 
in the current work are strongly dependent of the filter width, 
which is proportional to the local mesh size. 

The numerical results indicated that it is possible to achieve
good results using second-order spatial discretization. The
mesh ought be well resolved in order to overcome the truncation 
errors from the low order numerical scheme. Very fine meshes 
originates very small filter width. Consequently, the effects 
of the eddy viscosity calculated by the SGS models on the
solution become unimportant. The work of Li and Wang \cite{Li15} 
have presented similar conclusions for simplified problems.
The authors indicate that SGS closures introduce numerical 
dissipation that can be used as a stabilizing mechanism. 
However, this numerical dissipation does not necessarily add 
more physics of the turbulence to the LES solution. Simulations 
without the use of any SGS model are welcome and could reinforce 
the argument.


\section*{Acknowledgments}

The authors gratefully acknowledge the partial support for this research provided by
Conselho Nacional de Desenvolvimento Cient\'ifico e Tecnol\'ogico, CNPq, under the
Research Grants No.\ 309985/2013-7, No.\ 400844/2014-1 and No.\ 443839/2014-0\@.
The authors are also indebted to the partial financial support received from 
Funda\c{c}\~{a}o de Amparo \`{a} Pesquisa do Estado de S\~{a}o Paulo, FAPESP, 
under the Research Grants No.\ 2008/57866-1, No.\ 2013/07375-0 and No.\ 2013/21535-0.

\FloatBarrier




%


\bibliography{sources/references}
\bibliographystyle{aiaa}

\end{document}